\RequirePackage{lineno}
\documentclass[aps,prd,longbibliography,notitlepage,superscriptaddress,nofootinbib]{revtex4-1}[11pt]
\usepackage{graphicx}
\usepackage{amssymb,mathrsfs}
\usepackage{dsfont}
\usepackage{bm}
\usepackage{slashed}
\usepackage{dcolumn}
\usepackage{hyperref}
\usepackage{amsmath}
\usepackage{float}
\usepackage[utf8]{inputenc}

\begin{document}

\title{Inclusive prompt photon production in electron-nucleus scattering at small $x$}

\author{Kaushik Roy}
\email{kaushik.roy.1@stonybrook.edu}
\affiliation{Dept. of Physics and Astronomy, Stony Brook University, Stony Brook, NY 11794, USA}
\affiliation{Physics Department, Brookhaven National Laboratory, Bldg. 510A, Upton, NY 11973, USA}

\author{Raju Venugopalan}
\email{raju@bnl.gov}
\affiliation{Physics Department, Brookhaven National Laboratory, Bldg. 510A, Upton, NY 11973, USA}

\date{\today}

\begin{abstract}
We compute the differential cross-section for inclusive prompt photon production in deeply inelastic scattering (DIS) of electrons on nuclei at small $x$ in the framework of the Color Glass Condensate (CGC) effective theory. The leading order (LO) computation in this framework resums leading logarithms in $x$ as well as power corrections to all orders in $Q_{s,A}^2/Q^2$, where $Q_{s,A}(x)$ is the nuclear saturation scale. This LO result is proportional to universal dipole and quadrupole Wilson line correlators in the nucleus. In the soft photon limit, the Low-Burnett-Kroll theorem allows us to recover existing results on inclusive DIS dijet production. The $k_{\perp}$ and collinearly factorized expressions for prompt photon production in DIS are also recovered in a leading twist approximation to our result. In the latter case, our result corresponds to  the dominant next-to-leading order (NLO) perturbative QCD contribution at small $x$.  We next discuss the computation of the NLO corrections to inclusive prompt photon production in the CGC framework. In particular, we emphasize the advantages for higher order computations in inclusive photon production, and for fully inclusive DIS, arising from the simple momentum space structure of the dressed quark and gluon ``shock wave" propagators in the ``wrong" light cone gauge $A^-=0$ for a nucleus moving with $P^{+}_{N} \rightarrow \infty$. 

\end{abstract}

\maketitle

%%%%%%%%%%%%%%%%%%%%%%%%%%%%%%%%
\section{Introduction}  \label{sec:intro}
%%%%%%%%%%%%%%%%%%%%%%%%%%%%%%%%

The measurement of isolated prompt photons in deeply inelastic scattering (DIS) provides a unique precision test of perturbative QCD (pQCD) with two distinct hard scales that are well understood from QED: the exchanged photon virtuality, $Q^{2}$ in the initial state and the transverse energy of the emitted prompt photon, $E_{\perp}^{\gamma}$. Prompt photon cross-sections were measured early on by the H1 and ZEUS experiments \cite{Breitweg:1999su,Aktas:2004uv,Chekanov:2006un} for the case of photoproduction, where the negative 4-momentum transfer squared, $Q^{2}=-q^{2}$ of the exchanged virtual photon is close to zero. Subsequently, the first measurements of prompt photon production in $e+p$ DIS, isolated and accompanied by jets, were performed by ZEUS and H1 for a wide range of $Q^2$~\cite{Chekanov:2004wr,Aaron:2007aa,Chekanov:2009dq,Abramowicz:2012qt}.

Isolated photons in DIS have proven to be clean and well calibrated probes of QCD dynamics~\cite{Gehrmann-DeRidder:2006zbx,Schmidt:2015zda}. We will explore here their potential for uncovering a novel gluon saturation regime of QCD~\cite{Gribov:1984tu,Mueller:1985wy} at small Bjorken $x$. This regime is characterized by the many-body recombination and screening dynamics of gluons that competes with the perturbative bremsstrahlung of increasing numbers of soft gluons at small $x$. An emergent dynamical scale $Q_{s,A}(x)$ screens color charges at increasingly smaller distances with decreasing $x$ thereby ensuring that the squared field strengths do not exceed $1/\alpha_S$, where $\alpha_S$ is the QCD coupling constant. The dynamics in this regime of QCD is fully nonlinear. Nevertheless, at sufficiently small $x$, or large enough nuclear size $A$, where $Q_{s,A}^2(x) \gg \Lambda_{\rm QCD}^2$, $\alpha_S \equiv\alpha_S(Q_{s,A}) \ll 1$. Many-body weak coupling techniques can therefore be employed to perform systematic computations in this nonlinear QCD regime~\cite{McLerran:1993ni,McLerran:1993ka,McLerran:1994vd}.

The rich dynamics of gluon saturation is captured in an effective field theory (EFT), the Color Glass Condensate (CGC)~\cite{McLerran:1993ni,McLerran:1993ka,McLerran:1994vd,Iancu:2003xm,Gelis:2010nm,Kovchegov:2012mbw,Blaizot:2016qgz}, and we will apply it here to compute the inclusive photon cross-section in DIS off nuclei.  The CGC EFT is formulated in the infinite momentum frame \cite{Fubini:1964boa,Bardakci:1969dv,Weinberg:1966jm,Susskind:1967rg} of the nucleus with an appropriate choice of gauge. It relies on the separation, in the longitudinal momentum fraction $x$, of the degrees of freedom into static color sources at large $x$ coupled eikonally to dynamical ``wee" gluon fields at small $x$~\cite{Gelis:2010nm}.  Because a large number of large $x$ sources couple to the wee gluons, the color charge of the sources can lie in any one of a number of higher dimensional color representations of the $SU(3)$ algebra~\cite{McLerran:1993ni,Jeon:2004rk}. This results in a stochastic distribution of classical color sources over a gauge  invariant weight functional $W_{Y_{A}}[\rho_{A}]$ representing the probability density of the color charge density $\rho_{A}(x^{-},\mathbf{x}_{\perp})$. The subscript $Y_{A}=\text{ln}(x_0/x)$  denotes the spacetime rapidity separation of small $x$ target gluons relative to those at $x\equiv x_0$ corresponding to the nuclear beam rapidity, $Y_{beam}$.

The expectation value of any physical operator at rapidity $Y_{A}$ is determined by the average over all possible color charge configurations: 
\begin{equation}
\left \langle \mathcal{O} \right \rangle_{Y_{A}} =\int [\mathcal{D} \rho_{A} ] \enskip W_{Y_{A}} [\rho_{A}]   \mathcal{O} [\rho_{A}] \enskip,
\label{eq:expectation-value-O}
\end{equation}
where $\mathcal{O} [\rho_{A}]$ is the expectation value of the operator for a particular configuration $\rho_{A}$ of color sources. In the problem of interest, this operator is the  differential cross-section for inclusive photon production for a given $\rho_A$. 
Requiring that physical observables be independent of the arbitrary scale separation between sources and fields results in the JIMWLK functional renormalization group equation \cite{JalilianMarian:1997jx,JalilianMarian:1997dw,Iancu:2001ad,Iancu:2000hn,Ferreiro:2001qy}
\begin{equation}
\frac{\partial
W_{Y_{A}}[\rho_{A}]}{\partial Y_{A}} =\mathcal{H} \big[ \rho_{A},\frac
{\delta}{\delta \rho_{A}} \big] \enskip W_{Y_{A}} [\rho_{A}] \enskip . 
\label{eq:JIMWLK}
\end{equation}
This equation describes the evolution of the distribution of color charges in the nuclear wavefunction from its fragmentation region at large $x$ (or small $Y_{A}$) to the small $x$ (or large $Y_{A}$) of interest as determined by the kinematics of the process. Here $\mathcal{H}$ is the JIMWLK Hamiltonian; explicit expressions and properties of the Hamiltonian are discussed for instance in \cite{Iancu:2003xm,Weigert:2005us}. It is worthwhile to note here that the JIMWLK evolution equation can alternatively be expressed as the Balitsky-JIMWLK hierarchy~\cite{Balitsky:1995ub,Weigert:2000gi} of equations for expectation values of $n$-point Wilson line correlators. In the limit of large number of colors $N_c \rightarrow \infty$ and for large nuclei $A\rightarrow \infty$, the closed form Balitsky-Kovchegov (BK) equation is obtained for the two-point correlator of Wilson lines~\cite{Balitsky:1995ub,Kovchegov:1999yj}. The BK equation is a good approximation in many practical situations to the full Balitsky-JIMWLK hierarchy~\cite{Rummukainen:2003ns,Dumitru:2011vk} and is therefore extremely useful in phenomenological studies at collider experiments. 

An attractive feature of the CGC, as in any EFT, is that there is a well defined power counting for systematic higher order computations. As we will show explicitly later for our specific case, this power counting allows one to match results in the CGC framework to those in the collinear factorization and $k_\perp$ factorization perturbative frameworks in appropriate kinematic limits. For $e+A$ and $p+A$ collisions, the appropriate CGC power counting is in a so-called ``dilute-dense" limit~\cite{Gelis:2010nm} where  dilute color source density $\rho_p$ in the generic projectile is of order $g \ll  1$ in the QCD coupling, while the density of color sources $\rho_A\sim 1/g$ in the nuclear target. Strictly speaking, the dilute-dense power counting corresponds to computations that are lowest order in the dimensionless ratios $Q_{s,p}^2/k_{p,\perp}^2 \ll 1$, and all orders in $Q_{s,A}^2/k_{A,\perp}^2\sim 1$, where $k_{p,\perp}$ and $k_{A,\perp}$ correspond to the momentum transfer from the projectile and target respectively. The dilute-dense power counting was implemented recently by one of us and collaborators in computing inclusive photon production in $p+A$ collisions to next-to-leading order (NLO) accuracy~\cite{Benic:2016uku} extending earlier leading order (LO) computations in this context~\cite{Gelis:2002ki}. 

\begin{figure}[H]
\begin{center}
\includegraphics[scale=0.2]{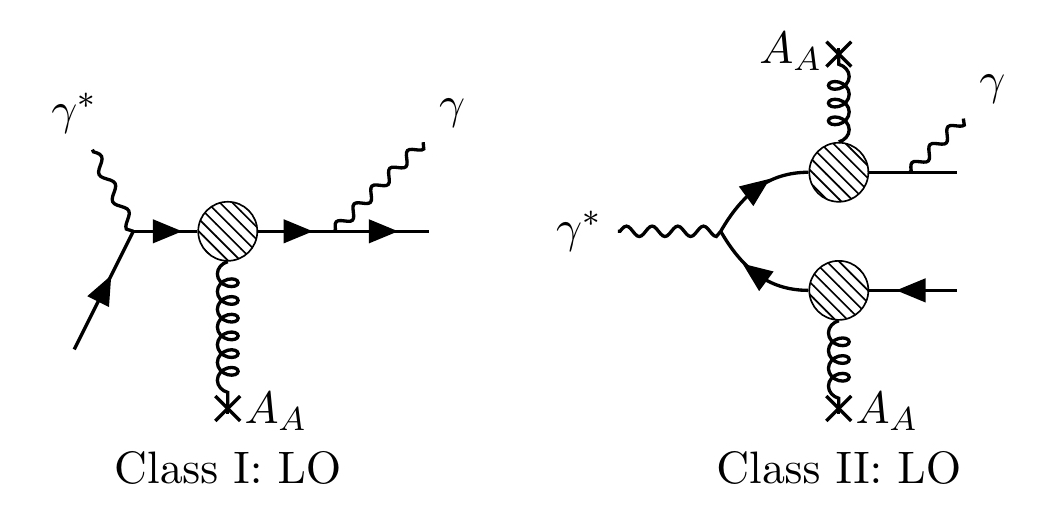}
\caption{Representative diagrams for leading order (LO) processes contributing to inclusive prompt photon production in DIS. For Class II processes, the photon can be emitted from either fermion line. The blobs represent the resummed eikonal interactions between the quark/antiquark and the classical color field of the nucleus. \label{fig:classes}} 
\end{center}
\end{figure}

The power counting in the CGC framework for the $e+A$ DIS case at hand is simpler than the $p+A$ case because we don't have a power counting in $\rho_p$ on account of the lepton probe. In general, at LO in the QCD coupling constant, we obtain the two classes of processes shown in Fig. \ref{fig:classes} that contribute to the inclusive prompt photon cross-section.

  The Class I processes represent the bremsstrahlung of a photon from a valence quark in the wavefunction of the target nucleus. Their contribution to the differential cross-section for inclusive photon production has been calculated \cite{Gelis:2002ki,Dominguez:2011wm} at small $x$ in the CGC framework albeit in the context of $p+A$ collisions. This result can however be straightforwardly adapted to the present problem. 

The Class II processes correspond to configurations that become important at small $x$ where the virtual photon emitted by the electron 
fluctuates into a long lived quark-antiquark dipole~\cite{Gribov:1965hf,Ioffe:1969kf} and the dipole subsequently has a nearly instantaneous ``shock wave" eikonal scattering off the gauge fields in the nuclear target~\cite{Bjorken:1970ah,Nikolaev:1990ja,Kovchegov:2012mbw}. In this dipole picture of $e+A$ DIS, the power counting is dictated by strong color sources $\rho_{A} \sim 1/g$ in the target, and their energy evolution with respect to the quark-antiquark dipole.  Attaching new sources to the diagram in Fig. \ref{fig:classes} does not change the order in strong coupling constant, because $g\rho_{A} \sim 1$. Independent powers of $g$ arise only when a vertex is not connected to the source. From this argument, it should be clear that we are performing an all-twist expansion in $g\rho_{A}$ at each order in $\alpha_{S}$ in this power counting scheme\footnote{Note that there is also an LO contribution analogous to that described in \cite{Benic:2016yqt} whereby the real photon is emitted from a quark loop that the virtual photon fluctuates into. Just as in the $p+A$ case, this contribution is highly suppressed relative to the Class II processes we will discuss.}. 

\begin{figure}[!htb]
 \begin{center}
 \includegraphics[scale=0.2]{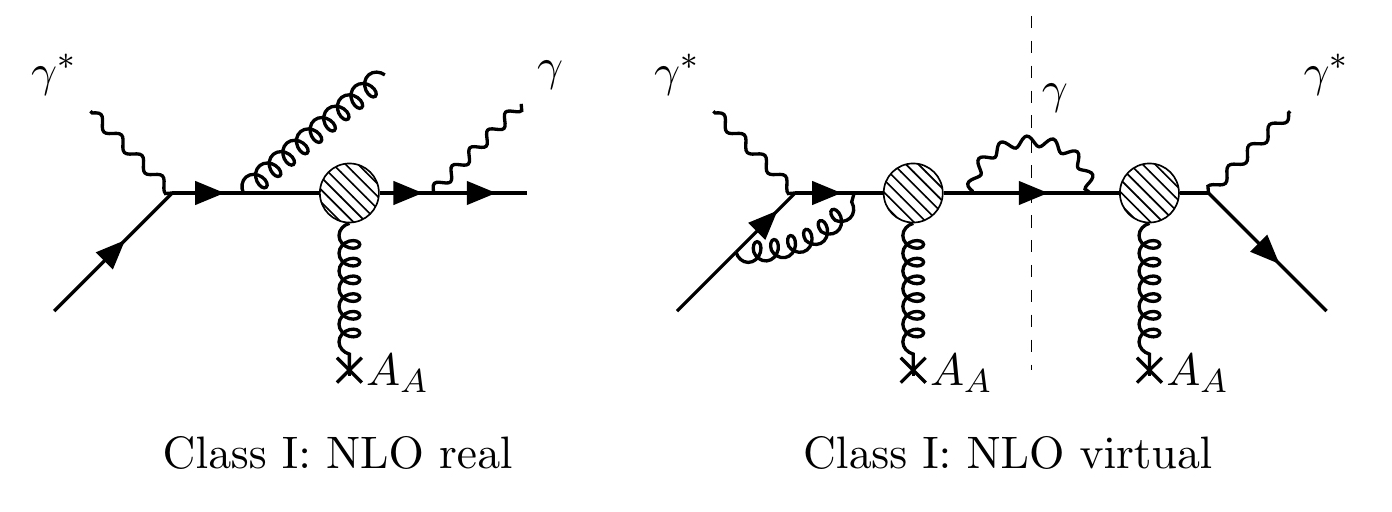}
 \caption{Representative diagrams for the Class I processes that contribute at NLO to inclusive photon production. Both real emission and interference contributions are shown.} \label{fig:classes-NLO}
 \end{center}
 \end{figure}

At NLO, typical contributions\footnote{For a recent computation of the real gluon emission contribution, see \cite{Altinoluk:2018uax}.} to Class I processes are shown in Fig. \ref{fig:classes-NLO}. For inclusive photon production, these possess the divergence structure inherent in NLO quark production containing i) dominant contributions from the large phase space in transverse momentum available to the emitted gluon ($\alpha_{S}\text{ln}(k_{\perp})\sim 1$), ii) subdominant contribution from logarithms sensitive to $x$ (since valence quark distributions are peaked at $x\sim 1$) and finally, iii) finite pieces that are not phase space enhanced. By an appropriate choice of scale and scheme for factorization, the finite pieces can be absorbed in the definition of the quark distribution function. These contributions are therefore effectively of order $\mathcal{O}(\alpha_{e})$ (where $\alpha_{e}$ is the QED fine structure constant) if we replace the bare valence quark distribution in the nucleus by one that absorbs the leading logarithmic contributions. These last contributions, to all orders in perturbation theory, are captured by the DGLAP~\cite{Dokshitzer:1977sg,Gribov:1972ri,Altarelli:1977zs} renormalization group (RG) evolution of the valence quark distributions.

At small $x$, the Class I contributions are strongly suppressed relative to the Class II contribution. The reason for this is that the former are sensitive to the valence quark distribution in the target while the latter are sensitive to the gluon distribution. At small $x$, as seen in the  $e+p$ HERA DIS data~\cite{Abt:1993cb,Ahmed:1995fd,Derrick:1993fta,Derrick:1994sz,Martin:1994kn,Lai:1994bb}, the gluon distribution clearly dominates the valence quark distribution. 
Since our interest is in inclusive photon production at small $x$, we shall not examine Class I type processes any further and shall focus our attention exclusively on Class II processes. 

 %%%%%%%%%%%%%%%%%%%%%%%%%%%%%%%%%%%%%%
In the first part of this paper, we will compute the Class II LO diagrams shown in Fig. \ref{fig:classes}. We will subsequently discuss the structure of the NLO computation of inclusive photon production. We  shall in particular discuss a simple formulation of the dressed gluon propagator in light cone gauge. The NLO computations are important for two reasons. Firstly, as noted in our discussion of Class I NLO contributions, there are logarithmically enhanced contributions of order $\alpha_S \ln(x_0/x)$ which can be as large as the LO contributions. Such contributions appear in each order of perturbation theory; they can be isolated and resummed in a so-called leading logarithmic (LLx) RG treatment. There are also pure $\alpha_S$ suppressed NLO contributions. These are important for precision measurements but can also lead to qualitative changes in spectra that impact the discovery potential of such measurements. In particular, they will be important for a quantitative extraction of the saturation scale $Q_{s,A}(x)$. 

These statements can be understood more clearly by considering the organization of the perturbative series for the argument of $\langle \mathcal{O}\rangle$ in Eq.~\ref{eq:expectation-value-O} as\footnote{An extended discussion along these lines for field theories with strong time dependent sources can be found in \cite{Gelis:2006yv,Gelis:2006cr}; for the CGC specifically, see~\cite{Gelis:2007kn,Gelis:2008rw,Gelis:2008ad}.}
\begin{equation}
\mathcal{O}[\rho_{A}]= \sum_{n=0} c_{n}\alpha_{S}^{n} \enskip.
\label{eq:power-counting}
\end{equation}
Here each coefficient $c_{n}=\sum_{j=1}^{\infty} \xi_{nj} (g\rho_{A})^{j}$, where the matrix elements $\xi_{nj}$ are numbers of order unity, resums the contributions obtained by adding extra sources of magnitude $\rho_A\sim 1/g$ to the allowed graphs.
Before we proceed any further, we must invoke the essential ingredient of the CGC effective field theory -- the separation of large $x$ static light cone sources from dynamical small $x$ gauge fields -- as represented by the structure of Eq.~\ref{eq:expectation-value-O}. 

\begin{figure}[!htb]
\begin{center}
\includegraphics[scale=0.25]{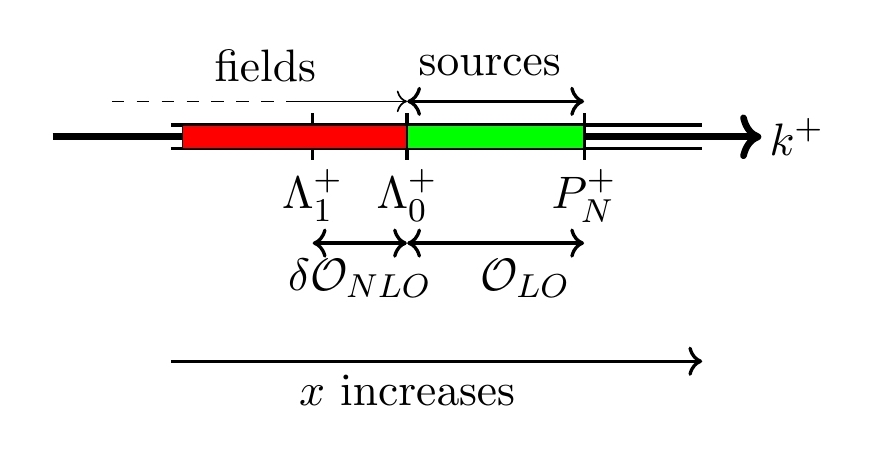}
\caption{Schematic illustration of sources and fields in the CGC effective theory in terms of the cutoff scale $\Lambda_{0}^{+}$ and equivalently rapidity, $Y_{0}$. At LO, this separation is arbitrary. At NLO, contribution from field modes in the range $\Lambda_1^+ < k^+ <\Lambda_{0}^{+}$, such that $\alpha_S\ln(\Lambda_0^+/\Lambda_1^+) \equiv \alpha_S\Delta Y < 1$, are integrated out and absorbed into the source densities at the scale $\Lambda_1^+$. This self-similar renormalization group (RG) pattern is repeated successively generating the JIMWLK RG equation for the source densities. See text for a detailed discussion. \label{fig:RG-cartoon}} 
\end{center}
\end{figure}

The starting point of any CGC computation therefore includes an initial cutoff scale $\Lambda_{0}^{+}$ (in the $`+$' longitudinal momentum) or $Y_{0}$ (in the rapidity) between sources and fields.  This is shown schematically in  Fig.~\ref{fig:RG-cartoon}. At leading order (LO), the cutoff scale $\Lambda_{0}^{+}$ (or $Y_{0}=\ln(\Lambda_{\rm beam}/\Lambda_0^+)$) distinguishing soft and hard partons is arbitrary and the fast quantum modes with $k^{+} \gg \Lambda^{+}_{0}$ or $Y < Y_{0}$ are represented by the classical color source density $\rho_{A}$ with the weight functional $W_{\Lambda_{0}^{+} (Y_{0})} [\rho_{A}]$.  The quantum evolution of sources and fields is described by the following renormalization group procedure (RG).  One first integrates out quantum fluctuations within the range $\Lambda_1^{+}< \vert k^{+} \vert < \Lambda_0^+$. Here $\Lambda_1^+$ is chosen such that $\alpha_S\ln(\Lambda_0^+/\Lambda_1^+) \equiv \alpha_S \Delta Y<1$, where $\Delta Y = Y_1-Y_0$.  

\begin{figure}[!htb]
 \begin{center}
 \includegraphics[scale=0.22]{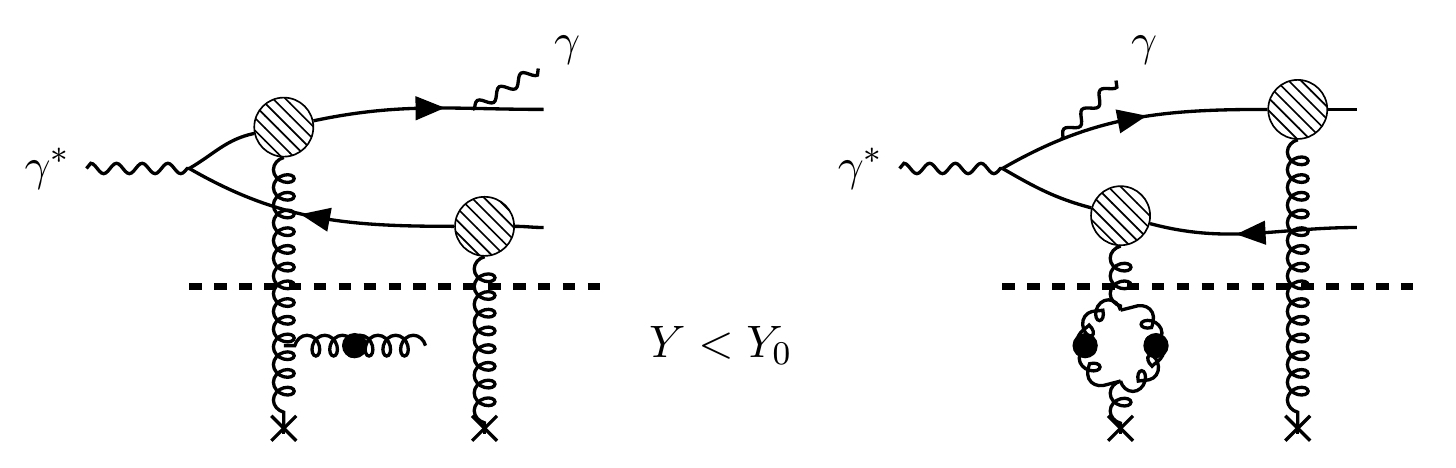}
 \caption{Representative Feynman graphs contributing at NLO to the evolution of color sources. In the CGC effective theory, the integration over gluon field modes with rapidity $Y < Y_{0}$ is given by contributions from these classes. These contributions are actually $O(1)$ in magnitude because of the presence of large logarithms in $x$ leading to $\alpha_{S}\text{ln} (1/x) \sim 1$.  \label{fig:NLO-LLx} }
 \end{center}
\end{figure}

For our computation of inclusive photon production at small $x$, typical NLO diagrams that generate such logarithms upon change of scale from $\Lambda_0^+$ to $\Lambda_1^+$ are shown in Fig.~\ref{fig:NLO-LLx}. Their contribution can be absorbed into the evolution of the weight functional describing the distribution of color sources:
\begin{align}
W_{\Lambda_{1}^{+}}[\rho_{A}]&=\Big( 1+\text{ln}(\Lambda_{0}^{+}/\Lambda_{1}^{+}). \mathcal{H} \Big) W_{\Lambda_{0}^{+}} [\rho_{A}] \quad \text{or equivalently,} \nonumber \\
W_{Y_{1}}[\rho_{A}]&=\Big( 1+\Delta Y . \mathcal{H} \Big) W_{Y_{0}} [\rho_{A}] \enskip.
\end{align}
Here $\mathcal{H}$ is the JIMWLK Hamiltonian we alluded to previously. Hence these particular NLO $\delta \mathcal{O}$ contributions generate a classical effective theory at this new scale expressed as
\begin{equation}
\int [\mathcal{D} \rho_{A} ] \enskip W_{\Lambda_{0}^{+}} [\rho_{A}] (\mathcal{O}_{LO}+\delta \mathcal{O}_{NLO})=\int   [\mathcal{D} \rho_{A} ] \enskip W_{\Lambda_{1}^{+}}[\rho_{A}] \mathcal{O} [\rho_{A}] \enskip ,
\end{equation} 
where the LLx contributions have been absorbed in the JIMWLK evolution of $W$. To leading order accuracy, the perturbative expansion in Eq.~\ref{eq:power-counting} then has the coefficients:
 \begin{equation}
 c_{n}=\sum_{j=0}^{\infty}  d_{nj} (g \rho_A)^j \Big( \text{ln}(1/x) \Big)^{n}, \qquad n=1,2, \ldots \enskip ,
 \end{equation}
 
\begin{figure}[!htb]
\begin{center}
\includegraphics[scale=0.2]{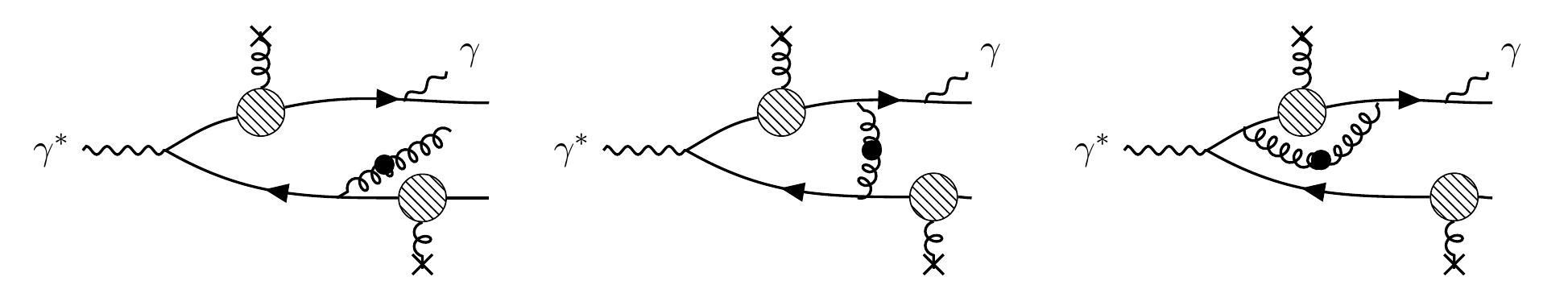}
\caption{Some of the Feynman graphs contributing at NLO to inclusive photon production in $e+A$ DIS at small $x$. These generate  O($\alpha_S$) corrections to inclusive photon production.} \label{fig:NLO-coeff}
\end{center}
\end{figure}
% \begin{figure}[H]
% \begin{center}
% \includegraphics[scale=0.25]{Fig29.jpg}
% \caption{Simplified distinction of sources and fields in the CGC effective theory in terms of longitudinal momentum cutoff scale $\Lambda_{0}^{+}$ and equivalently rapidity, $\eta_{0}$. There is no contribution, at leading order (LO), from the fluctuations of the background field to the quantum evolution of the color source distribution.}
% \end{center}
%\label{fig:RG-cartoon}
%\end{figure}

At NLO however, there are $\alpha_S$ contributions that do not come accompanied with logarithms in $x$. These would then correspond to coefficients expressed more generally in the expansion of the perturbative series as 
\begin{equation}
 c_{n}=\sum_{i=0}^n \sum_{j=0}^{\infty} f_{nj}^i (g\rho_A)^j \Big( \text{ln}(1/x) \Big)^{i}, \qquad n=1,2,\ldots \enskip ,
 \label{eq:coeff-expansion}
 \end{equation}
in particular the coefficients $f_{nj}^{n-1}$. For our process of interest, some of the the corresponding diagrams are shown in Fig.~\ref{fig:NLO-coeff}. These additional contributions are not rapidity ordered relative to the leading order contribution; the phase space integrals therefore do not generate logarithms in $x$. Thus while the NLO diagrams represented in Fig.~\ref{fig:NLO-LLx} can be absorbed into the LO order computation of photon production accompanied by leading logarithmic JIMWLK evolution of the weight functional $W$, the NLO graphs represented in Fig.~\ref{fig:NLO-coeff} should be combined with next-to-leading-log (NLLx) JIMWLK evolution to obtain the full NLO result for inclusive photon production in $e+A$ collisions at small $x$. Fortunately, the NLLx JIMWLK evolution equations are known~\cite{Balitsky:2013fea,Grabovsky:2013mba,Kovner:2013ona} as well as the NLLx BK equations~\cite{Kovchegov:2006vj,Balitsky:2008zza,Balitsky:2006wa}. Therefore with the computation of the NLO diagrams represented in Fig.~\ref{fig:NLO-coeff}, all the elements will be in place for quantitative predictions, to NLO accuracy, for inclusive photon production in $e+A$ collisions at small $x$. This paper represents the first step in this direction. While we will discuss key aspects of the structure of the NLO computation here, the full computation will be presented in future publications in preparation. 

 %%%%%%%%%%%%%%%%%%%%%%%%%%%%%%%%%%%%%%

The paper is organized as follows. In section \ref{sec:sectionII}, we begin by discussing the ingredients necessary for the computation of the amplitude for inclusive photon production; a compact expression for this amplitude is given in Eq.~\ref{eq:final-amplitude}. In section \ref{sec:sectionIII}, we will calculate the cross-section for the production of a direct photon accompanied by a quark-antiquark pair as well as the inclusive differential cross-section for direct photon production at LO. The former provides the rate for inclusive production of a photon with a dijet pair in small $x$ kinematics. Our results for these cross-sections are expressed respectively in Eqs.~\ref{eq:triple-differential} and \ref{eq:single-differential} as a convolution of the lepton tensor $L^{\mu \nu}$ that is familiar from inclusive DIS and a hadron tensor constituted of all-twist lightlike Wilson line correlators. In section \ref{sec:sectionIV},  which we divide into three subsections, we discuss the important properties of the photon production amplitude at leading order. We first examine the $k_\perp$ and collinear factorization limits of our computation in the limit of large transverse momentum $k_{\perp} \gg Q^{A}_{S}$.  In the former case, the cross-section is proportional to the unintegrated gluon distribution within the nucleus while the latter is proportional to the usual leading twist nuclear gluon distribution in perturbative QCD, evaluated at the scale $Q^2$ of the virtual photon. The corresponding expressions for the hadron tensor in this limit are given in Eqs.~\ref{eq:kT-gluon} and \ref{eq:collinear-gluon} respectively. We next demonstrate that the Low-Burnett-Kroll theorem \cite{Low:1958sn,Burnett:1967km,Bell:1969yw} is satisfied when $k_{\gamma} \rightarrow 0$: we see explicitly that the nonradiative DIS differential cross-section for the inclusive dijet case in DIS matches extant results in the literature for the same~\cite{Dominguez:2011wm}. In the final subsection, we show that the amplitude derived in Eq.~\ref{eq:final-amplitude} has a simple and efficient interpretation in terms of a modified fermion propagator in the classical background field of the nucleus. This will prove beneficial for higher order computations.
%Subsequently we derive the same result by perturbatively calculating the leading twist $k_{\perp}$-factorized amplitude. 

Section \ref{sec:sectionV} outlines the machinery for the computation of the amplitude at NLO. This includes a discussion of the motivation behind choosing the ``wrong" light cone gauge $A^{-}=0$ for the kinematics of our process as the efficient gauge in which to perform our computations. We then derive the corresponding momentum space Feynman rules for the small fluctuation gluon propagator and conclude this section with an analysis of the contributing processes at this order.  We end the paper with a brief summary and an outline of work in progress. 

Appendices \ref{appendixA}, \ref{appendixB} and  \ref{appendixC} supplement the material in the body of the paper. The notations and conventions are clarified in Appendix \ref{appendixA}. Appendix \ref{appendixB} includes the proof of gauge invariance by virtue of the Ward identity. It also contains computational detail for the subsection dealing with the soft photon factorization of the photon production amplitude. Appendix \ref{appendixC} includes a detailed discussion of the techniques used in determining kinematically allowed processes contributing to inclusive photon production at both LO and NLO. 
%%%%%%%%%%%%%%%%%%%%%%%%%%%%%%%%%%%%%
\section{Components of the LO amplitude computation} \label{sec:sectionII}
%%%%%%%%%%%%%%%%%%%%%%%%%%%%%%%%%%%%%
In this section, we will outline the components needed for the computation of the amplitude for Class II processes at LO in the CGC framework. In the effective theory of CGC, owing to their large occupancy $A\sim 1/g$, the dynamics of small $x$ gluons to LO is described by the classical Yang-Mills equations\footnote{Note that henceforth the factor of g is taken out of the color charge density--this differs from the notation in \cite{Iancu:2003xm} for instance.}
\begin{equation}
[D_{\mu},F^{\mu \nu}](x)=g \delta^{\nu +} \delta(x^{-}) \rho_{A}(\mathbf{x}_{\perp}) \enskip  .
\label{eq:Yang-Mills}
\end{equation}
The nucleus is assumed to move in the positive $z$-direction at nearly the speed of light with large light cone longitudinal momentum $P^{+}_{N}\rightarrow\infty$. (See Appendix \ref{appendixA} for the conventions adopted in this work.) We also assume in this frame that the virtual photon  has a large $q^-$ component of its momentum. 

An essential element in the computation of the LO amplitude is the fermion propagator in the background strong classical color field of the nucleus. In Lorenz gauge $\partial_\mu A^\mu=0$, this has the coordinate space representation~\cite{McLerran:1998nk}
\begin{align}
 S_{A}(x,y)  =S_{0}(x,y)+  \bigg[\theta(x^{-})\theta(-y^{-}) \int \mathrm{d}^{4}z \enskip \delta(z^{-}) \big(\tilde{U}(\mathbf{z}_{\perp})-1 \big)  -  \theta(- x^{-})\theta(y^{-}) & \int \mathrm{d}^{4}z \enskip \delta(z^{-}) \big(\tilde{U}^{\dagger}(\mathbf{z}_{\perp})-1 \big) \bigg] \nonumber \\
& \times S_{0}(x,z) \gamma^{-} S_{0}(z,y) \enskip , \label{eq:effective-propagator-quark-coordinate}
\end{align}
where
\begin{equation}
S_{0}(x,y)= \int \frac{\mathrm{d}^{4}p}{(2 \pi)^{4}} e^{-ip.(x-y)} S_{0}(p), \quad S_{0}(p)=\frac{i(\slashed{p}+m)}{p^{2}-m^{2}+i \varepsilon}\enskip,
\label{eq:free-propagator}
\end{equation}
is the free fermion propagator. 
\begin{figure}[!htb]
\begin{center}
\includegraphics[scale=0.2]{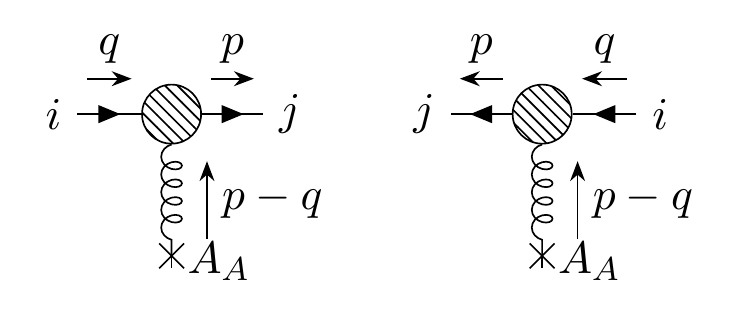}
\caption{Effective vertices for the quark and antiquark respectively. Here $i$ and $j$ represent color indices in the fundamental representation.} \label{fig:effective-vertex}
\end{center}
\end{figure}
This effective propagator appears in the momentum space Feynman rules as an effective vertex with the following factors,
\begin{equation}
\mathcal{T}_{ji}(q,p)= \pm (2 \pi)\delta(p^{-}-q^{-}) \gamma^{-} \int \mathrm{d}^{2} \mathbf{x}_{\perp} \enskip e^{i(\mathbf{q}_{\perp} - \mathbf{p}_{\perp}).\mathbf{x}_{\perp}} \big[\tilde{U}^{(\pm)}(\mathbf{x}_{\perp}) -1 \big]_{ji} \enskip ,
\label{eq:effective-vertex}
\end{equation}
where the plus (minus) sign corresponds to insertions on a quark (antiquark) line respectively and the Wilson line $\tilde{U}$ is written in the fundamental representation of $SU(N_{c})$ as
\begin{equation}
\tilde{U}(\mathbf{x_{ \perp}})=\mathcal{P}_{-} \text{exp}\left[-ig^{2} \int_{-\infty}^{+ \infty} \mathrm{d}z^{-} \frac{1}{\nabla^{2}_{\perp}}  \rho_{A}^{a} (z^{-},\mathbf{x_{ \perp}})t^{a} \right] .
\label{eq:wilson-fund}
\end{equation}
In Feynman diagrams, this effective vertex insertion is drawn as depicted in Fig.~\ref{fig:effective-vertex}. 
These effective vertices do not change the order of the diagrams parametrically by any power of $g$. In principle therefore there can be diagrams with multiple insertions with the emitted photon sandwiched between any two such vertices. However there is a kinematic constraint that restricts many such possibilities: \textit{On the same fermion line(quark or antiquark), there cannot be a photon sandwiched between two effective vertices.}  This has the physical meaning that an outgoing fermion (with a definite sign of $p^{-}$) can get scattered off of the nucleus and subsequently emit a photon but does not suffer a secondary scattering because the time scale governing the scattering is of order $1/P^{+}_{N}$, which is nearly instantaneous in the infinite momentum frame\footnote{Mathematically, this is manifest in the contour integration over the `$+$' component of the undetermined momentum with all the poles being on the same side of the real axis. This is discussed in further detail in Appendix \ref{appendixC}.}.
%%%%%%%%%%%%%%%%%%%%%%%%%%%%%%%%%%%%%%%%%%%%%%%%%%%
\subsection{Derivation of the amplitude} 
The amplitude for the process
\begin{equation}
e(\tilde{l})+A(P) \rightarrow e(\tilde{l'})+Q(k)+\overline{Q}(p)+\gamma(k_{\gamma})+X \enskip ,
\end{equation}
with $X$ denoting any other particle produced in the collision, can be written as 

%%% any other particle
\begin{equation}
\mathcal{M}(\mathbf{\tilde{l}}, \mathbf{\tilde{l'}},\mathbf{q},\mathbf{k},\mathbf{p},\mathbf{k}_{\gamma})=\frac{e}{Q^{2}} \overline{u}(\tilde{l'}) \gamma^{\mu} u(\tilde{l}) \mathcal{M}_{\mu}(\mathbf{q},\mathbf{k},\mathbf{p},\mathbf{k}_{\gamma};\lambda) \enskip,
\label{eq:amplitude-master}
\end{equation}
where
\begin{equation}
\mathcal{M}_{\mu}(\mathbf{q},\mathbf{k},\mathbf{p},\mathbf{k}_{\gamma};\lambda)= \epsilon^{* \alpha}(\mathbf{k}_{\gamma},\lambda) \mathcal{M}_{\mu \alpha}(\mathbf{q},\mathbf{k},\mathbf{p},\mathbf{k}_{\gamma}) \enskip, \label{eq:hadronic-amp-master}
\end{equation}
represents the amplitude for the hadronic subprocess and is the quantity of interest. Here $\epsilon^{ \alpha}(\mathbf{k}_{\gamma},\lambda) $ is the polarization vector for the outgoing photon. The momentum assignments are summarized in Table~\ref{tab:example} and boldface letters stand for the corresponding 3-momenta. Squaring the expression for the amplitude in Eq.~\ref{eq:amplitude-master}, and performing the necessary averaging and sum over electron spins\footnote{We use here the identity
\begin{equation*}
\sum_{\lambda}  \epsilon^{ \beta}(\mathbf{k}_{\gamma},\lambda) \epsilon^{* \alpha}(\mathbf{k}_{\gamma},\lambda)=-g^{\alpha \beta} \enskip ,
\end{equation*} 
as the sum over outgoing photon polarizations.}, we can write 
\begin{equation}
\frac{1}{2} \sum_{\text{spins}, \lambda} \vert \mathcal{M} \vert^{2}=L^{\mu \nu} X_{\mu \nu} \enskip.
\label{eq:squared_amp-master}
\end{equation}
Here 
\begin{equation}
L^{\mu \nu}=\frac{2e^{2}}{Q^{4}}\Big[(\tilde{l}^{\mu} \tilde{l}'^{\nu}+ \tilde{l}^{\nu} \tilde{l}'^{\mu})-\frac{Q^{2}}{2} g^{\mu \nu}     \Big] \enskip,
\label{eq:L-tensor}
\end{equation} 
is the lepton tensor that is identical to that obtained in inclusive DIS. The hadron tensor
\begin{equation}
X_{\mu \nu}= - \sum_{\text{spins}} \mathcal{M}_{\mu \alpha}^{*} (\mathbf{q},\mathbf{k},\mathbf{p},\mathbf{k}_{\gamma}) {\mathcal{M}_{\nu}}^{\alpha} (\mathbf{q},\mathbf{k},\mathbf{p},\mathbf{k}_{\gamma}) \enskip ,
\label{eq:H-tensor}
\end{equation}
is of course different because it allows for the emission of a photon from the quark-antiquark pair. After these bare preliminaries, we shall proceed to compute $\mathcal{M}_{\mu \alpha}$. 
\begin{table*}[!htbp]
\caption{\label{tab:example}4-momentum assignments used in the calculation}
\begin{ruledtabular}
\begin{tabular}{lll}
$\tilde{l}$: Incoming electron & ${\tilde l}^\prime$: Outgoing electron & $q$: Exchanged virtual photon\\
$k$: Quark, directed outward  & $p$: Antiquark, directed outward  & $k_{\gamma}$:  Outgoing photon \\
$l$: Nucleus to antiquark line for multiple insertions\\
$P$: Total momentum of final state = $p+k+k_{\gamma}$\\
\end{tabular}
\end{ruledtabular}
\end{table*}
\subsection{Classifying contributions to the amplitude} 
\begin{figure}[!htb]
\begin{center}
\includegraphics[scale=0.2]{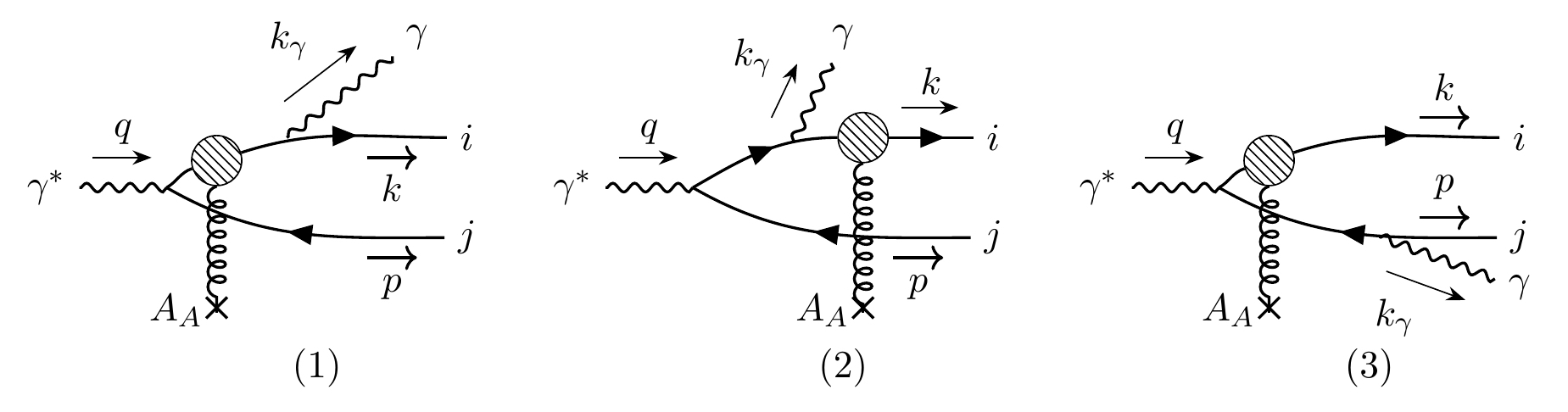}
\caption{Contributions to the amplitude with one Wilson line insertion on the quark line. The momentum labels and their directions are clearly shown. $i$ and $j$ stand for color indices in the fundamental representation of $SU(N_{c})$. \label{fig:Group-I}}
\end{center}
\end{figure}

There are a total of ten contributions, which we will group into diagrams with gluon field insertions on the quark or antiquark line or both. The diagrams for the first group are shown in Fig.~\ref{fig:Group-I}.
The contribution to the amplitude from the diagram labeled (1), representing the emission of the photon by the quark subsequent to its multiple scatterings with the nucleus, is given by
\begin{equation}
\mathcal{M}^{(1)}_{\mu \alpha}(\mathbf{q},\mathbf{k},\mathbf{p},\mathbf{k}_{\gamma})=  - (eq_{f})^{2}  \enskip \overline{u}(\mathbf{k}) \gamma_{\alpha} S_{0}(k+k_{\gamma}) \mathcal{T}_{ij}(q-p,k+k_{\gamma})  S_{0}(q-p) \gamma_{\mu} v(\mathbf{p}) \enskip,
\end{equation}
where $q_{f}$ is the charge of a quark/antiquark of a given flavor. Employing Eqs.~\ref{eq:free-propagator} and \ref{eq:effective-vertex}, and choosing a frame where $\mathbf{q}_{\perp}=0$ for the virtual photon, we can write the above expression as\footnote{The notation here and henceforth closely follows that employed previously in the discussion of photon production in $p+A$ collisions in the CGC 
framework~\cite{Benic:2016uku}.}
\begin{equation}
\mathcal{M}^{(1)}_{\mu \alpha}(\mathbf{q},\mathbf{k},\mathbf{p},\mathbf{k}_{\gamma}) =2\pi (eq_{f})^{2} \delta(P^{-}-q^{-}) \int_{\mathbf{x}_{\perp}}e^{-i\mathbf{P}_{\perp}.\mathbf{x}_{\perp}} \enskip \overline{u}(\mathbf{k}) R^{(1)}_{\mu \alpha}(\mathbf{P}_{\perp}) \Big[\tilde{U}(\mathbf{x}_{\perp})-1 \Big]_{ij}v(\mathbf{p}) \enskip,
\end{equation}
where $\int_{\mathbf{x}_{\perp}}$ is a shorthand notation for $\int \mathrm{d}^{2} \mathbf{x}_{\perp}$, $i$ ($j$) represent the color indices of the quark (antiquark) respectively and
\begin{align}
R^{(1)}_{\mu \alpha}(\mathbf{P}_{\perp})& =\gamma_{\alpha} \frac{  (\slashed{k}+\slashed{k}_{\gamma}+m) }{(k+k_{\gamma})^{2}-m^{2}+i\varepsilon}\gamma^{-}\frac{\slashed{q}-\slashed{p}+m}{(q-p)^{2}-m^{2}+i\varepsilon}\gamma_{\mu}  \nonumber \\
&= \frac{\gamma_{\alpha} \slashed{k}_{\gamma}+2k_{\alpha}}{(k+k_{\gamma})^{2}-m^{2}+i\varepsilon} \gamma^{-} \frac{\slashed{q}\gamma_{\mu} -2 p_{\mu} }{(q-p)^{2}-m^{2}+i\varepsilon} \enskip .
\label{eq:R1}
\end{align}

We use the following relations
\begin{align}
(\slashed{p}-m) \gamma_{\mu} v(\mathbf{p})& =2p_{\mu } v(\mathbf{p}) \enskip , \nonumber \\
\overline{u}(\mathbf{k}) \gamma_{\alpha} (\slashed{k}+m)& = 2k_{\alpha} \overline{u}(\mathbf{k}) \enskip,
\label{eq:relation-22}
\end{align}
to obtain the second line of Eq.~\ref{eq:R1}. Note that the $\mathbf{P}_{\perp}$-dependence appearing in $R_{\mu \alpha}^{(1)}$ includes an implicit dependence on the three transverse momenta constituting $\mathbf{P}_{\perp}=\mathbf{k}_{\perp}+\mathbf{k}_{\gamma \perp}+\mathbf{p}_{\perp}$.

The contributions from diagrams (2) and (3) in Fig.~\ref{fig:Group-I} can be computed similarly; the sum of the contributions from these three diagrams can then be written as 
\begin{equation}
\sum_{\beta=1}^{3}  \mathcal{M}^{(\beta)}_{\mu \alpha}( \mathbf{q}, \mathbf{k},\mathbf{p},\mathbf{k}_{\gamma}) =2\pi (eq_{f})^{2} \delta(P^{-}-q^{-}) \int_{\mathbf{x}_{\perp}}e^{-i\mathbf{P}_{\perp}.\mathbf{x}_{\perp}} \enskip \overline{u}(\mathbf{k}) T^{(q)}_{\mu \alpha}(\mathbf{P}_{\perp}) \Big[\tilde{U}(\mathbf{x}_{\perp})-1 \Big]_{ij}v(\mathbf{p}) \enskip,
\label{eq:Group-I}
\end{equation}
where 
\begin{equation}
T^{(q)}_{\mu \alpha}(\mathbf{P}_{\perp})=\sum_{\beta=1}^{3} R^{(\beta)}_{\mu \alpha}(\mathbf{P}_{\perp}) \enskip ,
\end{equation}
with  
\begin{align}
R^{(2)}_{\mu \alpha}(\mathbf{P}_{\perp})&= \frac{\gamma^{-}(\slashed{q}-\slashed{P})+2k^{-} }{(q-p-k_{\gamma})^{2}-m^{2}+i\varepsilon } \gamma_{\alpha} \frac{\slashed{q}\gamma_{\mu}-2p_{\mu} } {(q-p)^{2}-m^{2}+i\varepsilon }\enskip , \nonumber \\
R^{(3)}_{\mu \alpha}(\mathbf{P}_{\perp})&= - \frac{\gamma^{-}(\slashed{q}-\slashed{P})+2k^{-} }{(q-p-k_{\gamma})^{2}-m^{2}+i\varepsilon } \gamma_{\mu} \frac{\slashed{k}_{\gamma} \gamma_{\alpha}+2p_{\alpha}      }{(p+k_{\gamma})^{2}-m^{2}+i\varepsilon } \enskip .
\label{eq:R2-R3}
\end{align}
\begin{figure}[!htb]
\begin{center}
\includegraphics[scale=0.2]{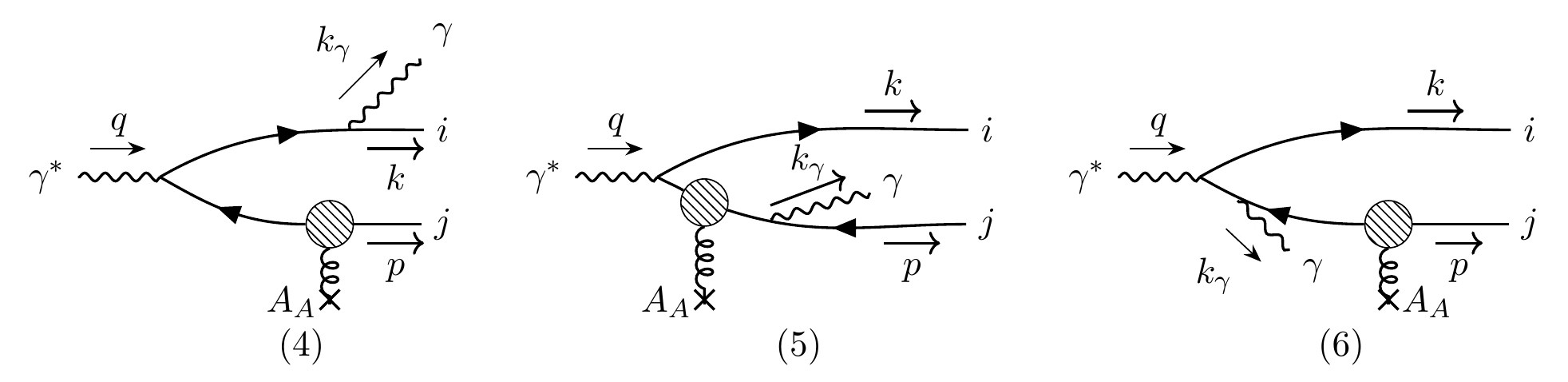}
\caption{Contributions to the amplitude with one Wilson line insertion on the antiquark line. The momentum labels and their directions are clearly shown. $i$ and $j$ stand for color indices in the fundamental representation of $SU(N_{c})$. \label{fig:Group-II}}
\end{center}
\end{figure}
Similarly, the contribution from diagrams with insertions on the antiquark line are shown in Fig. \ref{fig:Group-II}. Their collective contribution can be written as 
\begin{equation}
\sum_{\beta=4}^{6}   \mathcal{M}^{(\beta)}_{\mu \alpha}( \mathbf{q}, \mathbf{k},\mathbf{p},\mathbf{k}_{\gamma}) =2\pi (eq_{f})^{2} \delta(P^{-}-q^{-}) \int_{\mathbf{y}_{\perp}}e^{-i\mathbf{P}_{\perp}.\mathbf{y}_{\perp}} \enskip \overline{u}(\mathbf{k}) T^{(\bar{q})}_{\mu \alpha}(\mathbf{P}_{\perp}) \Big[\tilde{U}^{\dagger}(\mathbf{y}_{\perp})-1 \Big]_{ij}v(\mathbf{p}) \enskip ,
\label{eq:Group-II}
\end{equation}
where 
\begin{equation}
T^{(\bar{q})}_{\mu \alpha}(\mathbf{P}_{\perp})=\sum_{\beta=4}^{6} R^{(\beta)}_{\mu \alpha}(\mathbf{P}_{\perp}) \enskip ,
\end{equation}
and
\begin{align}
R^{(4)}_{\mu \alpha}& (\mathbf{P}_{\perp})= \frac{\gamma_{\alpha} \slashed{k}_{\gamma}+2k_{\alpha} }{(k+k_{\gamma})^{2}-m^{2}+i\varepsilon} \gamma_{\mu} \frac{ ( \slashed{q}-\slashed{P})\gamma^{-}+2p^{-}  }{ (k+k_{\gamma}-q)^{2}-m^{2}+i\varepsilon  } \enskip , \nonumber \\
R^{(5)}_{\mu \alpha} & (\mathbf{P}_{\perp})= \frac{2k_{\mu}-\gamma_{\mu}\slashed{q} }{(q-k)^{2}-m^{2}+i\varepsilon } \gamma^{-} \frac{\slashed{k}_{\gamma} \gamma_{\alpha}+2p_{\alpha} }{(p+k_{\gamma})^{2}-m^{2}+i\varepsilon } \enskip , \nonumber \\
R^{(6)}_{\mu \alpha}& (\mathbf{P}_{\perp})= \frac{2k_{\mu}-\gamma_{\mu}\slashed{q} }{(q-k)^{2}-m^{2}+i\varepsilon } \gamma_{\alpha} \frac{(\slashed{q}-\slashed{P})\gamma^{-}+2p^{-} }{(k+k_{\gamma}-q)^{2}-m^{2}+i\varepsilon } \enskip .
\label{eq:R4-R5-R6}
\end{align}

\begin{figure}[!htb]
\begin{center}
\includegraphics[scale=0.2]{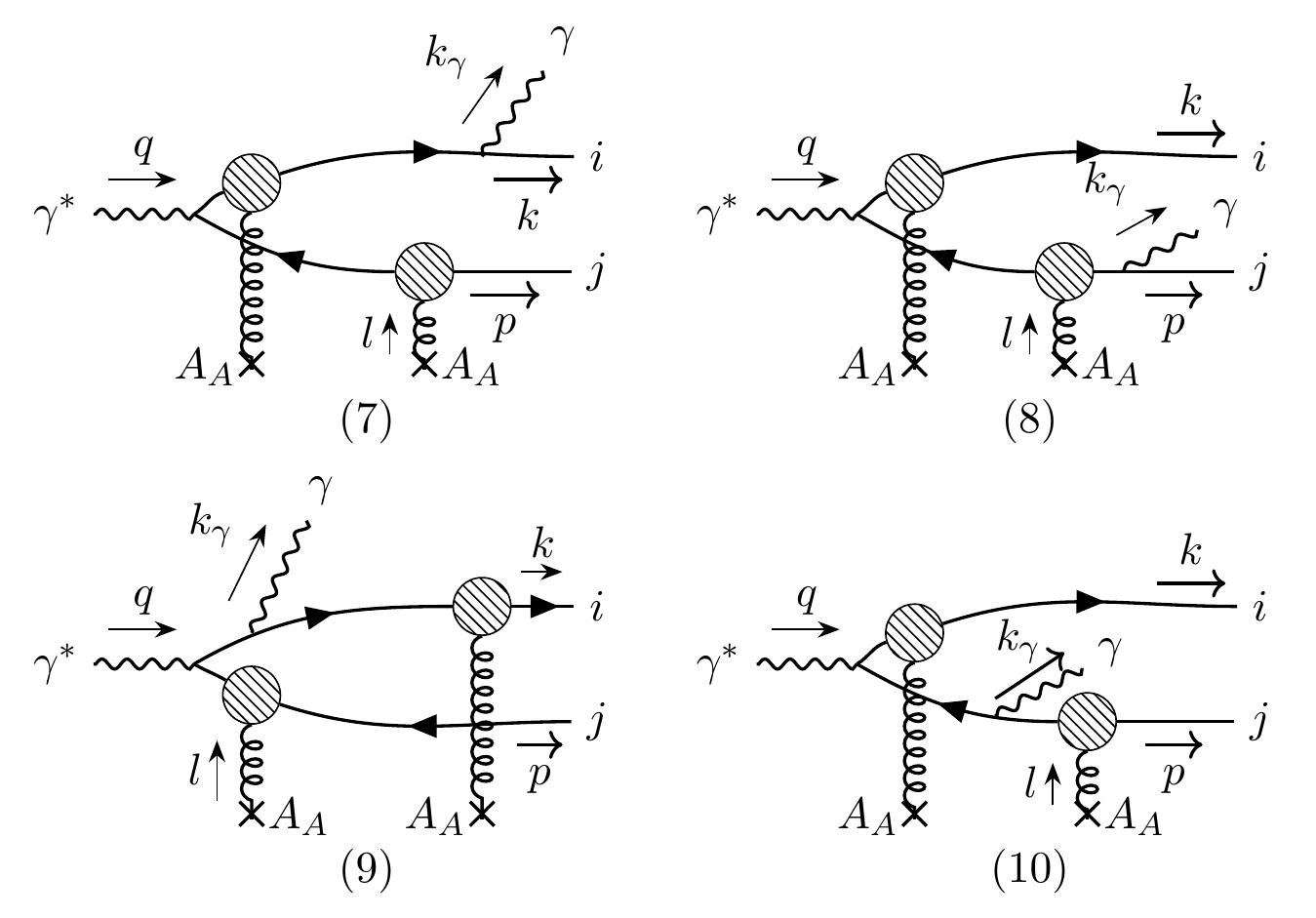}
\caption{Contributions to the amplitude with two Wilson line insertions.  The momentum labels and their directions are clearly shown. $i$ and $j$ stand for color indices in the fundamental representation of $SU(N_{c})$. \label{fig:Group-III}}
\end{center}
\end{figure}

Finally, we  need to consider the diagrams shown in  Fig.~\ref{fig:Group-III} with insertions on both lines. For these cases, we will need to integrate over the nuclear momentum transfer $l$. 
The contribution from the diagram labeled (7) can be written as
\begin{align}
\mathcal{M}^{(7)}_{\mu \alpha}(\mathbf{q},\mathbf{k},\mathbf{p},\mathbf{k}_{\gamma})  =  - (eq_{f})^{2} \int \frac{\mathrm{d}^{4}l}{(2\pi)^{4}} \enskip \overline{u}(\mathbf{k}) \gamma_{\alpha} S_{0}(k+k_{\gamma}) & \mathcal{T}_{ik}(q-p+l,k+k_{\gamma}) S_{0}(q-p+l) \nonumber \\
& \times     \gamma_{\mu}S_{0}(l-p) \mathcal{T}_{kj}(-p,l-p) v(\mathbf{p}) \enskip .
\end{align}
The integration over $l^{-}$ is trivial because of the $\delta(l^{-})$ factor from one of the effective vertices. The integration over $l^{+}$ can be done using the theorem of residues. The result can be again compactly written as
\begin{align}
 \mathcal{M}^{(7)}_{\mu \alpha}(\mathbf{q},\mathbf{k},\mathbf{p},\mathbf{k}_{\gamma})= 2\pi (eq_{f})^{2}\delta(P^{-}-q^{-}) \int_{\mathbf{x}_{\perp}} \int_{\mathbf{y}_{\perp}} \int_{\mathbf{l}_{\perp}} e^{-i\mathbf{P}_{\perp}.\mathbf{x}_{\perp}+i\mathbf{l}_{\perp}.\mathbf{x}_{\perp}} & e^{-i \mathbf{l}_{\perp}.\mathbf{y}_{\perp}}   \enskip \overline{u}(\mathbf{k}) R^{(7)}_{\mu \alpha}(\mathbf{l}_{\perp},\mathbf{P}_{\perp}) \nonumber \\
 &\times\Big[\big( \tilde{U}(\mathbf{x}_{\perp})-1\big) \big(\tilde{U}^{\dagger}(\mathbf{y}_{\perp})-1\big)   \Big]_{ij} v(\mathbf{p}) \enskip ,
\end{align}
where 
\begin{equation}
 R^{(7)}_{\mu \alpha}(\mathbf{l}_{\perp},\mathbf{P}_{\perp})=\frac{\gamma_{\alpha}\slashed{k}_{\gamma}+2k_{\alpha}}{(k+k_{\gamma})^{2}-m^{2}+i\varepsilon}\gamma^{-}\frac{\slashed{q}-\slashed{p}+\slashed{\mathbf{l}}_{\perp}+m }{N(\mathbf{l}_{\perp})} \gamma_{\mu} (\gamma^{-}\slashed{\mathbf{l}}_{\perp}+2p^{-}) \enskip .
 \label{eq:R7}
 \end{equation}
 The factor
 \begin{equation}
 N(\mathbf{l}_{\perp}) = 4p^{-}(q^{-}-p^{-})\Big[ q^{+}-\frac{M^{2}(\mathbf{l}_{\perp}-\mathbf{p}_{\perp})}{2p^{-}} -\frac{M^{2}(\mathbf{l}_{\perp}-\mathbf{p}_{\perp})}{2(q^{-}-p^{-})}\Big] \enskip ,
 \end{equation}
appearing in the denominator is deliberately written in a way to make contact with the expression in Eq.~\ref{eq:R1}; this will be exploited in a later part of the calculation. In the above, $M^{2}(\mathbf{p}_{\perp})=\mathbf{p}^{2}_{\perp}+m^{2}$ stands for the squared transverse mass and $\slashed{\mathbf{p}}_{\perp}=\gamma^{i}p_{i}$. Finally, $\int_{\mathbf{l}_{\perp}}$ is a shorthand notation for $\int \mathrm{d}^{2} \mathbf{l}_{\perp}/(2\pi)^{2}$.

The remaining three diagrams can be worked out similarly and and the combined result compactly written as
\begin{align}
 \sum_{\beta=7}^{10} \mathcal{M}^{(\beta)}_{\mu \alpha}(\mathbf{q},\mathbf{k},\mathbf{p},\mathbf{k}_{\gamma})= 2\pi (eq_{f})^{2}\delta(P^{-}-q^{-}) & \int_{\mathbf{x}_{\perp}} \int_{\mathbf{y}_{\perp}} \int_{\mathbf{l}_{\perp}} e^{-i\mathbf{P}_{\perp}.\mathbf{x}_{\perp}+i\mathbf{l}_{\perp}.\mathbf{x}_{\perp}} e^{-i \mathbf{l}_{\perp}.\mathbf{y}_{\perp}}  \nonumber \\
& \times \overline{u}(\mathbf{k}) T^{(q\bar{q})}_{\mu \alpha}(\mathbf{l}_{\perp},\mathbf{P}_{\perp})\Big[\big(\tilde{U}(\mathbf{x}_{\perp})-1\big)\big(\tilde{U}^{\dagger}(\mathbf{y}_{\perp})-1\big)   \Big]_{ij} v(\mathbf{p}) \enskip ,
\label{eq:Group-III}
\end{align}
where
\begin{equation}
T^{(q\bar{q})}_{\mu \alpha}(\mathbf{l}_{\perp},\mathbf{P}_{\perp}) =\sum_{\beta=7}^{10} R^{\beta}_{\mu \alpha} (\mathbf{l}_{\perp},\mathbf{P}_{\perp}) \enskip .
\end{equation}
The various R-factors obtained after the contour integration over $l^{+}$ are 
\begin{equation}
R^{(8)}_{\mu \alpha}(\mathbf{l}_{\perp},\mathbf{P}_{\perp})= -\frac{\gamma^{-}(\slashed{q}-\slashed{P}+\slashed{\mathbf{l}}_{\perp})+2k^{-} }{(p+k_{\gamma})^{2}-m^{2}+i\varepsilon}  \gamma_{\mu} (\slashed{p}+\slashed{k}_{\gamma}-\slashed{\mathbf{l}}_{\perp}-m ) \gamma^{-}  \frac{(\slashed{k}_{\gamma}\gamma_{\alpha}+2p_{\alpha})}{S(\mathbf{l}_{\perp})}  \enskip ,
\label{eq:R8}
\end{equation}
\begin{equation}
R^{(9)}_{\mu \alpha}(\mathbf{l}_{\perp},\mathbf{P}_{\perp})= \Big( \gamma^{-}(\slashed{q}-\slashed{P}+\slashed{\mathbf{l}}_{\perp})+2k^{-}\Big) \gamma_{\alpha} (\slashed{q}-\slashed{p}+\slashed{l}_{a}+m) \gamma_{\mu} \frac{(\gamma^{-}\slashed{\mathbf{l}}_{\perp}+2p^{-})}{V(\mathbf{l}_{\perp})}  \enskip ,
\label{eq:R9}
\end{equation}
and 
\begin{equation}
R^{(10)}_{\mu \alpha}(\mathbf{l}_{\perp},\mathbf{P}_{\perp})= \Big( \gamma^{-}(\slashed{q}-\slashed{P}+\slashed{\mathbf{l}}_{\perp})+2k^{-}\Big) \gamma_{\mu} (\slashed{l}_{b} -\slashed{p}-\slashed{k}_{\gamma}+m) \gamma_{\alpha} \frac{(\gamma^{-}\slashed{\mathbf{l}}_{\perp}+2p^{-})}{W(\mathbf{l}_{\perp})} \enskip .
\label{eq:R10}
\end{equation}
In the expressions above, the compactly written functions are 
\begin{equation}
 S(\mathbf{l}_{\perp})  = 4(p^{-}+k_{\gamma}^{-}) (q^{-}-p^{-}-k_{\gamma}^{-}) \Big[q^{+}- \frac{M^{2}(\mathbf{p}_{\perp}+\mathbf{k}_{\gamma \perp}-\mathbf{l}_{\perp}) q^{-}}{2(p^{-}+k^{-}_{\gamma}) (q^{-}-k^{-}_{\gamma}-p^{-})} \Big] \enskip ,
\end{equation}
\begin{align}
V(\mathbf{l}_{\perp}) & = 8p^{-} (q^{-}-p^{-}) (q^{-}-p^{-}-k^{-}_{\gamma}) \Big[q^{+}-\frac{M^{2}(\mathbf{l}_{\perp}-\mathbf{p}_{\perp})q^{-}}{2p^{-}(q^{-}-p^{-})}       \Big]  \nonumber \\
& \times \Big[  q^{+}-k^{+}_{\gamma}-\frac{M^{2}(\mathbf{l}_{\perp}-\mathbf{p}_{\perp})}{2p^{-}} -  \frac{M^{2}(\mathbf{p}_{\perp}+\mathbf{k}_{\gamma \perp}-\mathbf{l}_{\perp})}{2(q^{-}-p^{-}-k^{-}_{\gamma})}         \Big]  \enskip ,
\end{align}
and 
\begin{align}
W(\mathbf{l}_{\perp})& = 8p^{-} (p^{-}+k^{-}_{\gamma}) (q^{-}-p^{-}-k^{-}_{\gamma}) \Big[ q^{+}-\frac{M^{2}(\mathbf{p}_{\perp}+\mathbf{k}_{\gamma \perp}-\mathbf{l}_{\perp})q^{-}}{2(p^{-}+k^{-}_{\gamma})(q^{-}-p^{-}- k^{-}_{\gamma}) }       \Big] \nonumber \\
& \times \Big[  q^{+}-k^{+}_{\gamma}-\frac{M^{2}(\mathbf{l}_{\perp}-\mathbf{p}_{\perp})}{2p^{-}} -  \frac{M^{2}(\mathbf{p}_{\perp}+\mathbf{k}_{\gamma \perp}-\mathbf{l}_{\perp})}{2(q^{-}-p^{-}- k^{-}_{\gamma})}         \Big] \enskip .
\end{align}
The factors $\slashed{l}_{a}$ and $\slashed{l}_{b}$ in Eqs.~\ref{eq:R9} and \ref{eq:R10} are obtained by evaluating $\slashed{l}$  at the enclosed poles depending on the choice of contours:  
\begin{align}
\slashed{l}_{a}& = \gamma^{-}\Bigg( p^{+}-\frac{M^{2}(\mathbf{l}_{\perp}-\mathbf{p}_{\perp})}{2p^{-}} \Bigg) + \slashed{\mathbf{l}}_{\perp} \enskip , \nonumber \\
\slashed{l}_{b} &= \gamma^{-} \Bigg( p^{+}+k^{+}_{\gamma}-q^{+}+\frac{M^{2}(\mathbf{p}_{\perp}-\mathbf{l}_{\perp}+\mathbf{k}_{\gamma \perp})}{2(q^{-}-p^{-}-k^{-}_{\gamma})} \Bigg) +\slashed{\mathbf{l}}_{\perp} \enskip .
\end{align}
%%%%%%%%%%%%%%%%%%%%%%%%%%%%%%%%%%%%%%%%%%%%
\subsection{Final expression for the amplitude}
The net contribution from all the diagrams is obtained by adding the expressions in Eqs.~\ref{eq:Group-I}, \ref{eq:Group-II} and \ref{eq:Group-III}.  Introducing a dummy integration $\int_{\mathbf{x}_{\perp}} \int_{\mathbf{k}_{\perp}} e^{-i\mathbf{k}_{\perp}.\mathbf{x}_{\perp}}=1$, this result can be expressed compactly as 
\begin{align}
 \mathcal{M}_{\mu \alpha}(\mathbf{q}, \mathbf{k},\mathbf{p},\mathbf{k}_{\gamma})& = \sum_{\beta=1}^{10} \mathcal{M}_{\mu \alpha}^{\beta} (\mathbf{q},\mathbf{k},\mathbf{p},\mathbf{k}_{\gamma})=2\pi (eq_{f})^{2}\delta(P^{-}-q^{-})  \int_{\mathbf{x}_{\perp}} \int_{\mathbf{y}_{\perp}} \int_{\mathbf{l}_{\perp}} e^{-i\mathbf{P}_{\perp}.\mathbf{x}_{\perp}+i\mathbf{l}_{\perp}.\mathbf{x}_{\perp}} e^{-i \mathbf{l}_{\perp}.\mathbf{y}_{\perp}} \nonumber \\
& \times  \overline{u}(\mathbf{k}) \Big[ T^{(q\bar{q})}_{\mu \alpha} (\mathbf{l}_{\perp},\mathbf{P}_{\perp})   \big[ \tilde{U}(\mathbf{x}_{\perp}) \tilde{U}^{\dagger} (\mathbf{y}_{\perp}) -1 \big] + \Big( T^{(q)}_{\mu \alpha} (\mathbf{P}_{\perp}) - T^{(q\bar{q})}_{\mu \alpha} (\mathbf{l}_{\perp},\mathbf{P}_{\perp}) \Big)  \big[ \tilde{U}(\mathbf{x}_{\perp}) -1 \big] \nonumber \\
& +  \Big( T^{(\bar{q})}_{\mu \alpha} (\mathbf{P}_{\perp}) - T^{(q\bar{q})}_{\mu \alpha} (\mathbf{l}_{\perp},\mathbf{P}_{\perp}) \Big)  \big[ \tilde{U}^{\dagger}(\mathbf{y}_{\perp}) -1 \big]   \Big] v(\mathbf{p})  \enskip .
\label{eq:amp-sum}
\end{align}
This expression can be further simplified by observing that the following relations holds for the various R-factors given in the previous section,
\begin{align}
R^{(1)}_{\mu \alpha} (\mathbf{P}_{\perp}) - R^{(7)}_{\mu \alpha} (\mathbf{0}_{\perp},\mathbf{P}_{\perp})& =0 \enskip ,\nonumber \\
R^{(2)}_{\mu \alpha} (\mathbf{P}_{\perp}) - R^{(9)}_{\mu \alpha} (\mathbf{0}_{\perp},\mathbf{P}_{\perp})& =0  \enskip ,\nonumber \\
R^{(3)}_{\mu \alpha} (\mathbf{P}_{\perp}) - (R^{(8)}_{\mu \alpha}+ R^{(10)}_{\mu \alpha} ) (\mathbf{0}_{\perp},\mathbf{P}_{\perp}) & =0 \enskip ,
\end{align}
which leads to 
\begin{equation}
T^{(q)}_{\mu \alpha} (\mathbf{P}_{\perp}) - T^{(q \bar{q})}_{\mu \alpha} (\mathbf{0}_{\perp},\mathbf{P}_{\perp})=0 \enskip .
\label{eq:identity-1}
\end{equation}
Since the second term in the sum of the three terms in Eq.~\ref{eq:amp-sum} is independent of $\mathbf{y}_{\perp}$, that sets $\mathbf{l}_{\perp}=0$ for this term. Hence the identity in Eq.~\ref{eq:identity-1} implies that the second term in Eq.~\ref{eq:amp-sum} vanishes.

There is an intuitive way of understanding this relation. For this, we identify that the transverse momentum kicks to the antiquark and quark lines are respectively $\mathbf{l}_{\perp}$ and $\mathbf{P}_{\perp}-\mathbf{l}_{\perp}$. In the limit of $\mathbf{l}_{\perp}$ going to zero, diagrams (1) and (7) give identical contributions with the transverse momentum transfer being $\mathbf{P}_{\perp}$ for both cases. The same argument holds for processes (2) and (9). For processes (8) and (10), the insertions on the antiquark line are on either side of the photon and hence it is their sum which equals the contribution from process (3) for vanishing $\mathbf{l}_{\perp}$. The same argument holds for processes (4), (5) and (6) with the corresponding limit being $\mathbf{l}_{\perp}=\mathbf{P}_{\perp}$ so that in this case the transverse momentum kick to the quark line vanishes. One thus gets
\begin{align}
R^{(5)}_{\mu \alpha} (\mathbf{P}_{\perp}) - R^{(8)}_{\mu \alpha} (\mathbf{P}_{\perp},\mathbf{P}_{\perp})& =0 \enskip , \nonumber \\
R^{(6)}_{\mu \alpha} (\mathbf{P}_{\perp}) - R^{(10)}_{\mu \alpha} (\mathbf{P}_{\perp},\mathbf{P}_{\perp})& =0 \enskip , \nonumber \\
R^{(4)}_{\mu \alpha} (\mathbf{P}_{\perp}) - (R^{(7)}_{\mu \alpha}+ R^{(9)}_{\mu \alpha} ) (\mathbf{P}_{\perp},\mathbf{P}_{\perp}) & =0 \enskip ,
\end{align}
leading to
\begin{equation}
T^{(\bar{q})}_{\mu \alpha} (\mathbf{P}_{\perp}) - T^{(q \bar{q})}_{\mu \alpha} (\mathbf{P}_{\perp},\mathbf{P}_{\perp})=0 \enskip .
\label{eq:identity-2}
\end{equation}
For the same reason articulated previously, this identity implies that the third term in Eq.~\ref{eq:amp-sum} vanishes as well. 
The relations in Eqs.~\ref{eq:identity-1} and \ref{eq:identity-2} should be compared to the second line of Eq.~ 2.46 in \cite{Benic:2016uku} for the  $p+A$  NLO calculation. Such qualitative similarities will appear throughout the LO discussion in this paper.   

Since the second and third terms in Eq.~\ref{eq:amp-sum} vanish, the LO amplitude therefore reduces to the expression: 
\begin{align}
 \mathcal{M}_{\mu \alpha}(\mathbf{q}, \mathbf{k},\mathbf{p},\mathbf{k}_{\gamma}) = \sum_{\beta=1}^{10} \mathcal{M}_{\mu \alpha}^{\beta} (\mathbf{q},\mathbf{k},\mathbf{p},\mathbf{k}_{\gamma}) & =2\pi (eq_{f})^{2}\delta(P^{-}-q^{-})  \int_{\mathbf{x}_{\perp}} \int_{\mathbf{y}_{\perp}} \int_{\mathbf{l}_{\perp}} e^{-i\mathbf{P}_{\perp}.\mathbf{x}_{\perp}+i\mathbf{l}_{\perp}.\mathbf{x}_{\perp}} e^{-i \mathbf{l}_{\perp}.\mathbf{y}_{\perp}}  \nonumber \\
& \times  \overline{u}(\mathbf{k}) \Big[ T^{(q\bar{q})}_{\mu \alpha} (\mathbf{l}_{\perp},\mathbf{P}_{\perp})   \big[ \tilde{U}(\mathbf{x}_{\perp}) \tilde{U}^{\dagger} (\mathbf{y}_{\perp}) -1 \big] \Big] v(\mathbf{p}) \enskip .
\label{eq:final-amplitude}
\end{align}

%%%%%%%%%%%%%%%%%%%%%%%%%%%%%%%%%%%%%%%%%%%%%%%%%%
\section{The inclusive photon cross-section at LO} \label{sec:sectionIII}
In proceeding to write down the expression for the inclusive photon differential cross-section, we note that since Eq.~\ref{eq:final-amplitude} contains a delta function prefactor, the squared amplitude will naively contain a squared delta function. However this potential problem is resolved by realizing that the photon plane wave hitherto considered should be replaced by a properly normalized wave packet for the incoming virtual photon\footnote{A nice discussion of this very question can be found in Sec. IV of \cite{Gelis:2002ki} and in pages 99-107 of \cite{Peskin:1995ev}.}.  Following a similar procedure to that described in  \cite{Gelis:2002ki} , and working in a frame where the lepton and the nucleus are moving towards each other at near light speed, we can write the differential probability as
\begin{equation}
\frac{\mathrm{d} \sigma}{\mathrm{d}x   \mathrm{d}Q^{2}} = \frac{2 \pi y^{2}}{64\pi^{3} Q^{2}} \frac{\mathrm{d}^{3} \mathbf{k}}{(2\pi)^{3} 2E_{k}} \frac{\mathrm{d}^{3} \mathbf{p}}{(2\pi)^{3} 2E_{p}} \frac{\mathrm{d}^{3} \mathbf{k}_{\gamma}}{(2\pi)^{3} 2E_{k_{\gamma}}} \frac{1}{2q^{-}}\enskip \Big( \frac{1}{2} \sum_{\text{spins},
\lambda} 
\left \langle \vert \overset{\sim}{\mathcal{M}} \vert^{2} \right\rangle_{Y_{A}} \Big) (2\pi) \delta(P^{-}-q^{-}) \enskip ,
\end{equation}
where $\langle \ldots \rangle_{Y_{A}}$ is used to denote the CGC color average, defined in Eq.~\ref{eq:expectation-value-O}, of the modulus squared of the amplitude which is the prescription~\cite{Blaizot:2004wu} used for inclusive processes. In this expression,  $s=(\tilde{l}+P_{N})^{2}=Q^{2}/xy=4EE_{N}$ is the $e+A$ squared center-of-mass energy, where $E$ is the incoming electron energy and $E_{N}$ the energy per nucleon of the incoming nucleus. Further, $x=Q^{2}/2 P_{N}\cdot q$ and $y=q\cdot P_{N}/\tilde{l}.P_{N}$ respectively denote the  familiar DIS Lorentz invariant variables Bjorken $x$ and the inelasticity. Finally, $E_{k}$, $E_{p}$ and $E_{k_{\gamma}}$ are the respective relativistic energies of the quark, antiquark and the outgoing photon. The amplitude squared, averaged over colors, spins and polarizations, can be expressed as
\begin{equation}
 \frac{1}{2} \sum_{\text{spins},
\lambda} 
\left \langle \vert \overset{\sim}{\mathcal{M}} \vert^{2} \right \rangle_{Y_{A}}=16 \pi^{2} \alpha^{2} q^{4}_{f} N_{c} \enskip  L^{\mu \nu} \overset{\sim}{X}_{\mu \nu} \enskip ,
\end{equation}
where $L^{\mu \nu}$ is given by Eq.~\ref{eq:L-tensor} and
\begin{align}
\overset{\sim}{X}_{\mu \nu}= \int_{\mathbf{x}_{\perp},\mathbf{y}_{\perp},\mathbf{x'}_{\perp},\mathbf{y'}_{\perp},\mathbf{l}_{\perp},\mathbf{l'}_{\perp}} \!\!\!\!\!\!\!\!\!\! \!\!\!\!\!\!e^{-i (\mathbf{P}_{\perp}-\mathbf{l}_{\perp}).\mathbf{x}_{\perp}-i\mathbf{l}_{\perp}.\mathbf{y}_{\perp}+i(\mathbf{P}_{\perp}-\mathbf{l'}_{\perp}).\mathbf{x'}_{\perp}+i\mathbf{l'}_{\perp}.\mathbf{y'}_{\perp}}  \enskip  {\tau_{\mu \nu}}^{q\bar{q},q\bar{q}}(\mathbf{l}_{\perp},\mathbf{l'}_{\perp}\vert\mathbf{P}_{\perp}) \enskip  \Xi(\mathbf{x}_{\perp},\mathbf{y}_{\perp};\mathbf{x'}_{\perp},\mathbf{y'}_{\perp}) \enskip .
\label{eq:H-tensor2}
\end{align}
In the above expression, the Dirac trace is written in the compact notation introduced in \cite{Blaizot:2004wu}:
\begin{equation}
{\tau_{\mu \nu}}^{q\bar{q},q\bar{q}}(\mathbf{l}_{\perp},\mathbf{l'}_{\perp} \vert \mathbf{P}_{\perp}) = \text{Tr}\Big[ (\slashed{k}+m) {T_{\nu }^{(q\bar{q})}}^{\alpha} (\mathbf{l}_{\perp},\mathbf{P}_{\perp}) (m-\slashed{p}) \hat{\gamma}^{0} {T^{(q\bar{q})}_{\mu \alpha}}^{\dagger}(\mathbf{l'}_{\perp},\mathbf{P}_{\perp}) \hat{\gamma}^{0} \Big] \enskip . \label{eq:Dirac-trace-general}
\end{equation}
The nonperturbative information on strongly correlated gluons in the nucleus is entirely contained in the term 
\begin{align}
\Xi(\mathbf{x}_{\perp},\mathbf{y}_{\perp};\mathbf{x'}_{\perp},\mathbf{y'}_{\perp})= 1-D(\mathbf{x}_{\perp},\mathbf{y}_{\perp}) -D(\mathbf{y'}_{\perp},\mathbf{x'}_{\perp})+Q(\mathbf{x}_{\perp},\mathbf{y}_{\perp};\mathbf{y'}_{\perp},\mathbf{x'}_{\perp}) \enskip ,
\label{eq:dipole-quadrupole-correlators}
\end{align}
where $D$  and $Q$ are respectively the dipole and quadrupole Wilson line correlators defined as 
\begin{align}
D(\mathbf{x}_{\perp},\mathbf{y}_{\perp}) &= \frac{1}{N_{c}}  \langle {\rm Tr}\left(\tilde{U}(\mathbf{x}_{\perp}) \tilde{U}^{\dagger}(\mathbf{y}_{\perp})\right)\rangle_{Y_{A}} \enskip , \nonumber \\
Q(\mathbf{x}_{\perp},\mathbf{y}_{\perp};\mathbf{y'}_{\perp},\mathbf{x'}_{\perp})&= \frac{1}{N_{c}}  \langle \text{Tr} \left(\tilde{U}(\mathbf{y'}_{\perp}) \tilde{U}^{\dagger}(\mathbf{x'}_{\perp}) \tilde{U}(\mathbf{x}_{\perp}) \tilde{U}^{\dagger}(\mathbf{y}_{\perp}) \right) \rangle_{Y_{A}}   \enskip .
\end{align}
These are universal, gauge invariant quantities that appear in many processes in both $p+A$ and $e+A$ collisions~\cite{JalilianMarian:2005jf,Dominguez:2011wm}.

Be defining these correlators in terms of nuclear unintegrated distributions~\cite{Blaizot:2004wu,Fujii:2006ab,Benic:2016uku}, the expression for $\overset{\sim}{X}_{\mu \nu}$ can be given an explicit momentum space `look'. For this, we first introduce a dummy integration over $\mathbf{l}_{1\perp}$ and a $\delta$-function representing overall transverse momentum conservation, $(2\pi)^{2} \delta^{(2)}(\mathbf{P}_{\perp}-\mathbf{l}_{1\perp})$. Now, define
\begin{equation}
\int_{\mathbf{x}_{\perp},\mathbf{y}_{\perp},\mathbf{x'}_{\perp},\mathbf{y'}_{\perp}}\!\!\!\! \!\!\!\!e^{-i(\mathbf{l}_{1\perp}-\mathbf{l}_{\perp}).\mathbf{x}_{\perp} -i\mathbf{l}_{\perp}.\mathbf{y}_{\perp} + i(\mathbf{l}_{1\perp}-\mathbf{l'}_{\perp}).\mathbf{x'}_{\perp} + i\mathbf{l'}_{\perp}.\mathbf{y'}_{\perp}} D(\mathbf{x}_{\perp},\mathbf{y}_{\perp})
 =\frac{2\alpha_{S}N_c}{\mathbf{l}_{1\perp}^{2}} \phi_{Y_A}^{D}(\mathbf{l}_{1\perp}-\mathbf{l}_{\perp},\mathbf{l}_{\perp}) \enskip , \label{eq:unintegrated-dipole}
\end{equation}
%\begin{equation}
%\int_{\mathbf{x}_{\perp},\mathbf{y}_{\perp},\mathbf{x'}_{\perp},\mathbf{y'}_{\perp}} e^{-i(\mathbf{l}_{1\perp}-\mathbf{l}_{\perp}).\mathbf{x}_{\perp}}\!\!\!\!  e^{-i\mathbf{l}_{\perp}.\mathbf{y}_{\perp}} e^{i(\mathbf{l}_{1\perp}-\mathbf{l'}_{\perp}).\mathbf{x'}_{\perp}}  e^{i\mathbf{l'}_{\perp}.\mathbf{y'}_{\perp}} D(\mathbf{y}_\perp^\prime,\mathbf{y}_\perp^\prime)
% =\frac{2\alpha_{S}N_c}{\mathbf{l}_{1\perp}^{2}} \phi_{Y_A}^{D}(\mathbf{l}_{1\perp}-\mathbf{l'}_{\perp},\mathbf{l'}_{\perp}) \enskip ,
%\end{equation}
\begin{align}
\int_{\mathbf{x}_{\perp},\mathbf{y}_{\perp},\mathbf{x'}_{\perp},\mathbf{y'}_{\perp}} \!\!\!\!\!\!\!\!e^{-i(\mathbf{l}_{1\perp}-\mathbf{l}_{\perp}).\mathbf{x}_{\perp} -i\mathbf{l}_{\perp}.\mathbf{y}_{\perp} + i(\mathbf{l}_{1\perp}-\mathbf{l'}_{\perp}).\mathbf{x'}_{\perp} + i\mathbf{l'}_{\perp}.\mathbf{y'}_{\perp}} Q(\mathbf{x}_{\perp},\mathbf{y}_{\perp};\mathbf{y'}_{\perp},\mathbf{x'}_{\perp})
 = \frac{2\alpha_{S}N_c}{\mathbf{l}_{1\perp}^{2}} \phi_{Y_A}^{Q}(\mathbf{l}_{1\perp}-\mathbf{l}_{\perp},\mathbf{l}_{\perp}; \mathbf{l}_{1\perp}-\mathbf{l'}_{\perp},\mathbf{l'}_{\perp}) \enskip . \label{eq:unintegrated-quadrupole}
\end{align}
where $\phi^{D}_{Y_{A}}$ and $\phi^{Q}_{Y_{A}}$ should be contrasted with the correlators appearing in the $p+A$ NLO calculation (see Eqs. 3.11-3.13 of \cite{Benic:2016uku}) for inclusive photon production. Eq.~\ref{eq:H-tensor2} therefore becomes
\begin{align}
\overset{\sim}{X}_{\mu \nu}= & \int_{\mathbf{l}_{1\perp}} \int_{\mathbf{l}_{\perp},\mathbf{l'}_{\perp}} (2\pi)^{2} \delta^{(2)}(\mathbf{P}_{\perp}-\mathbf{l}_{1\perp}) {\tau_{\mu \nu}}^{q\bar{q},q\bar{q}}(\mathbf{l}_{\perp},\mathbf{l'}_{\perp} \vert \mathbf{l}_{1 \perp}) \nonumber \\
& \times \Big[\phi_{0} -\frac{2\alpha_{S}}{N_{c}\mathbf{l}_{1\perp}^{2}}        \Big(\phi_{Y_A}^{D}(\mathbf{l}_{1\perp}-\mathbf{l}_{\perp},\mathbf{l}_{\perp})+ \phi_{Y_A}^{D'}(\mathbf{l}_{1\perp}-\mathbf{l'}_{\perp},\mathbf{l'}_{\perp})+  \phi_{Y_A}^{Q}(\mathbf{l}_{1\perp}-\mathbf{l}_{\perp},\mathbf{l}_{\perp}; \mathbf{l}_{1\perp}-\mathbf{l'}_{\perp},\mathbf{l'}_{\perp}) \Big)   \Big] \enskip ,
\label{eq:H-tensor3}
\end{align}
where
\begin{equation}
\phi_{0}=(2\pi)^{8} \delta^{(2)}(\mathbf{l}_{1\perp}-\mathbf{l}_{\perp})  \delta^{(2)}(\mathbf{l}_{\perp})  \delta^{(2)}(\mathbf{l}_{1\perp}-\mathbf{l'}_{\perp})  \delta^{(2)}(\mathbf{l'}_{\perp})   \enskip .
\end{equation}

The form of Eq.~\ref{eq:H-tensor3} becomes particularly helpful in the next section where we will obtain $k_{\perp}$ and collinear factorized limits of these results and compare them to the corresponding leading twist pQCD results.

Following \cite{Benic:2016uku}, we define the products $\mathrm{d}^{6} K_{\perp}=\mathrm{d}^{2} \mathbf{p}_{\perp}  \mathrm{d}^{2} \mathbf{k}_{\perp} \mathrm{d}^{2} \mathbf{k}_{\gamma \perp}$ and $\mathrm{d}^{3} \eta_{K}=\mathrm{d} \eta_{p}  \mathrm{d} \eta_{k} \mathrm{d} \eta_{k_{\gamma}}$ and write the final form of the triple differential cross-section for inclusive photon production at LO as
\begin{equation}
\frac{\mathrm{d} \sigma}{\mathrm{d}x \mathrm{d} Q^{2} \mathrm{d}^{6} K_{\perp}\mathrm{d}^{3} \eta_{K}}= \frac{\alpha^{2}q^{4}_{f}y^{2}N_{c}}{512 \pi^{5} Q^{2}}  \frac{1}{(2\pi)^{4}}     \frac{1}{2q^{-}}  L^{\mu \nu}  \overset{\sim}{X}_{\mu \nu} (2\pi) \delta(P^{-}-q^{-}) \enskip ,
\label{eq:triple-differential}
\end{equation}
where $L^{\mu \nu}$ and $\overset{\sim}{X}_{\mu \nu}$ are given by Eqs.~\ref{eq:L-tensor} and \ref{eq:H-tensor3} respectively.
In $e+A$ DIS data, the above results can be applied to measurements of isolated photons accompanied by two jets or a quarkonium state, such as the $J/\Psi$ meson. Alternatively, one may integrate over the quark or antiquark to obtain the differential cross-section for direct photon plus jet production. Such measurements were performed for $e+p$ DIS at HERA \cite{Abramowicz:2012qt} for a wide range of $Q^{2}$ and transverse energy of the final state photon $E_{\gamma \perp}$, and are also feasible at future electron-nucleus collider facilities. 

The single inclusive differential cross-section for inclusive prompt photon production. This is obtained by integrating over the quark and antiquark momenta and rapidities. We obtain
\begin{equation}
\frac{\mathrm{d}\sigma}{\mathrm{d}x \mathrm{d}Q^{2} \mathrm{d}^{2}\mathbf{k}_{\gamma \perp} \mathrm{d}\eta_{k_{\gamma}}} = \frac{\alpha^{2}q^{4}_{f}y^{2}N_{c}}{512 \pi^{5} Q^{2}} \frac{1}{2q^{-}} \int_{0}^{+ \infty} \frac{\mathrm{d}k^{-}}{k^{-}} \int_{0}^{+ \infty} \frac{\mathrm{d}p^{-}}{p^{-}} \int_{\mathbf{k}_{\perp},\mathbf{p}_{\perp}} L^{\mu \nu}  \overset{\sim}{X}_{\mu \nu} (2\pi) \delta(P^{-}-q^{-}) \enskip ,
\label{eq:single-differential}
\end{equation}
where $\overset{\sim}{X}_{\mu \nu}$ is given by Eq.~\ref{eq:H-tensor3}.

%%%%%%%%%%%%%%%%%%%%%%%%%%%%%%%%%%%%%%%%%%%%%%%%%%

%%%%%%%%%%%%%%%%%%%%%%%%%%%%%%%%%%%%%%%%%%%%%%%%%%%%
\section{Properties of the photon production amplitude at LO} \label{sec:sectionIV}

\subsection{$k_{\perp}$-factorization and collinear factorization limits}
%%%%%%%%%%%%%%%%%%%%%%%%%%%%%%%%%%%%%%%%%%%%%%%%%%%%
In this section, we shall consider the large transverse momentum, $\mathbf{l}_{1 \perp}$ limit of the unintegrated distributions defined in Eqs.~\ref{eq:unintegrated-dipole} - \ref{eq:unintegrated-quadrupole}. This is equivalent to expanding the Wilson line $\tilde{U}(\mathbf{x}_{\perp})$ defined in Eq.~\ref{eq:wilson-fund}, to lowest nontrivial order in $\rho_{A}/\nabla^{2}_{\perp}$ and using the relations
\begin{equation}
\left \langle \rho_{A}^{a} (x^{-},\mathbf{x}_{\perp}) \rho_{A}^{b} (y^{-},\mathbf{y}_{\perp}) \right \rangle = \delta^{ab} \delta(x^{-}-y^{-}) \delta^{(2)} (\mathbf{x}_{\perp}-\mathbf{y}_{\perp}) \lambda_{A}(x^{-}) \enskip ,
\end{equation}
where
\begin{equation}
\int \mathrm{d} x^{-} \lambda_{A}(x^{-}) =\mu^{2}_{A}  \enskip,
\end{equation}
for Gaussian random color sources in a large nucleus  \cite{McLerran:1993ni,McLerran:1993ka,McLerran:1994vd,Iancu:2003xm}. This follows from the arguments outlined in the McLerran-Venugopalan (MV) model describing the nonperturbative gluon distribution at the onset of quantum evolution. In the MV model, $\mu^{2}_{A}=A/2\pi R^{2} \sim A^{1/3}$ is the average color charge squared of the valence quarks per color and per unit transverse area of a nucleus having mass number, A. Although there is no transverse coordinate dependence in $\mu^{2}_{A}$ in the MV model, explicit numerical solutions \cite{Rummukainen:2003ns,Dumitru:2011vk} of the Balitsky-JIMWLK hierarchy demonstrate that  a nonlocal $\mu^{2}_{A} \rightarrow \mu^{2}_{A}(\mathbf{x}_{\perp})$ well-approximates these numerical solutions when   
\begin{equation}
\phi_{Y_{A}}(\mathbf{l}_{1\perp})=\frac{2\pi N_{c}C_{F}g^{2}}{\mathbf{l}_{1\perp}^{2}} \int_{\mathbf{x}_{\perp}} \mu^{2}_{A}(Y_{A},\mathbf{x}_{\perp}) \enskip ,
\end{equation}
satisfies the BK equation.  In the limit of large transverse momentum, $\phi_{Y_{A}}(\mathbf{l}_{1\perp})$ evolves according to the BFKL equation~\cite{Kuraev:1977fs,Balitsky:1978ic}. 

Employing these ingredients, we obtain the following leading twist expression for the hadronic tensor:
\begin{equation}
 \overset{\sim}{X}_{\mu \nu}^{\text{LT}} =\frac{2 \alpha_{S}}{N_{c}} \frac{\phi_{Y_{A}}(\mathbf{P}_{\perp})}{\mathbf{P}_{\perp}^{2}} \enskip \Theta_{\mu \nu}(\mathbf{P}_{\perp}) \enskip ,
\label{eq:kT-gluon}
\end{equation}
where
\begin{equation}
\Theta_{\mu \nu}(\mathbf{P}_{\perp})= \text{Tr} \Big[ (\slashed{k}+m) {\big( T^{(\bar{q})}(\mathbf{P}_{\perp})-T^{(q)}(\mathbf{P}_{\perp})\big)_{\nu}}^{\alpha} (m-\slashed{p}) \hat{\gamma}^{0} \big( {T^{(\bar{q})}}^{\dagger}(\mathbf{P}_{\perp})-{T^{(q)}}^{\dagger}(\mathbf{P}_{\perp})\big)_{\mu \alpha} \hat{\gamma}^{0}  \Big] \enskip 
\end{equation}
is obtained using Eqs.~\ref{eq:identity-1} and \ref{eq:identity-2} respectively in the Dirac trace defined by Eq.~\ref{eq:Dirac-trace-general}.

\begin{figure}[!htb]
\begin{center}
\includegraphics[scale=0.2]{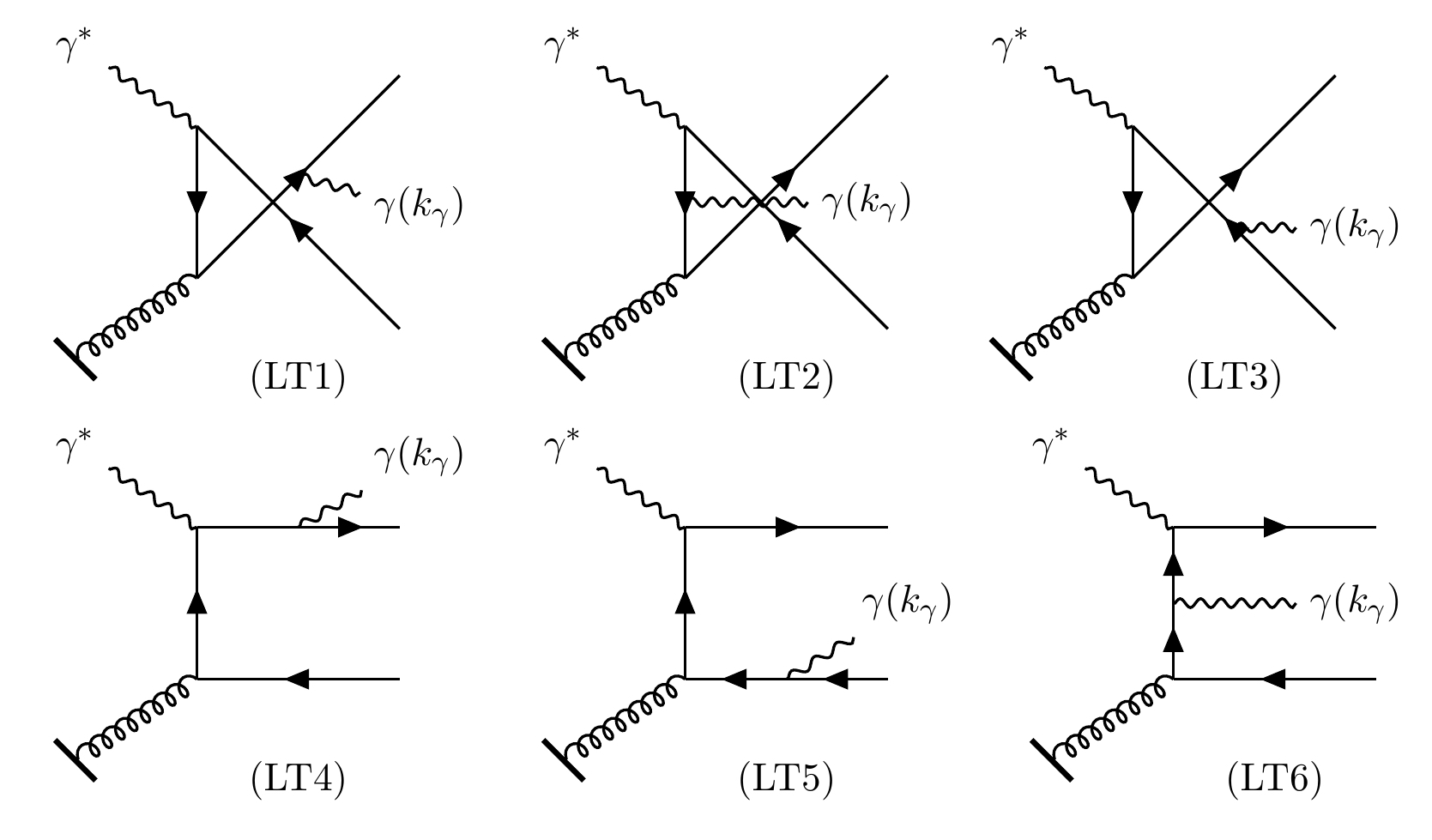}
\caption{Photon-gluon fusion diagrams contributing to the amplitude at leading twist, $\mathcal{O}(\rho_{A})$. The nuclear source is represented by the lower bold line. \label{fig:leading-twist}}
\end{center}
\end{figure}

An alternate way to arrive at the above expression is to compute the amplitude perturbatively. This can be done straightforwardly in Lorenz gauge by expanding the Wilson lines appearing in the amplitude expressions corresponding to processes (1)-(10). However the leading twist contribution is of $\mathcal{O}(\rho_{A})$; therefore, processes (7)-(10) containing two Wilson line insertions will not be considered here. These leading twist diagrams are shown in Fig. \ref{fig:leading-twist}.
The leading twist amplitude can be written as
\begin{equation}
\mathcal{M}^{\text{LT}}(\mathbf{q},\mathbf{k},\mathbf{p},\mathbf{k}_{\gamma})= \epsilon^{* \alpha}(\mathbf{k}_{\gamma},\lambda) \epsilon^{\mu}(\mathbf{q},\lambda') \mathcal{M}^{\text{LT}}_{\mu \alpha} (\mathbf{q},\mathbf{k},\mathbf{p},\mathbf{k}_{\gamma})  \enskip ,
\end{equation}
where
\begin{equation}
\mathcal{M}^{\text{LT}}_{\mu \alpha} (\mathbf{q},\mathbf{k},\mathbf{p},\mathbf{k}_{\gamma})= A^{\nu}_{a,A}(P-q) m^{a}_{\alpha \mu \nu} (\mathbf{q},\mathbf{k},\mathbf{p},\mathbf{k}_{\gamma}) \enskip . \label{eq:leading-twist-amplitude}
\end{equation}
The Fourier transform of the gluon field is given by
\begin{equation}
 A^{\nu}_{a,A}(P-q)= 2\pi g \delta^{\nu +} \delta(P^{-}-q^{-}) \enskip \frac{\rho_{a,A}(\mathbf{P}_{\perp})}{\mathbf{P}_{\perp}^{2}} \enskip ,
 \label{eq:classical-color-field-FT}
\end{equation}
and 
\begin{equation}
m^{a}_{\alpha \mu \nu} (\mathbf{q},\mathbf{k},\mathbf{p},\mathbf{k}_{\gamma}) = \sum_{\beta=1}^{6} m^{\beta, a}_{\alpha \mu \nu} (\mathbf{q},\mathbf{k},\mathbf{p},\mathbf{k}_{\gamma}) \enskip ,
\end{equation}
corresponds to the contributions from the leading twist diagrams (LT1)-(LT6), which can be individually written as
\begin{align}
m^{\beta,a}_{\alpha \mu +} (\mathbf{q},\mathbf{k},\mathbf{p},\mathbf{k}_{\gamma}) =  
\begin{cases}
 -ie^{2} g q^{2}_{f} \enskip \overline{u}(\mathbf{k}) R^{(\beta)}_{\alpha \mu} (\mathbf{P}_{\perp}) t^{a} v(\mathbf{p}) \enskip , \quad \beta=1,2,3  \\
 + ie^{2} g q^{2}_{f} \enskip  \overline{u}(\mathbf{k}) R^{(\beta)}_{\alpha \mu} (\mathbf{P}_{\perp}) t^{a} v(\mathbf{p}) \enskip , \quad \beta=4,5,6 \enskip ,
\end{cases}
\label{eq:leading-twist-individual}
\end{align}
where the various R-factors are given by Eqs.~\ref{eq:R1}, \ref{eq:R2-R3} and \ref{eq:R4-R5-R6} respectively. Using these relations~\ref{eq:classical-color-field-FT} and \ref{eq:leading-twist-individual} in Eq.~\ref{eq:leading-twist-amplitude}, we get 
\begin{equation}
\mathcal{M}^{\text{LT}}_{\mu \alpha} (\mathbf{q},\mathbf{k},\mathbf{p},\mathbf{k}_{\gamma})= -ie^{2}g^{2}q^{2}_{f}(2\pi) \delta(P^{-}-q^{-})\,\frac{\rho_{a,A}(\mathbf{P}_{\perp})}{\mathbf{P}_{\perp}^{2}} \,  \Big[ \overline{u}(\mathbf{k}) \big(T^{(\bar{q})}(\mathbf{P}_{\perp})-T^{(q)}(\mathbf{P}_{\perp}) \big)_{\mu \alpha}t^{a} v(\mathbf{p})  \Big]  \enskip .
\end{equation}

Taking the modulus squared and using 
\begin{equation}
\left \langle \rho_{A}^{a} (\mathbf{P}_{\perp}) \rho_{A}^{b} (\mathbf{P}_{\perp}) \right \rangle_{Y_{A}} = \frac{\delta^{ab} \mathbf{P}_{\perp}^{2}}{2\pi N_{c}C_{F}g^{2}} \enskip \phi_{Y_{A}}(\mathbf{P}_{\perp}) \enskip ,
\end{equation}
for the correlator of color sources, we obtain
\begin{equation}
X^{\text{LT}}_{\mu \nu}   = - \left \langle \sum_{spins} {\mathcal{M}_{\mu \alpha}^{\text{LT}}}^{*} (\mathbf{q},\mathbf{k},\mathbf{p},\mathbf{k}_{\gamma}) {\mathcal{M}_{\nu}^{\text{LT}}}^{\alpha} (\mathbf{q},\mathbf{k},\mathbf{p},\mathbf{k}_{\gamma}) \right \rangle = 16 \pi^{2} \alpha^{2} N_{c} q^{4}_{f} \enskip \overset{\sim}{X}_{\mu \nu}^{\text{LT}} \enskip ,
\end{equation}
where is identical to the expression in Eq.~\ref{eq:kT-gluon}.

Finally, we can take the collinear limit of the leading twist expressions. In our case, this is equivalent to taking the limit $\mathbf{P}_{\perp} \rightarrow 0$ of $\Theta_{\mu \nu}(\mathbf{P}_{\perp}) / \mathbf{P}^{2}_{\perp}$. In analogy to the discussion in the $p+A$ case ~\cite{Gelis:2003vh,Benic:2016uku}, we verify that the Ward identity 
\begin{equation}
(P-q)^{\nu} m^{a}_{\alpha \mu \nu} (\mathbf{q},\mathbf{k},\mathbf{p},\mathbf{k}_{\gamma}) = 0 \enskip ,
\end{equation}
is indeed satisfied\footnote{This follows from the relations
\begin{align*}
(P-q)^{\nu} ( m^{4,a}_{\alpha \mu \nu} ) & = - (P-q)^{\nu} ( m^{2,a}_{\alpha \mu \nu} + m^{3,a}_{\alpha \mu \nu} ) \enskip , \nonumber \\
(P-q)^{\nu} ( m^{1,a}_{\alpha \mu \nu} ) & = - (P-q)^{\nu} ( m^{5,a}_{\alpha \mu \nu} + m^{6,a}_{\alpha \mu \nu} ) \enskip .
\end{align*} This Ward identity should not be confused with that for the final state photon; the corresponding proof is given in Appendix ~\ref{appendixB1}. }.
Since $P^{-}-q^{-}=0$, the Ward identity implies
\begin{equation}
m^{a}_{\alpha \mu +} = \frac{P_{i} m^{a}_{\alpha \mu i} }{P^{+}-q^{+}} \enskip ,
\end{equation}
so that $m^{a}_{\alpha \mu +}$ vanishes linearly with $\mathbf{P}_{\perp}$. Thus the limit
\begin{equation}
\lim_{\mathbf{P}_{\perp} \rightarrow 0} \frac{\Theta_{\mu \nu}(\mathbf{P}_{\perp})}{\mathbf{P}^{2}_{\perp}} \enskip ,
\end{equation}
is well-defined, leading to the collinear perturbative QCD result for the hadron tensor:
\begin{equation}
\overset{\sim}{X}_{\mu \nu}^{\rm{coll.}} = \frac{2 \alpha_{S}\pi^{2}}{N_{c}} \enskip (2\pi)^{2} \delta^{(2)} (\mathbf{p}_{\perp}+\mathbf{k}_{\perp}+\mathbf{k}_{\gamma \perp})\enskip  x_{A} f_{g,A}(x_{A},Q^{2}) \lim_{\mathbf{P}_{\perp} \rightarrow 0} \frac{\Theta_{\mu \nu}(\mathbf{P}_{\perp})}{\mathbf{P}^{2}_{\perp}} \enskip .
\label{eq:collinear-gluon}
\end{equation}
Note that the l.h.s is proportional to the nuclear gluon distribution at small $x$,
\begin{equation}
x_{A} f_{g,A} (x_{A},Q^{2}) = \frac{1}{\pi^{2}} \int_{\mathbf{l}_{1\perp}} \phi_{Y_{A}}(\mathbf{l}_{1 \perp})  \enskip .
\end{equation}

The above result should be compared to inclusive photon production at NLO or $\mathcal{O}(\alpha_{S})$ in the power counting of pQCD, first analyzed in detail by Aurenche \textit{et. al.} \cite{Aurenche:1983hc} and later for photon plus jet cross-sections in~\cite{GehrmannDeRidder:2000ce, Kramer:1997nb, Michelsen:1995ag}. Excluding photoproduction, the NLO contributions come from two classes of inelastic diagrams; the photon-gluon fusion processes with a prompt photon in the final state and bremsstrahlung processes with a valence quark interacting directly with the virtual photon.  In the power counting of CGC, the latter form a subset of the Class I processes. As clearly shown in Eq. (10) of \cite{Aurenche:1983hc}, the cross-section for photon-gluon fusion processes clearly dominate the other inelastic processes at small $x$ by virtue of being proportional to the nucleon/nuclear gluon distribution in $e+p$/$e+A$ inclusive direct photon production. 

This correspondence between leading twist approximation of CGC  and conventional collinearly factorized pQCD results is useful. Besides corroborating our framework in a well-defined limit, it allows one to extract subleading power corrections to the nuclear gluon distribution.  Further, one recovers higher order results (in the collinear pQCD power counting)  that are the dominant NLO pQCD corrections to inclusive prompt photon production in the limit when both energies and virtualities are large even though $x \ll 1$. 

\subsection{Inclusive dijet cross-section in the soft photon limit} \label{sec:soft-photon-limit}
We will show here that in the infrared limit, $k_{\gamma} \rightarrow 0$, the Low-Burnett-Kroll theorem is satisfied by the inclusive photon production amplitude. Towards this end, we shall consider only the hadronic subprocesses for which the amplitude is given by Eqs. \ref{eq:hadronic-amp-master} and \ref{eq:final-amplitude}. In this limit, only the processes (7) and (8) (see Fig. \ref{fig:Group-III}) correspond to the photon being radiated after scattering from the nucleus and hence possess the desired structures
\begin{equation}
\frac{\slashed{k}+\slashed{k}_{\gamma}+m}{(k+k_{\gamma})^{2}-m^{2}} \quad \text{or} \quad \frac{\slashed{p}+\slashed{k}_{\gamma}-m}{(p+k_{\gamma})^{2}-m^{2}}  \enskip ,
\end{equation}
where the denominators linearly diverge as $k_{\gamma} \rightarrow 0$
\begin{equation}
(p+k_{\gamma})^{2}-m^{2} \rightarrow 2p.k_{\gamma} \enskip, \qquad (k+k_{\gamma})^{2}-m^{2} \rightarrow 2k.k_{\gamma} \enskip ,
\end{equation}
and the numerators stay finite
\begin{align}
\overline{u}(\mathbf{k}) \gamma_{\alpha}( \slashed{k}+\slashed{k}_{\gamma}+m) & \rightarrow  2k_{\alpha} \overline{u}(\mathbf{k}) \enskip,  \nonumber \\
(\slashed{p}+\slashed{k}_{\gamma}-m) \gamma_{\alpha} v(\mathbf{p}) & \rightarrow 2p_{\alpha} v(\mathbf{p}) \enskip ,
\end{align}
The contributions from diagrams (9) and (10) are of $\mathcal{O}(k_{\gamma}^{0})$ in the soft photon limit and are therefore subleading compared to those from diagrams (7) and (8). Combining the two leading contributions in the soft photon limit\footnote{As we shall show in the next subsection, and in greater detail in Appendix \ref{appendixC}, the structures in diagrams (1)--(6) are automatically included in 
diagrams (7)--(10) by a slight modification of the effective vertex. It is therefore sufficient to consider here diagrams (7)--(10) alone.}, we get 
\begin{equation}
\mathcal{M}_{\mu} (\mathbf{q},\mathbf{k}, \mathbf{p}, \mathbf{k}_{\gamma})  \rightarrow - (eq_{f}) \epsilon^{* }_{\alpha} (\mathbf{k}_{\gamma}, \lambda) \Big( \frac{p^{\alpha}}{p.k_{\gamma}} -\frac{k^{\alpha}}{k.k_{\gamma}} \Big) \mathcal{M}_{\mu}^{NR} (\mathbf{q},\mathbf{k}, \mathbf{p} )  \enskip ,
\end{equation}
where $\mathcal{M}_{\mu}^{NR} (\mathbf{q},\mathbf{k}, \mathbf{p} ) $ is the amplitude for inclusive dijet production computed in the CGC framework at small $x$ in \cite{Gelis:2002nn,Dominguez:2011wm}. The explicit expression is given by Eq. \ref{eq:nonradiative-DIS-amp}. It is then straightforward to calculate the differential cross-section for inclusive dijet production and express it in the familiar dipole factorized form. Details of this computation are given in Appendix \ref{appendixB2}. 

The final result can be expressed as 
\begin{align}
\frac{\mathrm{d}\sigma^{L,T}}{\mathrm{d}^{3}k \mathrm{d}^{3}p} & = \alpha\, q_{f}^{2} N_{c} \delta(q^{-}-p^{-}-k^{-}) \int \frac{\mathrm{d}^{2}\mathbf{x}_{\perp}}{(2\pi)^{2}} \frac{\mathrm{d}^{2}\mathbf{x'}_{\perp}}{(2\pi)^{2}} \frac{\mathrm{d}^{2}\mathbf{y}_{\perp}}{(2\pi)^{2}} \frac{\mathrm{d}^{2}\mathbf{y'}_{\perp}}{(2\pi)^{2}} \enskip e^{-i\mathbf{k}_{\perp}.(\mathbf{x}_{\perp}-\mathbf{x'}_{\perp})}  e^{-i\mathbf{p}_{\perp}.(\mathbf{y}_{\perp}-\mathbf{y'}_{\perp})} \nonumber \\
& \times \sum_{\alpha, \beta} \psi^{L,T}_{\alpha \beta}(q^{-},z,\vert \mathbf{x}_{\perp}-\mathbf{y}_{\perp} \vert ) \enskip \psi^{L,T *}_{\alpha \beta}(q^{-},z,\vert \mathbf{x'}_{\perp}-\mathbf{y'}_{\perp} \vert ) \times  \Xi(\mathbf{x}_{\perp},\mathbf{y}_{\perp};\mathbf{x'}_{\perp},\mathbf{y'}_{\perp}) \enskip ,
\end{align}
where $\Xi$ is given by Eq. \ref{eq:dipole-quadrupole-correlators}. Our result exactly matches that obtained previously in \cite{Dominguez:2011wm}  (see Eq. (22) of the same)  for inclusive dijet production at small $x$. This is a versatile result, because as argued in \cite{Dominguez:2011wm,Dominguez:2011br}, measurements of this cross-section in $e+A$ collisions provides a direct channel to extract the nuclear Weizs\"{a}cker-Williams distribution~\cite{McLerran:1998nk,Kovchegov:1998bi}. 

\subsection{The dressed fermion propagator} \label{sec:dressed-fermion}
In this subsection, we will outline a convenient modification of the momentum space Feynman rules for the fermion propagator in the background classical color field which enables one to efficiently recover our results thus far. As we shall also shortly discuss, this modification is valid for the dressed gluon propagator as well.

\begin{figure}[!htbp]
\begin{center}
\includegraphics[scale=0.2]{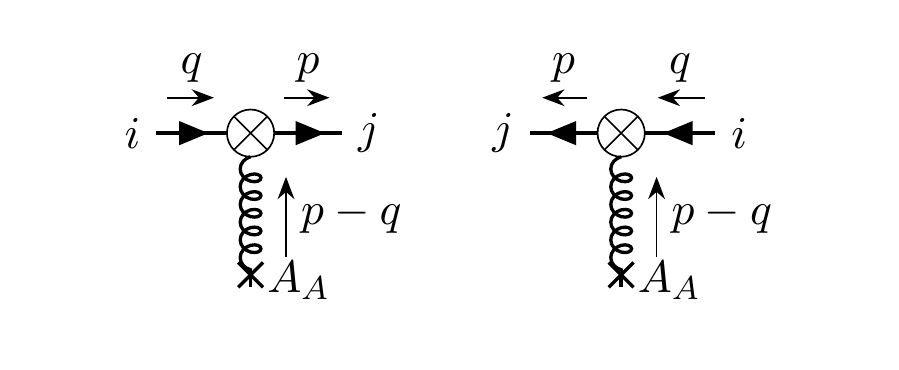}
\caption{Diagrammatic representation of the modified effective vertices for the quark and antiquark respectively. The crossed dots represent all possible scatterings off the classical color field of the nucleus including the case of no scattering. The vertex factors are given by Eq. \ref{eq:mod-effective-quark-vertex}.
\label{fig:effective-vertex_new}}
\end{center}
\end{figure}

We begin by observing that the factor of unity subtracted from the Wilson line in the effective vertex for the dressed fermion propagator (given in Eq. \ref{eq:effective-vertex}) imposes the requirement that the quark (antiquark) has to necessarily scatter off the nucleus. If we now redefine the effective vertex as
\begin{equation}
\overset{\sim}{\mathcal{T}}_{ji}(q,p)= \pm (2 \pi)\delta(p^{-}-q^{-}) \gamma^{-} \int \mathrm{d}^{2} \mathbf{x}_{\perp} e^{i(\mathbf{q}_{\perp} - \mathbf{p}_{\perp}).\mathbf{x}_{\perp}} \tilde{U}^{(\pm)}(\mathbf{x}_{\perp})_{ji} \enskip , \label{eq:mod-effective-quark-vertex}
\end{equation}
where the plus (minus) signs again correspond respectively to insertions on quark (antiquark) line, this includes the possibility of ``no scattering'' off the quark or antiquark line.  The LO amplitude expression in Eq. \ref{eq:final-amplitude} then has a simple physical interpretation. The first part proportional to $\tilde{U}(\mathbf{x}_{\perp}) \tilde{U}^{\dagger} (\mathbf{y}_{\perp}) $ represents all possible scatterings of the quark and antiquark off the nucleus, {\it including the case in which they do not scatter at all}. We can represent this diagrammatically as the sum of contributions from processes labeled (7)-(10) in Fig. \ref{fig:Group-III} but with the blobs replaced by crossed dots as shown in Fig. \ref{fig:effective-vertex_new}.
The second part which is proportional to unity gives the ``no scattering'' contribution and must be subtracted to get the physical result.\footnote{This can be shown by using Eq. \ref{eq:identity-1} and the following relations
\begin{align*}
R^{(1)}_{\mu \alpha}(\mathbf{0}_{\perp}) + R^{(2)}_{\mu \alpha}(\mathbf{0}_{\perp}) & = \frac{1}{q^{+}-P^{+}+\frac{i \varepsilon}{2k^{-}}} \frac{2k_{\alpha}+\gamma_{\alpha}\slashed{k}_{\gamma}}{(k+k_{\gamma})^{2}-m^{2}+i\varepsilon} \gamma_{\mu} \enskip , \nonumber \\
R^{(3)}_{\mu \alpha}(\mathbf{0}_{\perp}) & =  \frac{1}{q^{+}-P^{+}+\frac{i \varepsilon}{2k^{-}}} \gamma_{\mu} \frac{-(2p_{\alpha}+\slashed{k}_{\gamma}\gamma_{\alpha})}{(p+k_{\gamma})^{2}-m^{2}+i\varepsilon} \enskip .
\end{align*} 
to decompose the second part as the sum of two processes in which both quark and antiquark do not scatter off the nucleus. }  
As we will demonstrate in further detail in Appendix \ref{appendixC}, this seemingly simple modification offers a significant advantage because it is sufficient to compute fewer diagrams in the LO computation than previously. An identical argument holds for the effective vertex for the dressed gluon propagator. These modified quark and gluon vertices will prove very valuable in computations at NLO and higher orders.
 
\section{Components of the NLO computation} 
\label{sec:sectionV}
\subsection{Advantages of the ``wrong" light cone gauge}
Albeit computations of physical observables are gauge independent, an appropriate gauge choice can result in a significant simplification of our path to the final result. In DIS, an attractive gauge choice, in the infinite momentum frame (IMF) where $P^{+}_{N} \rightarrow \infty$,  is the light cone gauge $A^+=0$ . This follows from the operator definition of parton distribution functions (PDFs), which are defined respectively for quark and gluon fields as~\cite{Diehl:2003ny}
\begin{align}
F^{q} &= \frac{1}{2} \int \frac{\mathrm{d}z^{-}}{2 \pi} \enskip e^{i xP^{+}_{N}z^{-}} \langle P_{N} \vert  \enskip  \bar{q}\Big(-\frac{z}{2} \Big) \gamma^{+} W \Big( -\frac{z^{-}}{2}, \frac{z^{-}}{2} \Big) q\Big( \frac{z}{2} \Big) \enskip  \vert P_{N} \rangle {\Bigg|}_{z^{+}=0, \mathbf{z}_{\perp}=0}  \enskip , \nonumber \\
F^{g} &= \frac{1}{P^{+}_{N}} \int \frac{\mathrm{d}z^{-}}{2 \pi} \enskip e^{i x P^{+}_{N} z^{-}} \langle P_{N} \vert \enskip  G^{+ \mu} \Big( -\frac{z}{2} \Big) W\Big( -\frac{z^{-}}{2}, \frac{z^{-}}{2} \Big) {G_{\mu}}^{+} \Big(\frac{z}{2}\Big) \enskip  \vert P_{N} \rangle {\Bigg|}_{z^{+}=0, \mathbf{z}_{\perp}=0}  \enskip .
\end{align}
In the IMF, the $W$'s are lightlike gauge links which are defined to be 
\begin{equation}
W(a,b)= \mathcal{P}_{-} \text{exp} \Big( ig \int_{b}^{a} \mathrm{d}x^{-} A^{+} \Big) \enskip .
\end{equation}
These are of course unity in in $A^{+}=0$ gauge. In a covariant representation, with off-shell partons and on-shell hadrons, a simple interpretation of the PDFs in terms of Green functions is possible in this gauge~\cite{Diehl:2003ny} . Similarly, in a noncovariant representation with on-shell partons, or in other words, light cone quantization \cite{Kogut:1969xa,Bjorken:1970ah,Brodsky:1995rn,Brodsky:1997de,Venugopalan:1998zd}, a number density interpretation of the quark and gluon PDFs is manifest in $A^{+}=0$ gauge.

Though $A^+=0$ is the right gauge choice for the reasons outlined, it turns out that the ``wrong"  gauge choice of $A^-=0$ is the right one for computations in shock wave background fields. Firstly, we note that in the IMF $P^{+}_{N} \rightarrow \infty$,  the only large component of the current density is $J^{+}$ with other components suppressed by $1/P^{+}_{N}$. The current conservation equation is therefore equivalent to 
\begin{equation}
D_{+} J^{+} =0 \enskip .
\end{equation}
It ensures that  $J^{+}$ is independent of the light cone time $x^{+}$, or equivalently, is a static charge on the light cone.  This is implicitly assumed in Eq. \ref{eq:Yang-Mills} and in the  subsequent analysis. The solution of the classical Yang-Mills equations in $A^-=0$ gauge gives~\cite{McLerran:1993ni,McLerran:1993ka} 
\begin{equation}
A^{+}=0, \enskip F^{ij}=0 \qquad \text{with} \enskip A^{+}, A^{i}\,\, x^+\,\,\text{independent} \enskip ,
\end{equation}
a solution that is also of course valid in the right light cone gauge $A^+=0$. 

However unlike $A^+=0$ gauge, $A^-=0$ gauge is related to the Lorenz gauge $\partial_\mu A^\mu=0$ by a simple gauge transformation. Note that our LO computations thus far were in the latter gauge, where the fermion propagator had a particularly simple structure~\cite{McLerran:1998nk,Venugopalan:1999wu}. Most importantly, as noted previously~\cite{Ayala:1995kg,Ayala:1995hx}, the structure of the small fluctuation propagator in $A^-=0$ gauge is far simpler than that in $A^+=0$ gauge. Further, this dressed gluon propagator, when transformed to Lorenz gauge, has a structure that is identical to that of the fermion propagator. 

Leading order computations in shock wave backgrounds, comparing and contrasting results in the Lorenz and $A^-=0$ light cone gauge have been performed previously in the context of $p+A$ collisions. While it was noted that results for gluon production in $A^-=0$ gauge~\cite{Gelis:2005pt} are less cumbersome relative to those in Lorenz gauge~\cite{Blaizot:2004wu}, there is no particular virtue of one or the other gauge for quark pair production~\cite{Blaizot:2004wv} or inclusive photon production~\cite{Benic:2016uku}. Because the shock wave background fields in the two gauges are simply related, intermediate steps may be performed more efficiently in Lorenz gauge before transforming the final result to $A^-=0$ gauge. 

\subsection{Gluon ``shock wave'' propagator in $A^{-}=0$ gauge}

We will begin by rederiving~\cite{Ayala:1995kg} the shock wave gluon propagator in $A^-=0$ gauge.  We can split the gauge field into a classical background field piece (parametrically of order $1/g$) and a small fluctuation piece (of order unity) as 
\begin{equation}
A^{\mu}(x)=B^{\mu}(x)+b^{\mu}(x) \enskip .
\end{equation}
Here $B^\mu (x)$ is the classical background field with the components
\begin{align}
B^{+}(x)=0, \enskip B^{-}(x)=0, \enskip B^{i}&=\theta(x^{-}) \kappa^{i}(\mathbf{x}_{\perp}) \enskip,  \qquad i=1,2 \nonumber \enskip , \\
\text{where} \quad \kappa^{i} (\mathbf{x}_{\perp})&= \frac{i}{g} \enskip U(\mathbf{x}_{\perp}) \partial^{i} U^{\dagger}(\mathbf{x}_{\perp}) \enskip ,
\end{align}
and $U$ is the Wilson line in the adjoint representation of $SU(N_{c})$, 
\begin{equation}
U(\mathbf{x_{ \perp}})=\mathcal{P}_{-} \text{exp}\left[-ig^{2} \int_{-\infty}^{+ \infty} \mathrm{d}z^{-} \frac{1}{\nabla^{2}_{\perp}}  \rho_{A}^{a} (z^{-},\mathbf{x_{ \perp}})T^{a} \right] \enskip .
\label{eq:wilson-adj}
\end{equation}
The equations of motion for the small fluctuation field $b(x)$ are
\begin{equation}
\Big( D(B)^{2}g^{\mu \nu} -D^{\mu} (B) D^{\nu} (B) \Big) b_{\nu} - 2F^{\mu \nu} b_{\nu}=0 \enskip .
\label{eq:small-fluctuation-eom}
\end{equation}
Specifically, for the transverse fields, we have a sourceless Klein-Gordon equation
\begin{equation}
[2\partial^{+} \partial^{-}- \mathbf{D}_{\perp}^{2}(B)] \enskip b^{i}(x)=0 \enskip ,
\label{eq:b_i-eom}
\end{equation}
after fixing the residual gauge freedom with the Gauss' law constraint $D^{\mu} (B) b_{\mu}=0$ \footnote{The covariant derivative, $D^{\mu}(B)=\partial^{\mu}-igB^{\mu,a}T^{a}$}. Taking $\mu=-$ in Eq. \ref{eq:small-fluctuation-eom}, we get another sourceless equation 
\begin{equation}
\partial^{-} \Big( \partial^{-}b^{+}- \mathbf{D}_{\perp}.\mathbf{b}_{\perp} \Big)=0 \enskip ,
\end{equation}
which is the Gauss' law constraint imposed on initial field configurations~\cite{Gelis:2007kn,Gelis:2008rw,Gelis:2008ad}. 

Finally, using the above constraint to write
\begin{equation}
b^{+}=\frac{1}{\partial^{-}} \enskip  \mathbf{D}_{\perp}.\mathbf{b}_{\perp} \enskip ,
\label{eq:gauss-law-constraint}
\end{equation}
and keeping in mind the discontinuity of the spatial covariant derivative operator at $x^{-}=0$, 
we can verify that Eqs. \ref{eq:b_i-eom} and \ref{eq:gauss-law-constraint} are consistent with the final small fluctuation equation
\begin{equation}
[2\partial^{+}\partial^{-} - \mathbf{D}_{\perp}^{2}(B) ] \enskip b^{+}=-2\delta(x^{-})  \enskip   \kappa^{i} (\mathbf{x}_{\perp}) b_{i} \enskip .
\end{equation}
The transverse field is then assumed to be a simple plane wave for $x^{-}<0$ and a linear superposition of plane waves for $x^{-}>0$. This is done in a manner to ensure continuity of the fields across $x^{-}=0$, which again is possible because of the absence of delta function singularities in the above equations at $x^{-}=0$. The expressions for the Green's functions for the small fluctuations thus obtained are given in the next subsection.

In contrast, in the right light cone gauge $A^{+}=0$, the discontinuity of the fields at the origin and the corresponding singular structure of the electric fields do not allow one to simply  integrate the equations of motion. Hence it is convenient to first obtain the small fluctuation Green functions in $A^{-}=0$ gauge and then, if required,  gauge transform the results to $A^{+}=0$. The expressions for the propagator in $A^{+}=0$ gauge are complicated and not of relevance in our NLO computation.

Following the procedure outlined above and using the identity
\begin{equation}
\tilde{U}(\mathbf{x}_{\perp}) t^{a} \tilde{U}^{\dagger} (\mathbf{x}_{\perp})=t^{b} U^{ba}(\mathbf{x}_{\perp}) \enskip ,
\label{eq:fund-adjoint-relation}
\end{equation}
we can express the transverse components of the fluctuation field as \cite{McLerran:1994vd,Ayala:1995kg}, 
\begin{align}
b^{i}(x) =e^{ip.x} \eta^{i} \Big[ \theta(-x^{-})t^{a}+\theta(x^{-}) \int \frac{\mathrm{d}^{2}\mathbf{q}_{\perp}}{(2\pi)^{2}} \int \mathrm{d}^{2} \mathbf{z}_{\perp}  \enskip e^{-i(\mathbf{q}_{\perp}-\mathbf{p}_{\perp}).(\mathbf{x}_{\perp}-\mathbf{z}_{\perp})} e^{i(\mathbf{q}_{\perp}^{2}-\mathbf{p}_{\perp}^{2})\frac{x^{-}}{2 p^{-}}} \enskip t^{c} \Big(U(\mathbf{x}_{\perp}) U^{\dagger}(\mathbf{z}_{\perp}) \Big)^{ca} \Big] \enskip ,
\label{eq:small-fluctuation-comp-i}
\end{align}
where $\eta^{i}$ is a unit vector.
Using the above expression for $b^{i}$, we get 
\begin{align}
b^{+}(x)=\frac{1}{\partial_{y^{+}}} \enskip [\mathbf{D}_{\perp},\mathbf{b}_{\perp}]= \frac{1}{\partial_{y^{+}}} \enskip \Big[ e^{ip.x} \eta^{j} \Bigl\{ \theta(-x^{-}) ip^{j} t^{a} + &  \theta(x^{-}) \int \frac{\mathrm{d}^{2} \mathbf{q}_{\perp}}{(2\pi)^{2}} \int \mathrm{d}^{2} \mathbf{z}_{\perp} \enskip e^{-i(\mathbf{q}_{\perp}-\mathbf{p}_{\perp}).(\mathbf{x}_{\perp}-\mathbf{z}_{\perp})} e^{i(\mathbf{q}_{\perp}^{2}-\mathbf{p}_{\perp}^{2})\frac{x^{-}}{2 p^{-}}} \nonumber \\
& \times i(p^{j}+q^{j})   t^{c} \Big(U(\mathbf{x}_{\perp}) U^{\dagger}(\mathbf{z}_{\perp}) \Big)^{ca}  \Bigr\} \Big] \enskip .
\label{eq:small-fluctuation-comp-plus}
\end{align}
The small fluctuation gluon propagator in coordinate space is defined as
\begin{equation}
G^{\mu \nu;ab}(x,y) = \left \langle b^{\mu,a}(x) b^{\nu,b}(y) \right \rangle  \enskip . 
\label{eq:small-fluctuation-propagator-defn}
\end{equation}
The nonzero components of the propagator in this gauge are $G^{++}$, $G^{+i}$, $G^{i+}$ and $G^{ij}$. Using the expression for $G^{ij}$ in \cite{McLerran:1994vd,Ayala:1995kg} and Eq. \ref{eq:fund-adjoint-relation}, it is straightforward to derive the following alternate expressions useful for momentum space calculations:
\begin{align}
G^{ij;ab}(x,y)& =g(x^{-},\mathbf{x}_{\perp}) \enskip G^{ij;ab}_{0}(x,y) \enskip g^{\dagger}(y^{-},\mathbf{y}_{\perp}) -  \int \mathrm{d}^{4} z \enskip \delta(z^{-}) g(x^{-},\mathbf{x}_{\perp}) \Bigl\{ \theta(x^{-}) \theta(-y^{-}) [U^{\dagger}(\mathbf{z}_{\perp})-1] \nonumber \\
&  -\theta(-x^{-}) \theta(y^{-}) [U(\mathbf{z}_{\perp})-1] \Bigr\}_{cd} g^{\dagger}(y^{-},\mathbf{y}_{\perp}) \enskip G^{ik;ac}_{0}(x,z) (-i2\partial_{z^{+}}) G^{lj;db}_{0}(z,y)  \delta_{kl} \enskip ,
\label{eq:Gij-greenfunction}
\end{align}
where
\begin{equation}
G_{0}^{\mu \nu;ab}(x,y)=\int \frac{\mathrm{d}^{4}p}{(2\pi)^{4}} \enskip e^{ip.(x-y)}     \frac{i}{p^{2}+i\varepsilon} \Big( -g^{\mu \nu}+\frac{p^{\mu } n^{\nu}+p^{\nu}n^{\mu}}{n.p} \Big) \delta^{ab} , \qquad n^{\mu}=\delta^{\mu +} \enskip ,
\label{eq:gluon-propagator-LC}
\end{equation}
is the free gluon propagator in $A^{-}=0$ gauge.  Further,  
\begin{equation}
g(x^{-},\mathbf{x}_{\perp})=\theta(-x^{-})+\theta(x^{-}) U(\mathbf{x}_{\perp}) \enskip ,
\end{equation}
is a gauge transformation matrix that transforms the background field $B^\mu(x)$ to Lorenz gauge in which it is a singular field having only a `plus' component. The same matrix, albeit in the fundamental representation, appears while calculating the fermion propagator \cite{McLerran:1998nk} in the background classical field.

The momentum space expression for the gluon propagator in $A^- = 0$ gauge with the background field $B(x)$ in Lorenz gauge can be expressed as 
\begin{equation}
G^{ij;ab}(p,p')=(2\pi)^{4}\delta^{(4)}(p-p')  \enskip G^{ij;ab}_{0}(p)+G^{ik;ac}_{0}(p) \enskip {\mathcal{T}}_ {kl;cd}(p,p') \enskip G^{lj;db}_{0}(p') \enskip ,
\end{equation}
where
\begin{equation}
{\mathcal{T}}_ {ij;ab}(p,p') =2\pi \delta(p^{-}-p'^{-})\, (2p^{-}) \text{sign}(p^{-}) \int \mathrm{d}^{2} \mathbf{z}_{\perp} \enskip e^{i(\mathbf{p}_{\perp}-\mathbf{p'}_{\perp}).\mathbf{z}_{\perp} }[U^{\text{sign}(p^{-})}(\mathbf{z}_{\perp})-1]_{ab} \delta_{ij} \enskip .
\end{equation}
One can similarly derive the following momentum space relations  for the other nonzero components of $G^{\mu \nu}$:
\begin{align}
G^{++;ab}(p,p')& =(2\pi)^{4}\delta^{(4)}(p-p')  \enskip G^{++;ab}_{0}(p) + G^{+i;ac}_{0}(p) \enskip  {\mathcal{T}}_ {ij;cd}(p,p') \enskip G^{j+;db}_{0}(p') \enskip , \nonumber \\
G^{i+;ab}(p,p')& =(2\pi)^{4}\delta^{(4)}(p-p')  \enskip G^{i+;ab}_{0}(p) + G^{ij;ac}_{0}(p) \enskip {\mathcal{T}}_ {jk;cd}(p,p') \enskip G^{k+;db}_{0}(p')  \enskip.
\end{align}
The general form for the small fluctuation propagator in momentum space and in $A^{-}=0$  gauge can be compactly expressed as 
\begin{equation}
G^{\mu \nu;ab}(p,p')=(2\pi)^{4}\delta^{(4)}(p-p')  \enskip G^{\mu \nu;ab}_{0}(p)+G^{\mu \rho;ac}_{0}(p) \enskip  {\mathcal{T}}_ {\rho \sigma;cd}(p,p') \enskip G^{\sigma \nu;db}_{0}(p')  \enskip ,
\end{equation}
where the effective vertex of the dressed gluon shock wave is given by
\begin{equation}
{\mathcal{T}}_ {\mu \nu;ab}(p,p') =-2\pi \delta(p^{-}-p'^{-})\,(2p^{-}) g_{\mu \nu} \enskip \text{sign}(p^{-}) \int \mathrm{d}^{2} \mathbf{z}_{\perp} \enskip e^{i(\mathbf{p}_{\perp}-\mathbf{p'}_{\perp}).\mathbf{z}_{\perp} }[U^{\text{sign}(p^{-})}(\mathbf{z}_{\perp})-1]_{ab}  \enskip ,
\label{eq:effective-gluon-vertex}
\end{equation}
a form identical to (with the subsitution of the fundamental Wilson line with its adjoint counterpart) the dressed fermion effective vertex. 
While the gluon shock wave propagator in $A^-=0$ gauge was first derived in \cite{McLerran:1994vd,Ayala:1995kg}, this form for the propagator in momentum space is given in a later paper by Balitsky and Belitsky \cite{Balitsky:2001mr} albeit it is also implicit in \cite{Hebecker:1998kv,McLerran:1998nk}. 

The discussion in Sec. \ref{sec:dressed-fermion} however suggests that we should modify the effective gluon vertex to include the ``no scattering" contribution by omitting the unit matrix in Eq. \ref{eq:effective-gluon-vertex} above; the modified vertex we will use henceforth in computations is 
\begin{equation}
\overset{\sim}{\mathcal{T}}_ {\mu \nu;ab}(p,p') =-2\pi \delta(p^{-}-p'^{-}) \, (2p^{-}) g_{\mu \nu} \enskip \text{sign}(p^{-}) \int \mathrm{d}^{2} \mathbf{z}_{\perp} \enskip e^{i(\mathbf{p}_{\perp}-\mathbf{p'}_{\perp}).\mathbf{z}_{\perp} }U^{\text{sign}(p^{-})}_{ab} (\mathbf{z}_{\perp}) \enskip .
\label{eq:mod-effective-gluon-vertex}
\end{equation}

%\begin{figure}[H]
%\begin{center}
%\includegraphics[scale=0.2]{Fig9.jpg}
%\caption{Effective vertex for the gluon. a and b represent adjoint color labels. \label{fig:effective-gluon-vertex}}
%\end{center}
%\end{figure}

\begin{figure}[H]
\begin{center}
\includegraphics[scale=0.2]{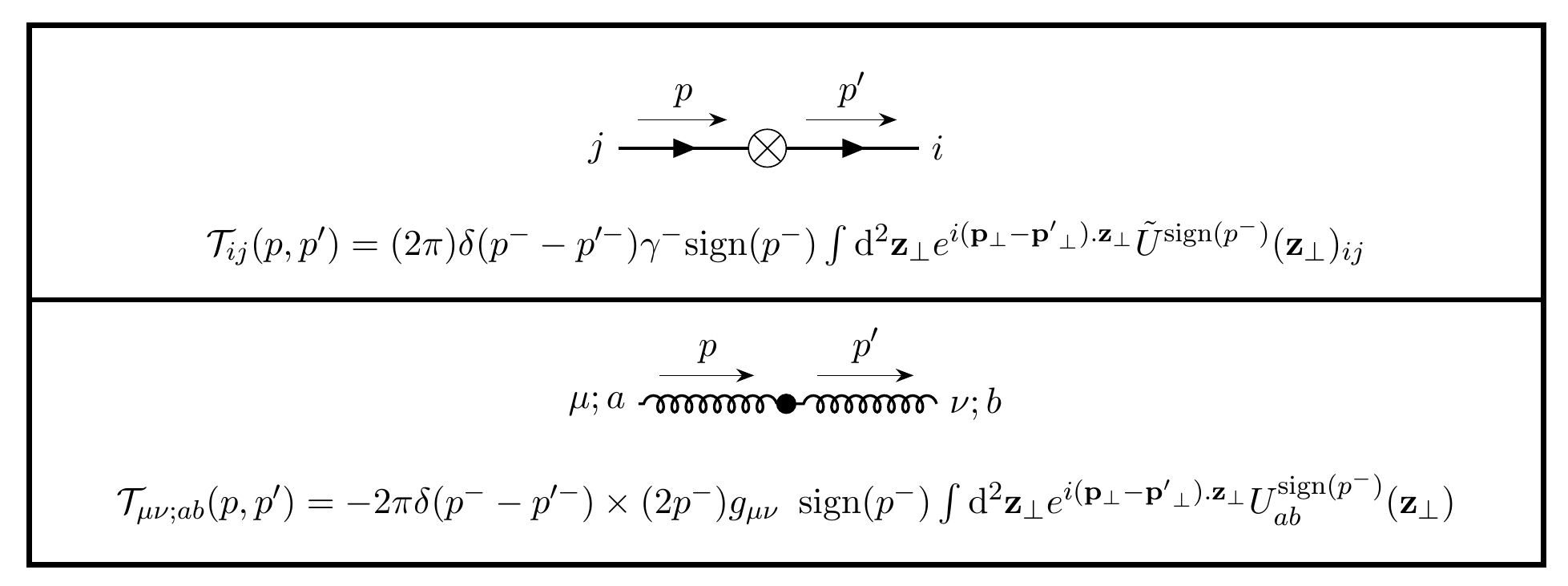}
\caption{Diagrammatic representation of the fermion (top) and gluon (bottom) effective vertices. Here, and henceforth, the `tilde' factors in Eqs.~\ref{eq:mod-effective-quark-vertex} and \ref{eq:mod-effective-gluon-vertex} are dropped.
\label{fig:fermion-gluon-together}
}
\end{center}
\end{figure}

To summarize, higher order all-twist computations at small $x$ can be performed with the conventional Feynman rules of covariant perturbation theory albeit with the dressed 
quark and gluon  effective vertices shown in Fig. \ref{fig:fermion-gluon-together}. 

\subsection{Processes contributing to the NLO amplitude} \label{sec:contributions-NLO}
Having outlined the necessary components for the computation of the inclusive photon amplitude to NLO, we will discuss here the relevant NLO processes. The general ideas were already presented in the introduction of this paper and we will elaborate further on that discussion. To begin with,  the perturbative expansion in Eq. \ref{eq:power-counting} of the expectation value of a general operator $\mathcal{O}$ calculated for a particular source charge density configuration $\rho_{A}$ can be expanded out to next-to-next-to-leading order (NNLO) as 
\begin{align}
\mathcal{O}[\rho_{A}]&= \underbrace{f^{0}_{0j} (g\rho_{A})^{j}}_\textrm{LO} +\underbrace{f^{0}_{1j}(g\rho_{A})^{j} \alpha_{S} +f^{1}_{1j}(g\rho_{A})^{j}  \alpha_{S}  \enskip \text{ln} (1/x) }_\textrm{NLO}  \nonumber \\
&+\underbrace{f^{0}_{2j}(g\rho_{A})^{j} \alpha_{S}^{2} +f^{1}_{2j}(g\rho_{A})^{j}\alpha_{S}^{2} \enskip  \text{ln}(1/x) +f^{2}_{2j}(g\rho_{A})^{j} \alpha^{2}_{S} \enskip \text{ln}^{2} (1/x) }_\textrm{NNLO} +\ldots  \enskip ,
\label{eq:expansion}
\end{align}
where the repeated indices are assumed to be summed over. For the differential cross-section for inclusive photon production, at each order in perturbation theory, there are contributions to the small $x$ evolution of the random color sources (encoded in the dipole and quadrupole Wilson line correlators given by Eq. \ref{eq:dipole-quadrupole-correlators}). These are enhanced by logarithms in $x$ and when $\alpha_S\ln(1/x)\sim 1$, they complicate the naive power counting. In addition, we will have non-log enhanced perturbative corrections to the light cone wavefunction for the virtual photon that fluctuates into a quark-antiquark dipole  (as well as ancillary complications due to the photon in the final state) that should be combined with the $\ln(1/x)$ contributions systematically. It is therefore efficient to reexpress the generic expansion in  Eq. \ref{eq:expansion} by factorizing the cross-section into a coefficient term and a matrix element of Wilson line correlators between nuclear wavefunctions, at each order in $\alpha_{S}$, as 
\begin{align}
\mathrm{d}\sigma [\rho_{A}]&= \underbrace{\mathcal{C}_{\text{LO}} \otimes \mathrm{d}\hat{\sigma}_{\text{LO}}}_{\alpha_{S}^{0}.\alpha_{S}^{0}} +\underbrace{\mathcal{C}_{\text{LO}} \otimes \mathrm{d} \hat{\sigma}_{\text{LLx}} }_{\alpha_{S}^{0}. \alpha_{S} \text{ln}(1/x)} + \underbrace{\mathcal{C}_{\text{NLO}} \otimes \mathrm{d} \hat{\sigma}_{\text{NLO}}}_{\alpha_{S}.\alpha_{S}^{0}}   \nonumber \\
& + \underbrace{\mathcal{C}_{\text{NLO}} \otimes \mathrm{d} \hat{\sigma}_{\text{LLx}}}_{\alpha_{S}. \alpha_{S} \text{ln}(1/x)} +\underbrace{ \mathcal{C}_{\text{LO}} \otimes \mathrm{d} \hat{\sigma}_{\text{NNLx}} }_{\alpha^{0}_{S} .\alpha^{2}_{S} \text{ln}(1/x)} +\ldots \enskip .
\label{eq:order-by-order}
\end{align}

We computed the first term in the r.h.s of the above equation in the previous sections. The second term is given by contributions from the diagrams of the type depicted in Fig. \ref{fig:NLO-LLx}. For $\alpha_{S}\,\text{ln}(1/x) \sim 1$, these contributions are as large as the LO contribution and can be absorbed in the LO result by a redefinition of the weight functional $W$. The resulting LLx RG equation (the JIMWLK equation) efficiently sums such leading contributions to all orders of perturbation theory while preserving the structure of the LO result. 

The third term appearing in Eq. \ref{eq:order-by-order} represents genuine $\alpha_{S}$ suppressed contributions that do not take into account the small $x$ evolution of the sources. In addition to the real emission diagrams shown in Figs. \ref{fig:NLO-coeff} and  \ref{fig:NLO-coeff-real}, there are contributions from loop graphs depicted by the second and third representative diagrams in Fig. \ref{fig:NLO-coeff}. Since the cross-section for inclusive gluon production in DIS was computed previously~\cite{Kovchegov:2001sc} at small $x$, taking the soft-photon limit to the real emission contributions will provide in principle a check of our results. The complete calculation of this term will give the next order correction $\mathcal{C}_{\text{NLO}}$ to the LO coefficient function or ``impact factor" $\mathcal{C}_{\text{LO}}$. It is worth mentioning here that the $\hat{\sigma}_{\text{NLO}}$ has a different correlator structure than $\hat{\sigma}_{\text{LO}}$ due to the additional adjoint Wilson lines and hence the distinct nomenclature. The overall contribution of the third term in Eq. \ref{eq:order-by-order} is strongly subleading at small $x$. 

\begin{figure}[H]
\begin{center}
\includegraphics[scale=0.2]{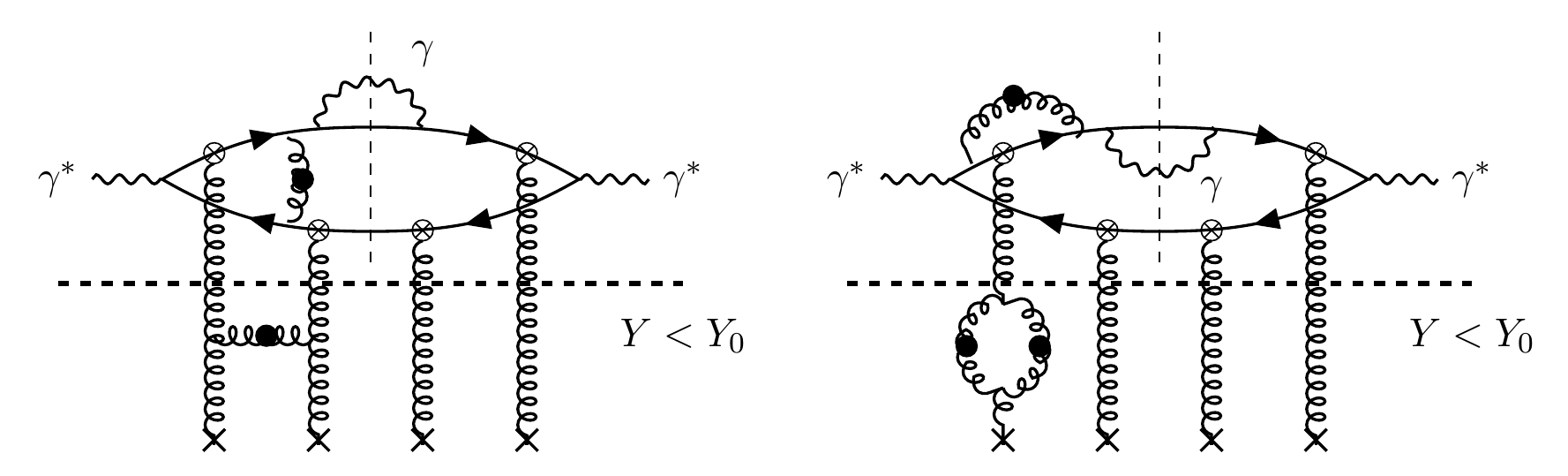}
\caption{ NNLO diagrams that are effectively $\alpha_{S}$ in ``size" for $\alpha_{S}\text{ln}(1/x) \sim 1$ at small $x$. The rapidity separation between fast sources and slow fields is shown clearly. Integration over modes with $Y > Y_{0} $ is represented by the loops in the upper part of the figure. The contributions below the rapidity cut are part of the LLx evolution of the color sources. See text for more details.
\label{fig:NLLx_one}
}
\end{center}
\end{figure}

The two terms in the second line of Eq. \ref{eq:order-by-order} are part of the NNLO contributions to inclusive photon production but are potentially quantitatively important because they are effectively of order $\alpha_{S}$ when $\alpha_S\ln(1/x)\sim 1$; these form the next-to-leading-log (NLLx) contribution to the cross-section. Under our RG ideology, the two terms can be considered separately. The first term in the second line of  Eq. \ref{eq:order-by-order} is represented by the diagrams in Fig. \ref{fig:NLLx_one}. The 
dashed horizontal line represents the rapidity scale $Y_0$ separating the classical fields that interact with the $q\bar{q}\gamma$ system from the static color sources RG evolved by the LLx JIMWLK equation from the target. The two diagrams shown are respectively the real and virtual contributions that are characteristic of the terms that generate the JIMWLK kernel. The computation of the NLO impact factor (above the $Y_0$ cut) using the techniques developed here is in progress and will be reported in a follow-up paper~\cite{Roy-Venugopalan-2}. A highly nontrivial check will be  to reproduce the NLO impact factor for fully inclusive DIS~\cite{Bartels:2000gt,Bartels:2002uz,Bartels:2001mv,Balitsky:2010ze,Beuf:2011xd,Beuf:2016wdz,Beuf:2017bpd,Hanninen:2017ddy} that should be recovered in the soft photon limit. It will also be important for the consistency of the framework to demonstrate JIMWLK factorization by explicit computation. 

\begin{figure}[!htbp]
\begin{center}
\includegraphics[scale=0.2]{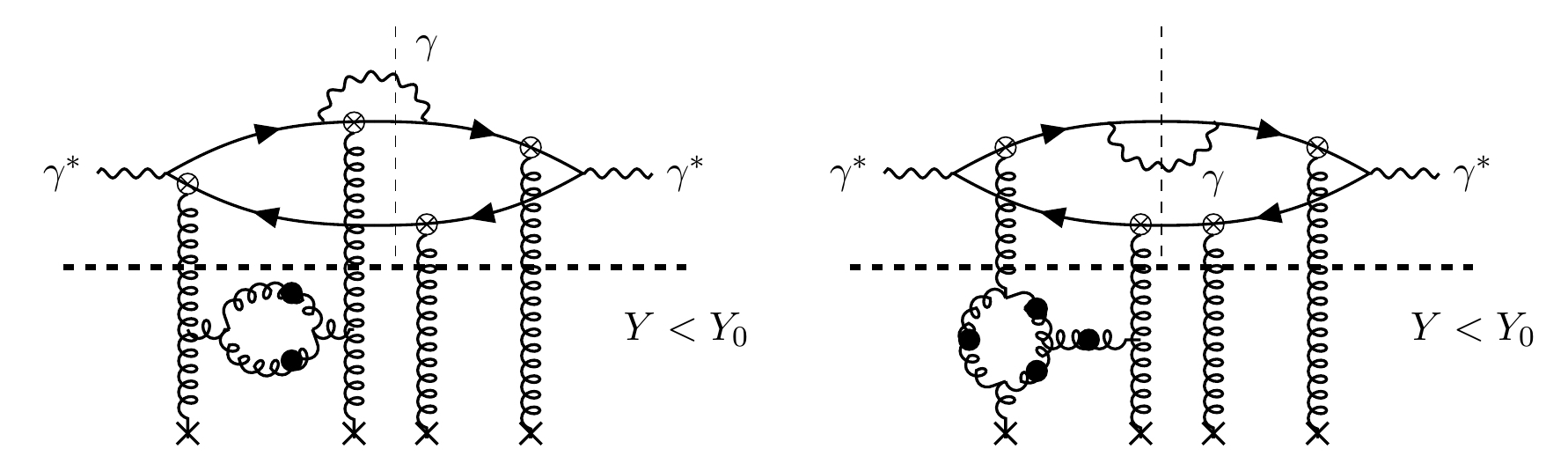}
\caption{NNLO diagrams that are effectively $\alpha_{S}$ in ``size" for $\alpha_{S}\text{ln}(1/x) \sim 1$ at small $x$. These diagrams contribute to the NLO JIMWLK kernel. 
\label{fig:NLLx_two}
}
\end{center}
\end{figure}

The final term in Eq. \ref{eq:order-by-order} is represented by the diagrams in Fig. \ref{fig:NLLx_two}. In this case, there are no radiative corrections above the rapidity cut and the dynamics is described by the LO impact factor. The contributions shown below the cut correspond to NLO contributions to the JIMWLK kernel. These have been computed previously~\cite{Balitsky:2008zza,Balitsky:2013fea} (see also \cite{Grabovsky:2013mba,Kovner:2013ona}) and can therefore be used to construct the NLLx result for inclusive photon production in $e+A$ collisions. We note that NLLx corrections have recently been implemented for numerical computations in fully inclusive DIS~\cite{Ducloue:2017ftk}.

An advantage of our approach is that it is fully implemented in momentum space; many shock wave computations use a mixture of momentum and coordinate space variables that is cumbersome. This is especially so in computing running coupling contributions~\cite{Balitsky:2008zza,Kovchegov:2006vj,Kovchegov:2007vf} where coordinate space prescriptions can be problematic as noted in \cite{Ducloue:2017dit}. In our framework, there is no need to switch back and forth; all computations can be realized fully in momentum space with our modified Feynman rules. These issues will be be addressed in future work. 
 
%%%%%%%%%%%%%%%%%%%%%%%%%%%%%%%%%%%%%%%%
\section{Summary and outlook}

We presented in this paper a first computation of inclusive photon production in $e+A$ DIS at small $x$ within the CGC EFT. At LO, the cross-section is directly proportional to 
universal gauge invariant dipole and quadrupole Wilson line correlators which are ubiquitous in final states that are measured in high energy $p+A$ collisions and potentially in $e+A$ collisions at a future Electron-Ion Collider (EIC). Indeed, since inclusive photon production in DIS has photons in both the initial and final states, it holds promise of being a clean  golden channel for unambiguous discovery of gluon saturation, complementary to other $e+A$ DIS measurements~\cite{Accardi:2012qut,Aschenauer:2017jsk}. In the soft photon limit, we recover the inclusive DIS dijet cross-section derived previously in  \cite{Dominguez:2011wm}. As argued there, this dijet $e+A$ channel may provide direct access to the nuclear Weizs\"{a}cker-Williams gluon distribution. 

We next discussed the structure of dominant small $x$ contributions  to the inclusive photon cross-section at next-to-leading order  and next-to-next-to-leading order. The essential ingredients here are the dressed quark and gluon propagators and the corresponding effective vertices in the shock wave classical background field of a nucleus at high energies. These effective vertices are proportional to the respective fundamental and adjoint Wilson lines that carry information about all-twist gluon correlations in the nucleus. The structure of the quark and gluon dressed propagators is remarkably simple in the ``wrong" light cone gauge $A^-=0$ and therefore permits efficient higher order computations. The computations are further simplified by exploiting the natural separation between static sources and dynamical gauge fields in the MV model and the JIMWLK RG treatment thereof. 
In particular, the cross-section can be factorized into impact factor contributions that are convoluted with the RG evolution of products of lightlike Wilson lines. The nontrivial ingredients that need to be computed are the NLO impact factor for inclusive photon production and the NLLx RG evolution of the Balitsky-JIMWLK hierarchy of Wilson line correlators. While the latter is known, the former needs to be determined; these computations are in progress and will be reported on separately~\cite{Roy-Venugopalan-2}. 

We note that, motivated by collider experiments, small $x$ computations in the gluon saturation regime are increasingly to next-to-leading order accuracy. Some examples of the studies being performed include, besides inclusive DIS~\cite{Ducloue:2017ftk}, DIS diffractive dijet production \cite{Boussarie:2014lxa,Boussarie:2016ogo} and exclusive light vector meson production  \cite{Boussarie:2016bkq}, single inclusive forward hadron production in $p+p$ and $p+A$ collisions \cite{Chirilli:2011km,Chirilli:2012jd,Altinoluk:2015vax,Iancu:2016vyg} and more recently inclusive photon production in $p+A$ collisions \cite{Benic:2016yqt,Benic:2016uku}. 
Looking further ahead, we believe that the momentum space methods discussed here can exploited to make progress in these and related computations. 

The forms of the shock wave propagators first derived in \cite{McLerran:1994vd,Ayala:1995kg}, and their expressions in terms of effective vertices~\cite{McLerran:1998nk,Balitsky:2001mr}, are identical to the quark-quark-reggeon and gluon-gluon-reggeon propagators~\cite{Caron-Huot:2013fea,Bondarenko:2017vfc,Hentschinski:2018rrf} in Lipatov's reggeon field theory~\cite{Lipatov:1996ts}. We have shown here that the slightly modified quark and gluon effective vertices in Eqs.~\ref{eq:mod-effective-quark-vertex} and \ref{eq:mod-effective-gluon-vertex} significantly simplify the computation of DIS inclusive photon production. 
Because of the form of the shock wave propagators, our results are equally valid for leading twist or all-twist computations. It would therefore be interesting to see if our modified Feynman rules are useful in simplifying multiloop leading twist computations in the Regge limit~\cite{Caron-Huot:2017zfo,DelDuca:2018hrv} or perhaps, conversely and more interestingly, results derived in those cases applied to advance computations in the saturation regime of high parton densities.

%%%%%%%%%%%%%%%%%%%%%%%%%%%%%%%%%%%%%%%%
\begin{acknowledgments}
R.V would like to thank Andrey Tarasov for a useful discussion.
This material is based on work supported by the U.S. Department of Energy, Office of Science, Office of Nuclear Physics, under Contracts No. DE-SC0012704 and within the framework TMD Theory Topical Collaboration. K. R is supported by an LDRD grant from Brookhaven Science Associates. 
\end{acknowledgments}

\section{Appendix}
\appendix
\section{Notations and conventions} \label{appendixA}
The metric used is the $-2$ metric, $\hat{g}=diag(+1,-1,-1,-1)$, where the `carat' denotes quantities in usual spacetime coordinates. The light cone coordinates are defined as 
\begin{equation*}
x^{+}=\frac{\hat{x}^{0}+\hat{x}^{3}}{\sqrt{2}}, \quad x^{-}=\frac{\hat{x}^{0}-\hat{x}^{3}}{\sqrt{2}} \enskip ,
\end{equation*}
with the transverse coordinates remaining the same and transforming as in Minkowski space. The same definition holds for the gamma matrices $\gamma^{+}$ and $\gamma^{-}$ with the Dirac algebra given by
\begin{equation}
\{\gamma^{\mu},\gamma^{\nu}\}=2g^{\mu \nu} \enskip ,
\label{eq:dirac-algebra}
\end{equation}
where $g^{+-}=g^{-+}=1$ and $g^{ij}=-\delta^{ij}$ ($i,j=1,2$) are the nonzero entries of the metric tensor. In this convention, $a.b=a^{+}b^{-}+a^{-}b^{+}-\mathbf{a}_{\perp}.\mathbf{b}_{\perp}$ and $a_{+}=a^{-}$, $a_{i}=-a^{i}$.
%%%%%%%%%%%%%%%%%%%%%%%%%%%%%%%%%%%%%%%%%%%%%%%%%%%%%%%%%%%%%%%%%%
\section{Gauge invariance and soft-photon factorization at LO} \label{appendixB}
The first part of this appendix provides a proof of the Ward identity for the final state real photon, thus establishing gauge invariance. In the second part, we will provide an explicit expression  for the nonradiative DIS amplitude recovered using the soft photon limit, $k_{\gamma} \rightarrow 0$. We will next derive the differential cross-section for inclusive dijet production in DIS for the case of an incoming longitudinally polarized virtual photon. The case of the transversely polarized virtual photon follows identically and will therefore not be shown.
%%%%%%%%%%%%%%%%%%%%%%%%%%%%%%%%%%%%%%%%%%%%%%%%%%%%%%%%%%%%%%%%%%%%
\subsection{Ward identity} \label{appendixB1}
The Ward identity for the final state photon requires that
\begin{equation}
k^{\alpha}_{\gamma} \mathcal{M}_{\mu \alpha} (\mathbf{q}, \mathbf{k},\mathbf{p},\mathbf{k}_{\gamma}) = 0  \enskip .
\end{equation}
Towards this end, we make the replacement $\epsilon^{* \alpha}(\mathbf{k}_{\gamma}, \lambda) \rightarrow k^{\alpha}_{\gamma}$ in the amplitude and check the effects of contracting  $k^{\alpha}_{\gamma}$ with $T^{(q \bar{q})} (\mathbf{l}_{\perp},\mathbf{P}_{\perp})$. Instead of using the final forms for the various R-factors constituting $T^{(q\bar{q})}$, we use their original forms in which the contour integrations are not performed. This simplifies the calculation. For example, we have
\begin{equation}
 \overline{u}(\mathbf{k}) k^{\alpha}_{\gamma} R^{(7)}_{\mu \alpha} (\mathbf{l}_{\perp}, \mathbf{P}_{\perp}) v(\mathbf{p}) = i^{3} \int \frac{\mathrm{d}l^{+}}{2\pi} \enskip \overline{u}(\mathbf{k}) \gamma^{-} \frac{\slashed{q}-\slashed{p}+\slashed{l}+m}{(q-p+l)^{2}-m^{2}+i\varepsilon  } \gamma_{\mu} \frac{\slashed{l}\gamma^{-}-2p^{-}}{(l-p)^{2}-m^{2}+i\varepsilon } v(\mathbf{p}) \enskip .
 \label{eq:B2}
\end{equation}
We can use the second identity in Eq. \ref{eq:relation-22} and the relation
\begin{equation}
\overline{u}(\mathbf{k}) \slashed{k}_{\gamma} (\slashed{k}+m)= (2k.k_{\gamma}) \overline{u}(\mathbf{k}) \enskip ,
\end{equation}
to derive Eq. \ref{eq:B2}. Likewise, we can write
\begin{align}
\overline{u}(\mathbf{k}) k^{\alpha}_{\gamma} R^{(9)}_{\mu \alpha} (\mathbf{l}_{\perp}, \mathbf{P}_{\perp}) v(\mathbf{p}) & = i^{3} \int \frac{\mathrm{d}l^{+}}{2\pi} \enskip \overline{u}(\mathbf{k}) \gamma^{-}  \frac{(\slashed{q}-\slashed{p}+\slashed{l}+m) \slashed{k}_{\gamma}(\slashed{q}-\slashed{p}+\slashed{l}+m) }{(q-p+l-k_{\gamma})^{2}-m^{2}+i\varepsilon }  \nonumber \\
& \times \gamma_{\mu } \frac{\slashed{l}\gamma^{-}-2p^{-}}{\big[(q-p+l)^{2}-m^{2}+i\varepsilon   \big] \big[(l-p)^{2}-m^{2}+i\varepsilon \big] } v(\mathbf{p}) \enskip .
\label{eq:B4}
\end{align}
Adding Eqs. \ref{eq:B2} and \ref{eq:B4}, and using the fact that $p^{2}-m^{2}= (\slashed{p}+m) (\slashed{p}-m)$, it is easy to show that one of the terms in square brackets in the denominator cancels out, giving
\begin{align}
 \overline{u}(\mathbf{k}) k^{\alpha}_{\gamma} & \Big[   R^{(7)}_{\mu \alpha} (\mathbf{l}_{\perp}, \mathbf{P}_{\perp}) + R^{(9)}_{\mu \alpha} (\mathbf{l}_{\perp}, \mathbf{P}_{\perp}) \Big] v(\mathbf{p})  \nonumber \\
& = i^{3} \int \frac{\mathrm{d}l^{+}}{2\pi} \enskip \overline{u}(\mathbf{k}) \gamma^{-} \frac{\slashed{q}-\slashed{p}+\slashed{l} -\slashed{k}_{\gamma}+m}{(q-p+l-k_{\gamma})^{2}-m^{2}+i\varepsilon   } \gamma_{\mu} \frac{\slashed{l}\gamma^{-} - 2p^{-} }{(l-p)^{2}-m^{2}+i\varepsilon} v(\mathbf{p}) \enskip .
\end{align}
Analogously, we can use the first identity in Eq. \ref{eq:relation-22} and 
\begin{equation}
(\slashed{p} - m) \slashed{k}_{\gamma} v(\mathbf{p})= (2p.k_{\gamma}) v(\mathbf{p})\,,
\end{equation}
to write 
\begin{equation}
 \overline{u}(\mathbf{k}) k^{\alpha}_{\gamma} R^{(8)}_{\mu \alpha} (\mathbf{l}_{\perp}, \mathbf{P}_{\perp}) v(\mathbf{p}) = i^{3} \int \frac{\mathrm{d}l^{+}}{2\pi} \enskip \overline{u}(\mathbf{k}) \gamma^{-}  \frac{\slashed{q}-\slashed{p}+\slashed{l}-\slashed{k}_{\gamma}+m}{(q-p+l-k_{\gamma})^{2}-m^{2}+i\varepsilon  } \gamma_{\mu} \frac{\slashed{l}\gamma^{-}-2p^{-}-\slashed{k}_{\gamma}\gamma^{-}}{(l-p-k_{\gamma})^{2}-m^{2}+i\varepsilon } v(\mathbf{p}) \enskip ,
 \label{eq:B7}
\end{equation}
and 
\begin{align}
\overline{u}(\mathbf{k}) k^{\alpha}_{\gamma} R^{(10)}_{\mu \alpha} (\mathbf{l}_{\perp}, \mathbf{P}_{\perp}) v(\mathbf{p}) & = i^{3} \int \frac{\mathrm{d}l^{+}}{2\pi} \enskip \overline{u}(\mathbf{k}) \gamma^{-} \frac{\slashed{q}-\slashed{p}+\slashed{l}-\slashed{k}_{\gamma}+m}{(q-p+l-k_{\gamma})^{2}-m^{2}+i\varepsilon  } \gamma_{\mu} \frac{\slashed{l}-\slashed{p}+m}{(l-p)^{2}-m^{2}+i\varepsilon } \nonumber \\
& \times \slashed{k}_{\gamma}   \frac{\slashed{l}\gamma^{-}-2p^{-}}{(l-p-k_{\gamma})^{2}-m^{2}+i\varepsilon} v(\mathbf{p}) \enskip .
\label{eq:B8}
\end{align}
Adding the expressions in Eqs. \ref{eq:B7} and \ref{eq:B8}, and using a similar reasoning used to simplify the previous sum, we get
\begin{equation}
\overline{u}(\mathbf{k}) k^{\alpha}_{\gamma} \Big[   R^{(7)}_{\mu \alpha} (\mathbf{l}_{\perp}, \mathbf{P}_{\perp}) + R^{(9)}_{\mu \alpha} (\mathbf{l}_{\perp}, \mathbf{P}_{\perp}) \Big] v(\mathbf{p}) = - \overline{u}(\mathbf{k}) k^{\alpha}_{\gamma} \Big[   R^{(8)}_{\mu \alpha} (\mathbf{l}_{\perp}, \mathbf{P}_{\perp}) + R^{(10)}_{\mu \alpha} (\mathbf{l}_{\perp}, \mathbf{P}_{\perp}) \Big] v(\mathbf{p})  \enskip .
\label{eq:B9}
\end{equation}
This implies that 
\begin{equation}
\overline{u}(\mathbf{k}) k^{\alpha}_{\gamma} T^{(q \bar{q})}_{\mu \alpha} (\mathbf{l}_{\perp}, \mathbf{P}_{\perp}) v(\mathbf{p}) = 0
\label{eq:B10}
\end{equation}
thereby showing that the Ward identity is indeed satisfied for the outgoing photon. 
%The striking similarity between the above Eqs. \ref{eq:B9} and \ref{eq:B10} in $e+A$ DIS at LO and relation (B.3) (and the preceeding identity) in the NLO calculation \cite{Benic:2016uku} for $p+A$ collisions should be noted. 
It can also be shown that the same relation holds for the case of the exchanged virtual photon if we are considering the hadronic subprocess. This gives us the freedom to consider only the $-g^{\mu \nu}$ part of the photon propagator which is implicitly assumed in writing Eq. \ref{eq:amplitude-master}.

%%%%%%%%%%%%%%%%%%%%%%%%%%%%%%%%%%%%%%%%%%
\subsection{Soft photon factorization} \label{appendixB2}
By taking the soft photon limit $k_{\gamma} \rightarrow 0$ of our LO amplitude expression in Eq. \ref{eq:final-amplitude}, we can recover the nonradiative DIS amplitude given by
\begin{align}
& \mathcal{M}^{NR}_{\mu} (\mathbf{q},\mathbf{k}, \mathbf{p} )  = 2\pi (eq_{f}) \delta(P^{-}-q^{-}) \int_{\mathbf{x}_{\perp}} \int_{\mathbf{y}_{\perp}} \int_{\mathbf{l}_{\perp}} e^{-i\mathbf{P}_{\perp}.\mathbf{x}_{\perp}+i\mathbf{l}_{\perp}.\mathbf{x}_{\perp} } e^{-i\mathbf{l}_{\perp}.\mathbf{y}_{\perp}} \nonumber \\
& \times - \frac{1}{2q^{-}} \enskip \overline{u}(\mathbf{k}) \gamma^{-} \frac{zq^{-}\gamma^{+}- (\slashed{\mathbf{p}}_{\perp} - \slashed{\mathbf{l}}_{\perp}) +m }{z(1-z) Q^{2}+M^{2}(\mathbf{l}_{\perp}-\mathbf{p}_{\perp})} \gamma_{\mu} (2p^{-}+\gamma^{-} \slashed{\mathbf{l}}_{\perp}) \times \Big( \tilde{U}(\mathbf{x}_{\perp}) \tilde{U}^{\dagger}(\mathbf{y}_{\perp}) -1 \Big)_{ij } v(\mathbf{p}) \enskip ,
\label{eq:nonradiative-DIS-amp}
\end{align}
where $P$ now equals $p+k$.

We will now show that the above expression can be used to factorize the amplitude squared in terms of products of light cone wavefunctions and a dipole scattering factor. In the soft photon limit, the amplitude for the subprocess 
\begin{equation}
\gamma^{*}(q) \rightarrow Q(k) + \overline{Q}(p) + \gamma(k_{\gamma})
\end{equation}
can be written as
\begin{equation}
\mathcal{M}(\mathbf{q},\mathbf{k}, \mathbf{p}, \mathbf{k}_{\gamma};\lambda,\lambda') = \epsilon^{\mu} (\mathbf{q},\lambda') \mathcal{M}_{\mu} (\mathbf{q},\mathbf{k}, \mathbf{p}, \mathbf{k}_{\gamma}) \enskip ,
\label{eq:non-radiative-masteramp}
\end{equation}
where $\epsilon^{\mu} (\mathbf{q},\lambda')$ is the polarization vector for the incoming virtual photon and $\mathcal{M}_{\mu}$ is given by Eq. \ref{eq:nonradiative-DIS-amp}. In order to identify the transverse and longitudinally polarized photon wavefunctions, it is convenient to parametrize the polarization vectors as
\begin{align}
\epsilon_{T}(\mathbf{q},\lambda'=+1)& = \Big( 0,0,-\frac{1}{\sqrt{2}},-\frac{i}{\sqrt{2}} \Big) \enskip , \nonumber \\
\epsilon_{T}(\mathbf{q},\lambda'=-1)& = \Big( 0,0,\frac{1}{\sqrt{2}},-\frac{i}{\sqrt{2}} \Big) \enskip , \nonumber \\
\epsilon_{L}(\mathbf{q},\lambda'=0)& = \Big( \frac{Q}{2q^{-}},\frac{q^{-}}{Q},0,0 \Big) \enskip ,
\end{align}
where $T$ and $L$ stand for transverse and longitudinal respectively. These vectors satisfy the relations
\begin{align}
& \epsilon^{2}_{T}(\mathbf{q},\lambda') =-1, \quad \epsilon^{2}_{L}(\mathbf{q},\lambda') =1, \quad \epsilon_{T}(\mathbf{q},\lambda') \epsilon_{T}^{*}(\mathbf{q},\lambda'')= -\delta_{\lambda',\lambda''} \enskip , \nonumber \\
& g_{\mu \nu} -\frac{q_{\mu}q_{\nu}}{q^{2}}= - \sum_{\lambda'= \pm 1} \epsilon_{\mu T}(\mathbf{q},\lambda') \epsilon^{*}_{\nu T}(\mathbf{q},\lambda') + \epsilon_{\mu L}(\mathbf{q},\lambda'=0) \epsilon^{*}_{\nu L}(\mathbf{q},\lambda'=0) \enskip .
\end{align}

We will now explicitly compute the cross-section for the longitudinally polarized case.  With the above choice of the $\lambda'=0$ polarization vector, Eq. \ref{eq:non-radiative-masteramp} gives
\begin{align}
 & \mathcal{M}^{NR} (\mathbf{q},\mathbf{k}, \mathbf{p};\lambda'=0 ) =\epsilon^{\mu}_{L}(\mathbf{q},\lambda'=0) \mathcal{M}^{NR}_{\mu}(\mathbf{q},\mathbf{k}, \mathbf{p} ) = 
2\pi (eq_{f})  \delta(P^{-}-q^{-}) \int_{\mathbf{x}_{\perp}} \int_{\mathbf{y}_{\perp}} \int_{\mathbf{l}_{\perp}} e^{-i\mathbf{P}_{\perp}.\mathbf{x}_{\perp}+i\mathbf{l}_{\perp}.\mathbf{x}_{\perp} } e^{-i\mathbf{l}_{\perp}.\mathbf{y}_{\perp}} \nonumber \\
& \times - \frac{1}{2q^{-}} \enskip \overline{u}(\mathbf{k}) \gamma^{-} \frac{zq^{-}\gamma^{+}- (\slashed{\mathbf{p}}_{\perp} - \slashed{\mathbf{l}}_{\perp}) +m }{z(1-z) Q^{2}+M^{2}(\mathbf{l}_{\perp}-\mathbf{p}_{\perp})} \Bigg(\frac{Q}{2q^{-}} \gamma^{-}+\frac{q^{-}}{Q} \gamma^{+} \Bigg) (2p^{-}+\gamma^{-} \slashed{\mathbf{l}}_{\perp}) \times \Big( \tilde{U}(\mathbf{x}_{\perp}) \tilde{U}^{\dagger}(\mathbf{y}_{\perp}) -1 \Big)_{ij } v(\mathbf{p}) \enskip .
\end{align}
Redefining $\mathbf{l}_{\perp}-\mathbf{p}_{\perp} \rightarrow \mathbf{l}_{\perp}$ and using the property \ref{eq:dirac-algebra} for gamma matrices along with the Dirac equation, we can simplify the above result to
\begin{align}
\mathcal{M}^{NR} (\mathbf{q},\mathbf{k}, \mathbf{p};\lambda'=0 ) & = 2\pi (eq_{f})  \delta(P^{-}-q^{-})  \int_{\mathbf{x}_{\perp}} \int_{\mathbf{y}_{\perp}} \int_{\mathbf{l}_{\perp}} e^{-i\mathbf{k}_{\perp}.\mathbf{x}_{\perp}-i\mathbf{p}_{\perp}.\mathbf{y}_{\perp} } e^{i\mathbf{l}_{\perp}.(\mathbf{x}_{\perp}-\mathbf{y}_{\perp})} \nonumber \\
& \times \Big( \tilde{U}(\mathbf{x}_{\perp}) \tilde{U}^{\dagger}(\mathbf{y}_{\perp}) -1 \Big)_{ij } \Bigg[ -\frac{2z(1-z)Q}{\mathbf{l}_{\perp}^{2}+\epsilon_{q}^{2}} +\frac{1}{Q} \Bigg] \overline{u}(\mathbf{k}) \gamma^{-} v(\mathbf{p}) \enskip ,
\end{align}
where $\epsilon_{q}^{2}=z(1-z)Q^{2}+m^{2}$. The term proportional to $1/Q$ vanishes because of the $\delta^{(2)}(\mathbf{x}_{\perp}-\mathbf{y}_{\perp})$ arising from the integration over $\mathbf{l}_{\perp}$ and the identity $\tilde{U}(\mathbf{x}_{\perp}) \tilde{U}^{\dagger} (\mathbf{x}_{\perp})=1$.\\

Finally using the formula
\begin{equation}
\int \frac{\mathrm{d}^{2}\mathbf{l}_{\perp}}{(2\pi)^{2}} \frac{e^{i\mathbf{l}_{\perp}.(\mathbf{x}_{\perp}-\mathbf{y}_{\perp})}}{\mathbf{l}_{\perp}^{2}+\epsilon_{q}^{2}} =\frac{1}{2\pi} K_{0}(\epsilon_{q}\vert \mathbf{x}_{\perp}-\mathbf{y}_{\perp} \vert ) \enskip ,
\end{equation}
where $K_{0}$ is the modified Bessel function of the second kind, we can recast $\mathcal{M}^{NR}$ as
\begin{equation}
\mathcal{M}^{NR}(\mathbf{q},\mathbf{k}, \mathbf{p};\lambda') =  2\pi  \delta(P^{-}-q^{-}) \tilde{\mathcal{M}}^{NR} (\mathbf{q},\mathbf{k}, \mathbf{p};\lambda') \enskip ,
\end{equation}
where
\begin{align}
\tilde{\mathcal{M}}^{NR} (\mathbf{q},\mathbf{k}, \mathbf{p};\lambda'=0) &= (eq_{f}) \int_{\mathbf{x}_{\perp}} \int_{\mathbf{y}_{\perp}} e^{-i\mathbf{k}_{\perp}.\mathbf{x}_{\perp}-i\mathbf{p}_{\perp}.\mathbf{y}_{\perp} }  \enskip \frac{K_{0} (\epsilon_{q}\vert \mathbf{x}_{\perp}-\mathbf{y}_{\perp} \vert )}{2\pi}  \Big( \tilde{U}(\mathbf{x}_{\perp}) \tilde{U}^{\dagger}(\mathbf{y}_{\perp}) -1 \Big)_{ij } \nonumber \\
& \times  \big(-2z(1-z)Q\big) \overline{u}(\mathbf{k}) \gamma^{-} v(\mathbf{p}) \enskip .
\end{align}

The differential cross-section for inclusive dijet production in DIS is given by
\begin{equation}
\frac{\mathrm{d}\sigma}{\mathrm{d}^{3}k \mathrm{d}^{3}p} = \frac{1}{2q^{-}} \frac{1}{(2\pi)^{3}2E_{k}} \frac{1}{(2\pi)^{3}2E_{p}} \left \langle \vert \overset{\sim}{\mathcal{M}}^{NR} \vert^{2} \right \rangle_{Y_{A}} (2\pi) \delta(P^{-}-q^{-}) \enskip,
\end{equation}
where the $\delta(0^{-})$ appearing in the amplitude squared is normalized as described earlier in Sec. \ref{sec:sectionIII}. Now using
\begin{equation}
\text{Tr} \big[(\slashed{k}+m) \gamma^{-} (\slashed{p}-m) \hat{\gamma}^{0} \gamma^{+} \hat{\gamma}^{0}   \big]=(2p^{-}) (2k^{-}) \enskip ,
\end{equation}
and the form of the $\lambda'=0$ photon wavefunction given in \cite{Dominguez:2011wm}
\begin{equation}
\psi_{\alpha \beta}^{L} (q^{-},z,r)= 2\pi \sqrt{\frac{4}{q^{-}}} z(1-z)Q K_{0}(\epsilon_{q}r) \delta_{\alpha \beta} \enskip ,
\end{equation}
it is a matter of straightforward algebra to show that 
\begin{align}
\frac{\mathrm{d}\sigma^{L}}{\mathrm{d}^{3}k \mathrm{d}^{3}p} & = N_{c}\alpha q_{f}^{2} \delta(q^{-}-p^{-}-k^{-}) \int \frac{\mathrm{d}^{2}\mathbf{x}_{\perp}}{(2\pi)^{2}} \frac{\mathrm{d}^{2}\mathbf{x'}_{\perp}}{(2\pi)^{2}} \frac{\mathrm{d}^{2}\mathbf{y}_{\perp}}{(2\pi)^{2}} \frac{\mathrm{d}^{2}\mathbf{y'}_{\perp}}{(2\pi)^{2}} \enskip e^{-i\mathbf{k}_{\perp}.(\mathbf{x}_{\perp}-\mathbf{x'}_{\perp})}  e^{-i\mathbf{p}_{\perp}.(\mathbf{y}_{\perp}-\mathbf{y'}_{\perp})} \nonumber \\
& \times \sum_{\alpha, \beta} \psi^{L}_{\alpha \beta}(q^{-},z,\vert \mathbf{x}_{\perp}-\mathbf{y}_{\perp} \vert ) \enskip \psi^{L *}_{\alpha \beta}(q^{-},z,\vert \mathbf{x'}_{\perp}-\mathbf{y'}_{\perp} \vert ) \times \Big[ 1-\frac{1}{N_{c}} \left \langle \text{Tr} \big( \tilde{U}(\mathbf{x}_{\perp}) \tilde{U}^{\dagger}(\mathbf{y}_{\perp}) \big) \right \rangle_{Y_{A}} \nonumber \\
& -\frac{1}{N_{c}} \left  \langle \text{Tr}\big( \tilde{U}(\mathbf{y'}_{\perp}) \tilde{U}^{\dagger}(\mathbf{x'}_{\perp}) \big) \right  \rangle_{Y_{A}} +\frac{1}{N_{c}} \left  \langle \text{Tr} \big( \tilde{U}(\mathbf{y'}_{\perp}) \tilde{U}^{\dagger}(\mathbf{x'}_{\perp}) \tilde{U}(\mathbf{x}_{\perp}) \tilde{U}^{\dagger}(\mathbf{y}_{\perp}) \big) \right  \rangle_{Y_{A}}    \Big] \enskip .
\end{align}
This result exactly matches Eq. (22) of \cite{Dominguez:2011wm} obtained for DIS dijet production at small $x$. The case of the transversely polarized photon proceeds in a similar fashion.

\section{Kinematically allowed processes} \label{appendixC}
In this appendix, we will explicitly demonstrate the topologies of the LO and NLO Feynman diagrams that are allowed by the kinematics of the process. The techniques described in this section are quite general and can be extended to find kinematically allowed diagrams beyond NLO.

\begin{figure}[H]
\begin{center}
\includegraphics[scale=0.25]{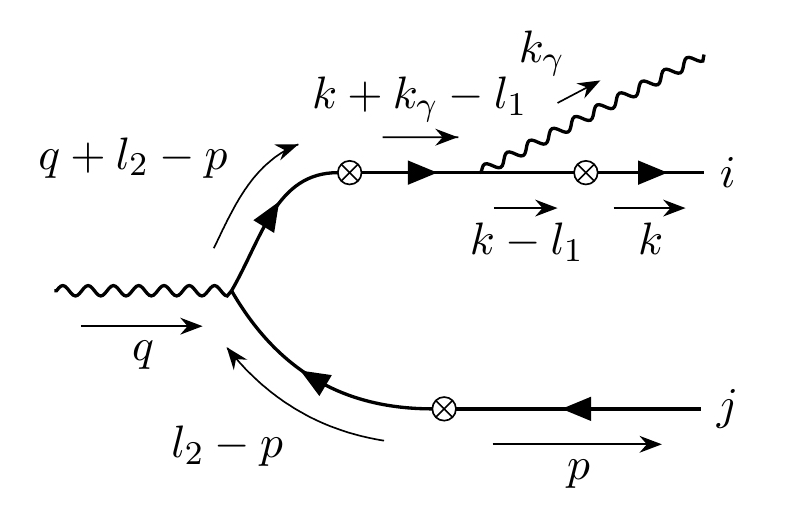}
\caption{Feynman diagram of the LO contribution to inclusive photon production with all dressed fermion lines. The momenta $l_{1}$ and $l_{2}$ are the momentum kicks from the nucleus to the quark and antiquark respectively. 
\label{fig:LO-general-diagram}}
\end{center}
\end{figure}

Let us first consider the most general diagram (see Fig. \ref{fig:LO-general-diagram}) for the LO process with all dressed fermion lines and photon emission from the quark line.  We will now show it is possible to find all allowed processes starting from this generic template provided we use the modified Feynman rules discussed in Sec.~\ref{sec:dressed-fermion}. An identical treatment follows for the case of photon emission from the antiquark line.
The amplitude for the hadronic subprocess is given by
\begin{align}
\mathcal{M}_{\mu \alpha}=-(eq_{f})^{2} \int_{l_{1},l_{2}} \overline{u}(\mathbf{k}) & \mathcal{T}_{im}(k-l_{1},k) S_{0}(k-l_{1}) \gamma_{\alpha}S_{0}(k+k_{\gamma}-l_{1}) \mathcal{T}_{mn}(q+l_{2}-p,k+k_{\gamma}-l_{1}) \nonumber \\
&\times S_{0}(q+l_{2}-p) \gamma_{\mu} S_{0}(l_{2}-p)\mathcal{T}_{nj}(-p,l_{2}-p)v(\mathbf{p}) \enskip ,
\label{eq:equationC1}
\end{align}
where $\int_{l_{i}}=\int  \mathrm{d}^{4}l_{i} / (2\pi)^{4}, i=1,2$ and $\mathcal{T}$ is given by Eq. \ref{eq:mod-effective-quark-vertex}.

Integrating out $l_{1}^{-}$ and $l_{2}^{-}$ using the delta functions embedded in the vertex factors, we are left with the delta function $\delta(q^{-}-P^{-})$ representing the overall longitudinal momentum conservation and integrations over transverse spatial and momentum coordinates. However the quantity of interest is the integral over $l_{1}^{+}$ and $l_{2}^{+}$ given by
\begin{equation}
\mathcal{I}=\int \mathrm{d}l_{1}^{+} \int \mathrm{d}l_{2}^{+} \enskip \frac{N}{D} \enskip ,
\end{equation} 
where 
\begin{align}
N=\overline{u}(\mathbf{k}) \gamma^{-} (\slashed{k}-\slashed{l_{1}}+m)\gamma_{\alpha} (\slashed{k}+\slashed{k}_{\gamma}-\slashed{l_{1}}+m)\gamma^{-}(\slashed{q}+\slashed{l_{2}}-\slashed{p}+m)\gamma_{\mu} (\slashed{l_{2}}-\slashed{p}+m) \gamma^{-} v(\mathbf{p}) \enskip ,
\end{align}
and 
\begin{align}
D& = \Bigg( k^{+}-l_{1}^{+}-\frac{M^{2}(\mathbf{k}_{\perp}-\mathbf{l}_{1\perp})}{2k^{-}}+\frac{i\varepsilon}{2k^{-}} \Bigg) \Bigg( k^{+}+k^{+}_{\gamma}-l_{1}^{+}-\frac{M^{2}(\mathbf{k}_{\perp}+\mathbf{k}_{\gamma \perp}-\mathbf{l}_{1\perp})}{2(k^{-}+k_{\gamma}^{-})}+\frac{i\varepsilon}{2(k^{-}+k_{\gamma}^{-})} \Bigg) \nonumber \\
& \times \Bigg( q^{+}+l_{2}^{+}-p^{+}-\frac{M^{2}(\mathbf{l}_{1\perp}-\mathbf{p}_{\perp})}{2(k^{-}+k_{\gamma}^{-})}+\frac{i\varepsilon}{2(k^{-}+k^{-}_{\gamma})} \Bigg) \Bigg( l_{2}^{+}-p^{+}+\frac{M^{2}(\mathbf{l}_{1\perp}-\mathbf{p}_{\perp})}{2p^{-}}-\frac{i\varepsilon}{2p^{-}} \Bigg) \enskip ,
\label{eq:equationC4}
\end{align}
denote the numerator and denominator respectively. Examining the structure of the poles in the propagator terms of the denominator, there are two $l_{1}^{+}$ poles on the positive side of the real axis and independently, two $l_{2}^{+}$ poles on either side of the real axis. Using the property $(\gamma^{-})^{2}=0$, it is easy to see that the numerator does not contain any term proportional to $l_{1}^{+}$ or $l_{2}^{+}$. Therefore the contour for the integration over $l_{1}^{+}$ can be closed below the real axis thereby giving a null result. 

\begin{figure}[!htbp]
\begin{center}
\includegraphics[scale=0.2]{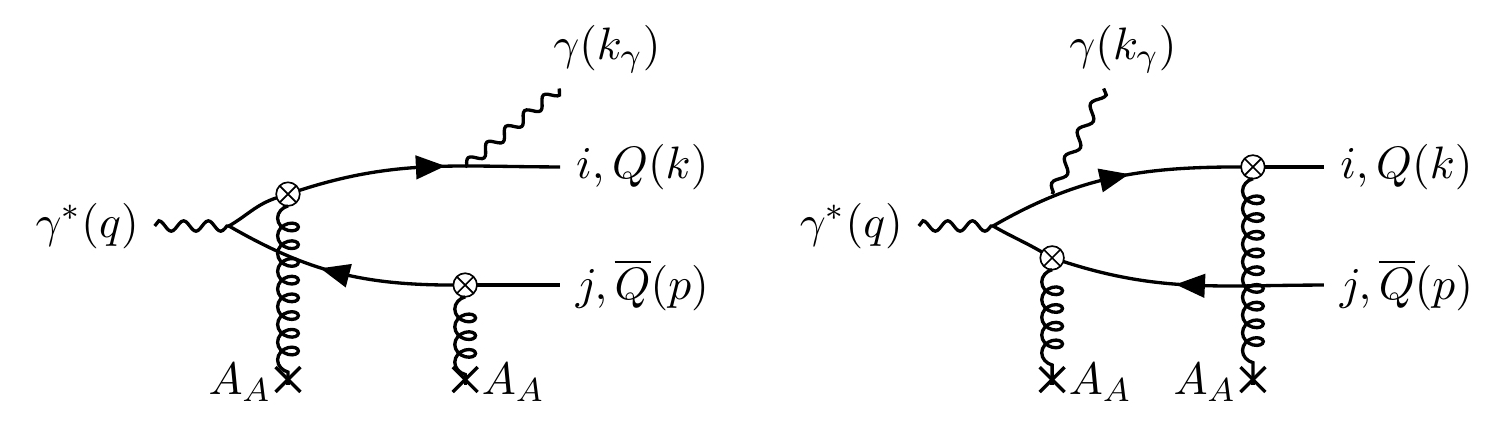}
\caption{Contributions to the photon production amplitude at LO with photon emission from the quark line. The remaining two diagrams can be obtained by interchanging the quark and antiquark lines. \label{fig:allowed-LO}}
\end{center}
\end{figure}

The arguments presented thus far does not invoke the possibility of ``no scattering'' in our definition of the effective vertices or equivalently the $\tilde{U}=\mathds{1}$ case. It can be shown easily that for either $\tilde{U}_{im}=\delta_{im}$ or $\tilde{U}_{mn}=\delta_{mn}$ in the vertex factors appearing in Eq.~\ref{eq:equationC1}, we will get a nonzero result from the above contour integration. Under these conditions, the two factors appearing in the first line of Eq.~\ref{eq:equationC4} resemble energy denominators that appear in light cone perturbation theory (LCPT) \cite{Kogut:1969xa,Bjorken:1970ah,Brodsky:1995rn,Brodsky:1997de,Venugopalan:1998zd} and the integration over $l_{2}^{+}$ can be done using the residue theorem. Corresponding to these conditions on the $\tilde{U}$'s, we get the two allowed diagrams with photon emission from the quark line as shown in Fig. \ref{fig:allowed-LO}; these are embedded in our generic diagram Fig. \ref{fig:LO-general-diagram}.

One can therefore start with the generic case and eventually deduce the impossibility of a secondary scattering subsequent to emission of the photon by an already scattered fermion. This is a consequence of the eikonal approximation in which the nucleus moving at near light speed interacts instantaneously with the quarks thereby removing the possibility of a second scattering. Once the allowed diagrams are computed, we simply need to deduct the contribution in which both quark and antiquark propagate unscattered. The latter have the same magnitude as the allowed diagrams modulo the Wilson line factors. The net contribution is given by Eq. \ref{eq:final-amplitude}.

The same physical principle discussed above for LO also applies to the NLO diagrams which are classified into three broad categories. For the diagrams contributing to the LLx and NLLx JIMWLK evolution, the allowed topologies can be easily extracted from existing literature\footnote{We should mention here that our diagrammatic representation is similar in spirit to the ``shock wave" approach used in these works.}  \cite{Balitsky:2013fea,Kovchegov:2006vj,Balitsky:2008zza}. Since the upper part of these diagrams has the same structure as LO diagrams, the rules discussed in the previous section apply trivially. In the following, we therefore discuss only the genuine $\alpha_{S}$ suppressed contributions in Figs. \ref{fig:NLO-coeff}.

\begin{figure}[!htbp]
\begin{center}
\includegraphics[scale=0.2]{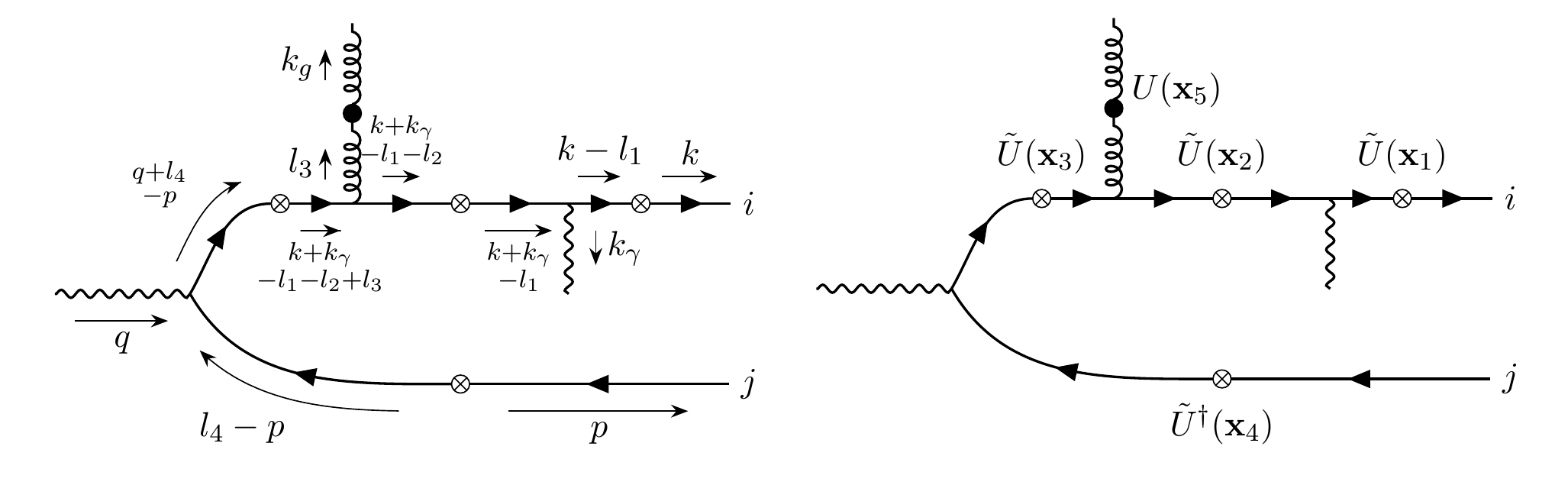}
\caption{Real emission graph for NLO with all fermion lines dressed.  $l_{1}$, $ l_{2}$ and $l_{4}$ represent the momentum transfer from the nucleus to the quark and antiquark line at different points in the scattering process. $l_{3}$ is the momentum carried by the gluon prior to scattering off the nucleus. $i$ and $j$ represent fundamental color indices. The Wilson line factors associated with each effective vertex are shown on the right. 
 \label{fig:NLO-example-A}}
\end{center}
\end{figure}

We consider one such representative generic diagram (see  Fig. \ref{fig:NLO-example-A}) which represents real emission of a gluon from the quark or antiquark in addition to the final state photon. We use the line of reasoning made for the LO case to deduce the allowed processes. The amplitude for this subprocess is given by

\begin{align}
\mathcal{M}_{\mu \alpha;a}& =-i(eq_{f})^{2}g \Big( \prod_{k=1}^{4} \int_{l_{k}} \Big) \enskip \overline{u}(\mathbf{k}) \mathcal{T}_{im} (k-l_{1},k) S_{0}(k-l_{1}) \gamma_{\alpha} S_{0}(k+k_{\gamma}-l_{1}) \mathcal{T}_{mn}(k+k_{\gamma}-l_{1}-l_{2},k+k_{\gamma}-l_{1}) \nonumber \\
& \times S_{0}(k+k_{\gamma}-l_{1}-l_{2})( t^{b})_{np} \gamma_{\beta}S_{0}(k+k_{\gamma}-l_{1}-l_{2}+l_{3}) \mathcal{T}_{pq}(q+l_{4}-p,k+k_{\gamma}-l_{1}-l_{2}+l_{3}) \nonumber \\
&\times  S_{0}(q+l_{4}-p) \gamma_{\mu} \mathcal{T}_{qj}(-p,l_{4}-p) v(\mathbf{p}) \times G_{0}^{\beta \nu;bc}(l_{3}) \mathcal{T}_{\nu \rho;ca}(l_{3},k_{g}) {\epsilon^{\rho}}^{*}(\mathbf{k}_{g}) \enskip ,
\end{align}
where the vertex factors for the fermion and gluon propagators are given respectively by Eqs. \ref{eq:mod-effective-quark-vertex} and \ref{eq:mod-effective-gluon-vertex}.

\begin{figure}[H]
\begin{center}
\includegraphics[scale=0.2]{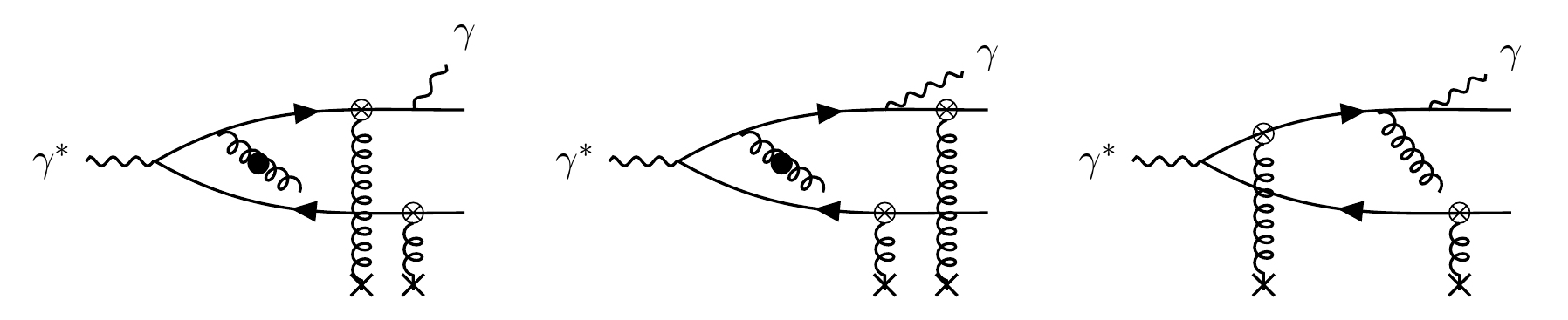}
\caption{The allowed set of diagrams embedded in Fig. \ref{fig:NLO-example-A}.  \label{fig:NLO-coeff-real}}
\end{center}
\end{figure}

\begin{figure}[H]
\begin{center}
\includegraphics[scale=0.2]{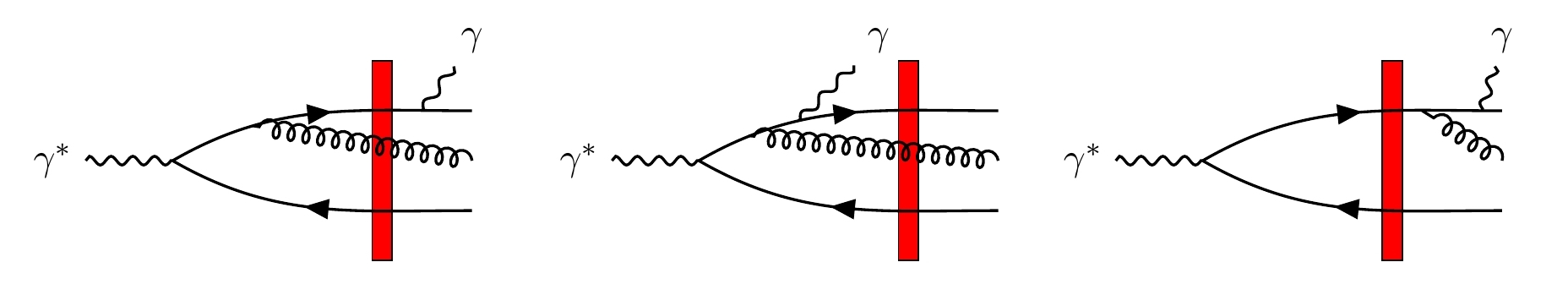}
\caption{An equivalent depiction of the allowed set of diagrams embedded in Fig. \ref{fig:NLO-example-A} with the Lorentz contracted nucleus shown here by the red rectangular wall.  \label{fig:shockwave-diagrams}}
\end{center}
\end{figure}

 By carefully integrating out the $l_{i}^{-}$'s ($i=1,\dots,4$) using the $\delta$-functions, the amplitude can be cast in terms of a momentum conserving delta function $\delta(q^{-}-p^{-}-k^{-}-k^{-}_{\gamma}-k^{-}_{g})$, integrations over transverse spatial and momentum coordinates and the following integral of interest.
\begin{equation}
\mathcal{I}_{1}=\Big( \prod_{k=1}^{4} \int \mathrm{d}l_{k}^{+} \Big) \enskip \frac{N_{1}}{D_{1}}
\end{equation}

where 
\begin{align}
N_{1}& = \overline{u}(\mathbf{k}) \gamma^{-} [\gamma^{+}k^{-}-\bm{\gamma}_{\perp}.(\mathbf{k}_{\perp}-\mathbf{l}_{1\perp})+m]\gamma_{\alpha}[\gamma^{+}(k^{-}+k^{-}_{\gamma})-\bm{\gamma}_{\perp}.(\mathbf{k}_{\perp}+\mathbf{k}_{\gamma \perp}-\mathbf{l}_{1\perp})+m]\gamma^{-} \nonumber \\
& \times [\gamma^{+}(k^{-}+k^{-}_{\gamma})-\bm{\gamma}_{\perp}.(\mathbf{k}_{\perp}+\mathbf{k}_{\gamma \perp}-\mathbf{l}_{1\perp}-\mathbf{l}_{2\perp})+m] \gamma_{\beta}  [\gamma^{+}(k^{-}+k^{-}_{\gamma}+k_{g}^{-})-\bm{\gamma}_{\perp}.(\mathbf{k}_{\perp}+\mathbf{k}_{\gamma \perp}-\mathbf{l}_{1\perp}-\mathbf{l}_{2\perp}+\mathbf{l}_{3\perp})+m] \nonumber \\
& \times \gamma^{-} [\gamma^{+}(k^{-}+k^{-}_{\gamma}+k_{g}^{-})-\bm{\gamma}_{\perp}.(\mathbf{l}_{4\perp}-\mathbf{p}_{\perp})+m]\gamma_{\mu} [\gamma^{+}p^{-}+\bm{\gamma}_{\perp}.(\mathbf{l}_{4\perp}-\mathbf{p}_{\perp})-m]\gamma^{-} \Bigg( -{\epsilon^{\beta}}^{*}(\mathbf{k}_{g})+\frac{(l_{3}.\epsilon^{*}(\mathbf{k}_{g}))n^{\beta}}{k^{-}_{g}} \Bigg) \enskip ,
\end{align}
is obtained using $n.\epsilon^{*}(\mathbf{k}_{g})=0$ and $(\gamma^{-})^{2}=0$ and
\begin{align}
D_{1}&= \Bigg( k^{+}-l_{1}^{+}-\frac{M^{2}(\mathbf{k}_{\perp}-\mathbf{l}_{1\perp})}{2k^{-}}+\frac{i\varepsilon}{2k^{-}} \Bigg) \Bigg( k^{+}+k_{\gamma}^{+}-l_{1}^{+}-\frac{M^{2}(\mathbf{k}_{\perp}+\mathbf{k}_{\gamma \perp}-\mathbf{l}_{1\perp})}{2(k^{-}+k^{-}_{\gamma})}+\frac{i\varepsilon}{2(k^{-}+k^{-}_{\gamma})} \Bigg) \nonumber \\
& \times \Bigg( k^{+}+k^{+}_{\gamma}-l_{1}^{+}-l^{+}_{2}-\frac{M^{2}(\mathbf{k}_{\perp}+\mathbf{k}_{\gamma \perp}-\mathbf{l}_{1\perp}-\mathbf{l}_{2\perp})}{2(k^{-}+k^{-}_{\gamma})}+\frac{i\varepsilon}{2(k^{-}+k^{-}_{\gamma})} \Bigg) \Bigg(l^{+}_{3}-\frac{\mathbf{l}_{3\perp}^{2}}{2k^{-}_{g}}+\frac{i\varepsilon}{2k^{-}_{g}} \Bigg) \nonumber \\
& \times \Bigg( k^{+}+k^{+}_{\gamma}-l_{1}^{+}-l^{+}_{2}+l_{3}^{+} -\frac{M^{2}(\mathbf{k}_{\perp}+\mathbf{k}_{\gamma \perp}-\mathbf{l}_{1\perp}-\mathbf{l}_{2\perp}+\mathbf{l}_{3\perp})}{2(k^{-}+k^{-}_{\gamma}+k^{-}_{g})}+\frac{i\varepsilon}{2(k^{-}+k^{-}_{\gamma}+k^{-}_{g})} \Bigg) \nonumber \\
& \times \Bigg( q^{+}+l^{+}_{4}-p^{+}-\frac{M^{2}(\mathbf{l}_{4\perp}-\mathbf{p}_{\perp})}{2(k^{-}+k^{-}_{\gamma}+k^{-}_{g})}+\frac{i\varepsilon}{2(k^{-}+k^{-}_{\gamma}+k^{-}_{g})} \Bigg)\Bigg( l^{+}_{4}-p^{+}+\frac{M^{2}(\mathbf{l}_{4\perp}-\mathbf{p}_{\perp})}{2p^{-}}-\frac{i\varepsilon}{2p^{-}} \Bigg)  \enskip.
\end{align}
 The expressions above clearly demonstrate that the numerator doesn't have any term proportional to $l_{1}^{+}$, $l_{2}^{+}$ or $l_{3}^{+}$ and the poles for all these variables are on the same side of the real axis. Hence the integration contours for any such variable can be deformed in a way so as not to enclose any pole giving a result of zero for general $U$ and $\tilde{U}$'s depicted in Fig. \ref{fig:NLO-example-A}. However it can be easily shown that for the following three cases
 \begin{itemize}
 \item $\tilde{U}(\mathbf{x}_{1})=\tilde{U}(\mathbf{x}_{3})=\mathds{1}$ with rest of the Wilson lines being general (including the identity matrix),
 \item $\tilde{U}(\mathbf{x}_{2})=\tilde{U}(\mathbf{x}_{3})=\mathds{1}$ with rest of the Wilson lines being general and
 \item $\tilde{U}(\mathbf{x}_{1})=\tilde{U}(\mathbf{x}_{2})=\mathds{1}$, $U(\mathbf{x}_{5})=\mathds{1}$ and remaining two being general, 
\end{itemize}  
we get a finite result. Diagrammatically, this corresponds to the processes shown respectively in Fig.~\ref{fig:NLO-coeff-real} or equivalently the diagrams in Fig. \ref{fig:shockwave-diagrams} in the ``shock wave" approach. Similar arguments can be applied to find the allowed virtual graphs at NLO.

%%%%%%%%%%%%%%%%%%%%%%%%%%%%%%%%%%%%%%%%%%%%%%%%%


%%%%%%%%%%%%%%%%%%%%%%%%%%%%%%%%%%%%%%%%%%%%%%%%%

\begin{thebibliography}{121}%
\makeatletter
\providecommand \@ifxundefined [1]{%
 \@ifx{#1\undefined}
}%
\providecommand \@ifnum [1]{%
 \ifnum #1\expandafter \@firstoftwo
 \else \expandafter \@secondoftwo
 \fi
}%
\providecommand \@ifx [1]{%
 \ifx #1\expandafter \@firstoftwo
 \else \expandafter \@secondoftwo
 \fi
}%
\providecommand \natexlab [1]{#1}%
\providecommand \enquote  [1]{``#1''}%
\providecommand \bibnamefont  [1]{#1}%
\providecommand \bibfnamefont [1]{#1}%
\providecommand \citenamefont [1]{#1}%
\providecommand \href@noop [0]{\@secondoftwo}%
\providecommand \href [0]{\begingroup \@sanitize@url \@href}%
\providecommand \@href[1]{\@@startlink{#1}\@@href}%
\providecommand \@@href[1]{\endgroup#1\@@endlink}%
\providecommand \@sanitize@url [0]{\catcode `\\12\catcode `\$12\catcode
  `\&12\catcode `\#12\catcode `\^12\catcode `\_12\catcode `\%12\relax}%
\providecommand \@@startlink[1]{}%
\providecommand \@@endlink[0]{}%
\providecommand \url  [0]{\begingroup\@sanitize@url \@url }%
\providecommand \@url [1]{\endgroup\@href {#1}{\urlprefix }}%
\providecommand \urlprefix  [0]{URL }%
\providecommand \Eprint [0]{\href }%
\providecommand \doibase [0]{http://dx.doi.org/}%
\providecommand \selectlanguage [0]{\@gobble}%
\providecommand \bibinfo  [0]{\@secondoftwo}%
\providecommand \bibfield  [0]{\@secondoftwo}%
\providecommand \translation [1]{[#1]}%
\providecommand \BibitemOpen [0]{}%
\providecommand \bibitemStop [0]{}%
\providecommand \bibitemNoStop [0]{.\EOS\space}%
\providecommand \EOS [0]{\spacefactor3000\relax}%
\providecommand \BibitemShut  [1]{\csname bibitem#1\endcsname}%
\let\auto@bib@innerbib\@empty
%</preamble>
\bibitem [{\citenamefont {Breitweg}\ \emph {et~al.}(2000)\citenamefont
  {Breitweg} \emph {et~al.}}]{Breitweg:1999su}%
  \BibitemOpen
  \bibfield  {author} {\bibinfo {author} {\bibfnamefont {J.}~\bibnamefont
  {Breitweg}} \emph {et~al.} (\bibinfo {collaboration} {ZEUS}),\ }\bibfield
  {title} {\enquote {\bibinfo {title} {{Measurement of inclusive prompt photon
  photoproduction at HERA}},}\ }\href {\doibase 10.1016/S0370-2693(99)01450-1}
  {\bibfield  {journal} {\bibinfo  {journal} {Phys. Lett.}\ }\textbf {\bibinfo
  {volume} {B472}},\ \bibinfo {pages} {175--188} (\bibinfo {year} {2000})},\
  \Eprint {http://arxiv.org/abs/hep-ex/9910045} {arXiv:hep-ex/9910045 [hep-ex]}
  \BibitemShut {NoStop}%
%%CITATION = HEP-EX/9910045;%%
\bibitem [{\citenamefont {Aktas}\ \emph {et~al.}(2005)\citenamefont {Aktas}
  \emph {et~al.}}]{Aktas:2004uv}%
  \BibitemOpen
  \bibfield  {author} {\bibinfo {author} {\bibfnamefont {A.}~\bibnamefont
  {Aktas}} \emph {et~al.} (\bibinfo {collaboration} {H1}),\ }\bibfield  {title}
  {\enquote {\bibinfo {title} {{Measurement of prompt photon cross sections in
  photoproduction at HERA}},}\ }\href {\doibase 10.1140/epjc/s2004-02085-x}
  {\bibfield  {journal} {\bibinfo  {journal} {Eur. Phys. J.}\ }\textbf
  {\bibinfo {volume} {C38}},\ \bibinfo {pages} {437--445} (\bibinfo {year}
  {2005})},\ \Eprint {http://arxiv.org/abs/hep-ex/0407018}
  {arXiv:hep-ex/0407018 [hep-ex]} \BibitemShut {NoStop}%
%%CITATION = HEP-EX/0407018;%%
\bibitem [{\citenamefont {Chekanov}\ \emph {et~al.}(2007)\citenamefont
  {Chekanov} \emph {et~al.}}]{Chekanov:2006un}%
  \BibitemOpen
  \bibfield  {author} {\bibinfo {author} {\bibfnamefont {S.}~\bibnamefont
  {Chekanov}} \emph {et~al.} (\bibinfo {collaboration} {ZEUS}),\ }\bibfield
  {title} {\enquote {\bibinfo {title} {{Measurement of prompt photons with
  associated jets in photoproduction at HERA}},}\ }\href {\doibase
  10.1140/epjc/s10052-006-0134-1} {\bibfield  {journal} {\bibinfo  {journal}
  {Eur. Phys. J.}\ }\textbf {\bibinfo {volume} {C49}},\ \bibinfo {pages}
  {511--522} (\bibinfo {year} {2007})},\ \Eprint
  {http://arxiv.org/abs/hep-ex/0608028} {arXiv:hep-ex/0608028 [hep-ex]}
  \BibitemShut {NoStop}%
%%CITATION = HEP-EX/0608028;%%
\bibitem [{\citenamefont {Chekanov}\ \emph {et~al.}(2004)\citenamefont
  {Chekanov} \emph {et~al.}}]{Chekanov:2004wr}%
  \BibitemOpen
  \bibfield  {author} {\bibinfo {author} {\bibfnamefont {S.}~\bibnamefont
  {Chekanov}} \emph {et~al.} (\bibinfo {collaboration} {ZEUS}),\ }\bibfield
  {title} {\enquote {\bibinfo {title} {{Observation of isolated high E(T)
  photons in deep inelastic scattering}},}\ }\href {\doibase
  10.1016/j.physletb.2004.05.033} {\bibfield  {journal} {\bibinfo  {journal}
  {Phys. Lett.}\ }\textbf {\bibinfo {volume} {B595}},\ \bibinfo {pages}
  {86--100} (\bibinfo {year} {2004})},\ \Eprint
  {http://arxiv.org/abs/hep-ex/0402019} {arXiv:hep-ex/0402019 [hep-ex]}
  \BibitemShut {NoStop}%
%%CITATION = HEP-EX/0402019;%%
\bibitem [{\citenamefont {Aaron}\ \emph {et~al.}(2008)\citenamefont {Aaron}
  \emph {et~al.}}]{Aaron:2007aa}%
  \BibitemOpen
  \bibfield  {author} {\bibinfo {author} {\bibfnamefont {F.~D.}\ \bibnamefont
  {Aaron}} \emph {et~al.} (\bibinfo {collaboration} {H1}),\ }\bibfield  {title}
  {\enquote {\bibinfo {title} {{Measurement of isolated photon production in
  deep-inelastic scattering at HERA}},}\ }\href {\doibase
  10.1140/epjc/s10052-008-0541-6} {\bibfield  {journal} {\bibinfo  {journal}
  {Eur. Phys. J.}\ }\textbf {\bibinfo {volume} {C54}},\ \bibinfo {pages}
  {371--387} (\bibinfo {year} {2008})},\ \Eprint
  {http://arxiv.org/abs/0711.4578} {arXiv:0711.4578 [hep-ex]} \BibitemShut
  {NoStop}%
%%CITATION = ARXIV:0711.4578;%%
\bibitem [{\citenamefont {Chekanov}\ \emph {et~al.}(2010)\citenamefont
  {Chekanov} \emph {et~al.}}]{Chekanov:2009dq}%
  \BibitemOpen
  \bibfield  {author} {\bibinfo {author} {\bibfnamefont {S.}~\bibnamefont
  {Chekanov}} \emph {et~al.} (\bibinfo {collaboration} {ZEUS}),\ }\bibfield
  {title} {\enquote {\bibinfo {title} {{Measurement of isolated photon
  production in deep inelastic ep scattering}},}\ }\href {\doibase
  10.1016/j.physletb.2010.02.045} {\bibfield  {journal} {\bibinfo  {journal}
  {Phys. Lett.}\ }\textbf {\bibinfo {volume} {B687}},\ \bibinfo {pages}
  {16--25} (\bibinfo {year} {2010})},\ \Eprint {http://arxiv.org/abs/0909.4223}
  {arXiv:0909.4223 [hep-ex]} \BibitemShut {NoStop}%
%%CITATION = ARXIV:0909.4223;%%
\bibitem [{\citenamefont {Abramowicz}\ \emph {et~al.}(2012)\citenamefont
  {Abramowicz} \emph {et~al.}}]{Abramowicz:2012qt}%
  \BibitemOpen
  \bibfield  {author} {\bibinfo {author} {\bibfnamefont {H.}~\bibnamefont
  {Abramowicz}} \emph {et~al.} (\bibinfo {collaboration} {ZEUS}),\ }\bibfield
  {title} {\enquote {\bibinfo {title} {{Measurement of isolated photons
  accompanied by jets in deep inelastic $ep$ scattering}},}\ }\href {\doibase
  10.1016/j.physletb.2012.07.031} {\bibfield  {journal} {\bibinfo  {journal}
  {Phys. Lett.}\ }\textbf {\bibinfo {volume} {B715}},\ \bibinfo {pages}
  {88--97} (\bibinfo {year} {2012})},\ \Eprint {http://arxiv.org/abs/1206.2270}
  {arXiv:1206.2270 [hep-ex]} \BibitemShut {NoStop}%
%%CITATION = ARXIV:1206.2270;%%
\bibitem [{\citenamefont {Gehrmann-De~Ridder}\ \emph
  {et~al.}(2006)\citenamefont {Gehrmann-De~Ridder}, \citenamefont {Gehrmann},\
  and\ \citenamefont {Poulsen}}]{Gehrmann-DeRidder:2006zbx}%
  \BibitemOpen
  \bibfield  {author} {\bibinfo {author} {\bibfnamefont {A.}~\bibnamefont
  {Gehrmann-De~Ridder}}, \bibinfo {author} {\bibfnamefont {T.}~\bibnamefont
  {Gehrmann}}, \ and\ \bibinfo {author} {\bibfnamefont {E.}~\bibnamefont
  {Poulsen}},\ }\bibfield  {title} {\enquote {\bibinfo {title} {{Isolated
  photons in deep inelastic scattering}},}\ }\href {\doibase
  10.1103/PhysRevLett.96.132002} {\bibfield  {journal} {\bibinfo  {journal}
  {Phys. Rev. Lett.}\ }\textbf {\bibinfo {volume} {96}},\ \bibinfo {pages}
  {132002} (\bibinfo {year} {2006})},\ \Eprint
  {http://arxiv.org/abs/hep-ph/0601073} {arXiv:hep-ph/0601073 [hep-ph]}
  \BibitemShut {NoStop}%
%%CITATION = HEP-PH/0601073;%%
\bibitem [{\citenamefont {Schmidt}\ \emph {et~al.}(2016)\citenamefont
  {Schmidt}, \citenamefont {Pumplin}, \citenamefont {Stump},\ and\
  \citenamefont {Yuan}}]{Schmidt:2015zda}%
  \BibitemOpen
  \bibfield  {author} {\bibinfo {author} {\bibfnamefont {Carl}\ \bibnamefont
  {Schmidt}}, \bibinfo {author} {\bibfnamefont {Jon}\ \bibnamefont {Pumplin}},
  \bibinfo {author} {\bibfnamefont {Daniel}\ \bibnamefont {Stump}}, \ and\
  \bibinfo {author} {\bibfnamefont {C.~P.}\ \bibnamefont {Yuan}},\ }\bibfield
  {title} {\enquote {\bibinfo {title} {{CT14QED parton distribution functions
  from isolated photon production in deep inelastic scattering}},}\ }\href
  {\doibase 10.1103/PhysRevD.93.114015} {\bibfield  {journal} {\bibinfo
  {journal} {Phys. Rev.}\ }\textbf {\bibinfo {volume} {D93}},\ \bibinfo {pages}
  {114015} (\bibinfo {year} {2016})},\ \Eprint
  {http://arxiv.org/abs/1509.02905} {arXiv:1509.02905 [hep-ph]} \BibitemShut
  {NoStop}%
%%CITATION = ARXIV:1509.02905;%%
\bibitem [{\citenamefont {Gribov}\ \emph {et~al.}(1983)\citenamefont {Gribov},
  \citenamefont {Levin},\ and\ \citenamefont {Ryskin}}]{Gribov:1984tu}%
  \BibitemOpen
  \bibfield  {author} {\bibinfo {author} {\bibfnamefont {L.~V.}\ \bibnamefont
  {Gribov}}, \bibinfo {author} {\bibfnamefont {E.~M.}\ \bibnamefont {Levin}}, \
  and\ \bibinfo {author} {\bibfnamefont {M.~G.}\ \bibnamefont {Ryskin}},\
  }\bibfield  {title} {\enquote {\bibinfo {title} {{Semihard Processes in
  QCD}},}\ }\href {\doibase 10.1016/0370-1573(83)90022-4} {\bibfield  {journal}
  {\bibinfo  {journal} {Phys. Rept.}\ }\textbf {\bibinfo {volume} {100}},\
  \bibinfo {pages} {1--150} (\bibinfo {year} {1983})}\BibitemShut {NoStop}%
%%CITATION = PRPLC,100,1;%%
\bibitem [{\citenamefont {Mueller}\ and\ \citenamefont
  {Qiu}(1986)}]{Mueller:1985wy}%
  \BibitemOpen
  \bibfield  {author} {\bibinfo {author} {\bibfnamefont {Alfred~H.}\
  \bibnamefont {Mueller}}\ and\ \bibinfo {author} {\bibfnamefont {Jian-wei}\
  \bibnamefont {Qiu}},\ }\bibfield  {title} {\enquote {\bibinfo {title} {{Gluon
  Recombination and Shadowing at Small Values of $x$}},}\ }\href {\doibase
  10.1016/0550-3213(86)90164-1} {\bibfield  {journal} {\bibinfo  {journal}
  {Nucl. Phys.}\ }\textbf {\bibinfo {volume} {B268}},\ \bibinfo {pages}
  {427--452} (\bibinfo {year} {1986})}\BibitemShut {NoStop}%
%%CITATION = NUPHA,B268,427;%%
\bibitem [{\citenamefont {McLerran}\ and\ \citenamefont
  {Venugopalan}(1994{\natexlab{a}})}]{McLerran:1993ni}%
  \BibitemOpen
  \bibfield  {author} {\bibinfo {author} {\bibfnamefont {Larry~D.}\
  \bibnamefont {McLerran}}\ and\ \bibinfo {author} {\bibfnamefont {Raju}\
  \bibnamefont {Venugopalan}},\ }\bibfield  {title} {\enquote {\bibinfo {title}
  {{Computing quark and gluon distribution functions for very large nuclei}},}\
  }\href {\doibase 10.1103/PhysRevD.49.2233} {\bibfield  {journal} {\bibinfo
  {journal} {Phys. Rev.}\ }\textbf {\bibinfo {volume} {D49}},\ \bibinfo {pages}
  {2233--2241} (\bibinfo {year} {1994}{\natexlab{a}})},\ \Eprint
  {http://arxiv.org/abs/hep-ph/9309289} {arXiv:hep-ph/9309289 [hep-ph]}
  \BibitemShut {NoStop}%
%%CITATION = HEP-PH/9309289;%%
\bibitem [{\citenamefont {McLerran}\ and\ \citenamefont
  {Venugopalan}(1994{\natexlab{b}})}]{McLerran:1993ka}%
  \BibitemOpen
  \bibfield  {author} {\bibinfo {author} {\bibfnamefont {Larry~D.}\
  \bibnamefont {McLerran}}\ and\ \bibinfo {author} {\bibfnamefont {Raju}\
  \bibnamefont {Venugopalan}},\ }\bibfield  {title} {\enquote {\bibinfo {title}
  {{Gluon distribution functions for very large nuclei at small transverse
  momentum}},}\ }\href {\doibase 10.1103/PhysRevD.49.3352} {\bibfield
  {journal} {\bibinfo  {journal} {Phys. Rev.}\ }\textbf {\bibinfo {volume}
  {D49}},\ \bibinfo {pages} {3352--3355} (\bibinfo {year}
  {1994}{\natexlab{b}})},\ \Eprint {http://arxiv.org/abs/hep-ph/9311205}
  {arXiv:hep-ph/9311205 [hep-ph]} \BibitemShut {NoStop}%
%%CITATION = HEP-PH/9311205;%%
\bibitem [{\citenamefont {McLerran}\ and\ \citenamefont
  {Venugopalan}(1994{\natexlab{c}})}]{McLerran:1994vd}%
  \BibitemOpen
  \bibfield  {author} {\bibinfo {author} {\bibfnamefont {Larry~D.}\
  \bibnamefont {McLerran}}\ and\ \bibinfo {author} {\bibfnamefont {Raju}\
  \bibnamefont {Venugopalan}},\ }\bibfield  {title} {\enquote {\bibinfo {title}
  {{Green's functions in the color field of a large nucleus}},}\ }\href
  {\doibase 10.1103/PhysRevD.50.2225} {\bibfield  {journal} {\bibinfo
  {journal} {Phys. Rev.}\ }\textbf {\bibinfo {volume} {D50}},\ \bibinfo {pages}
  {2225--2233} (\bibinfo {year} {1994}{\natexlab{c}})},\ \Eprint
  {http://arxiv.org/abs/hep-ph/9402335} {arXiv:hep-ph/9402335 [hep-ph]}
  \BibitemShut {NoStop}%
%%CITATION = HEP-PH/9402335;%%
\bibitem [{\citenamefont {Iancu}\ and\ \citenamefont
  {Venugopalan}(2003)}]{Iancu:2003xm}%
  \BibitemOpen
  \bibfield  {author} {\bibinfo {author} {\bibfnamefont {Edmond}\ \bibnamefont
  {Iancu}}\ and\ \bibinfo {author} {\bibfnamefont {Raju}\ \bibnamefont
  {Venugopalan}},\ }\bibfield  {title} {\enquote {\bibinfo {title} {{The Color
  glass condensate and high-energy scattering in QCD}},}\ }in\ \href {\doibase
  10.1142/9789812795533_0005} {\emph {\bibinfo {booktitle} {In *Hwa, R.C. (ed.)
  et al.: Quark gluon plasma* 249-3363}}}\ (\bibinfo {year} {2003})\ \Eprint
  {http://arxiv.org/abs/hep-ph/0303204} {arXiv:hep-ph/0303204 [hep-ph]}
  \BibitemShut {NoStop}%
%%CITATION = HEP-PH/0303204;%%
\bibitem [{\citenamefont {Gelis}\ \emph {et~al.}(2010)\citenamefont {Gelis},
  \citenamefont {Iancu}, \citenamefont {Jalilian-Marian},\ and\ \citenamefont
  {Venugopalan}}]{Gelis:2010nm}%
  \BibitemOpen
  \bibfield  {author} {\bibinfo {author} {\bibfnamefont {Francois}\
  \bibnamefont {Gelis}}, \bibinfo {author} {\bibfnamefont {Edmond}\
  \bibnamefont {Iancu}}, \bibinfo {author} {\bibfnamefont {Jamal}\ \bibnamefont
  {Jalilian-Marian}}, \ and\ \bibinfo {author} {\bibfnamefont {Raju}\
  \bibnamefont {Venugopalan}},\ }\bibfield  {title} {\enquote {\bibinfo {title}
  {{The Color Glass Condensate}},}\ }\href {\doibase
  10.1146/annurev.nucl.010909.083629} {\bibfield  {journal} {\bibinfo
  {journal} {Ann. Rev. Nucl. Part. Sci.}\ }\textbf {\bibinfo {volume} {60}},\
  \bibinfo {pages} {463--489} (\bibinfo {year} {2010})},\ \Eprint
  {http://arxiv.org/abs/1002.0333} {arXiv:1002.0333 [hep-ph]} \BibitemShut
  {NoStop}%
%%CITATION = ARXIV:1002.0333;%%
\bibitem [{\citenamefont {Kovchegov}\ and\ \citenamefont
  {Levin}(2012)}]{Kovchegov:2012mbw}%
  \BibitemOpen
  \bibfield  {author} {\bibinfo {author} {\bibfnamefont {Yuri~V.}\ \bibnamefont
  {Kovchegov}}\ and\ \bibinfo {author} {\bibfnamefont {Eugene}\ \bibnamefont
  {Levin}},\ }\href {http://www.cambridge.org/de/knowledge/isbn/item6803159}
  {\emph {\bibinfo {title} {{Quantum chromodynamics at high energy}}}},\
  Vol.~\bibinfo {volume} {33}\ (\bibinfo  {publisher} {Cambridge University
  Press},\ \bibinfo {year} {2012})\BibitemShut {NoStop}%
%%CITATION = CMMPE,33,;%%
\bibitem [{\citenamefont {Blaizot}(2017)}]{Blaizot:2016qgz}%
  \BibitemOpen
  \bibfield  {author} {\bibinfo {author} {\bibfnamefont {Jean-Paul}\
  \bibnamefont {Blaizot}},\ }\bibfield  {title} {\enquote {\bibinfo {title}
  {{High gluon densities in heavy ion collisions}},}\ }\href {\doibase
  10.1088/1361-6633/aa5435} {\bibfield  {journal} {\bibinfo  {journal} {Rept.
  Prog. Phys.}\ }\textbf {\bibinfo {volume} {80}},\ \bibinfo {pages} {032301}
  (\bibinfo {year} {2017})},\ \Eprint {http://arxiv.org/abs/1607.04448}
  {arXiv:1607.04448 [hep-ph]} \BibitemShut {NoStop}%
%%CITATION = ARXIV:1607.04448;%%
\bibitem [{\citenamefont {Fubini}\ and\ \citenamefont
  {Furlan}(1965)}]{Fubini:1964boa}%
  \BibitemOpen
  \bibfield  {author} {\bibinfo {author} {\bibfnamefont {S.}~\bibnamefont
  {Fubini}}\ and\ \bibinfo {author} {\bibfnamefont {G.}~\bibnamefont
  {Furlan}},\ }\bibfield  {title} {\enquote {\bibinfo {title} {{Renormalization
  effects for partially conserved currents}},}\ }\href@noop {} {\bibfield
  {journal} {\bibinfo  {journal} {Physics}\ }\textbf {\bibinfo {volume} {1}},\
  \bibinfo {pages} {229--247} (\bibinfo {year} {1965})}\BibitemShut {NoStop}%
%%CITATION = PYCSA,1,229;%%
\bibitem [{\citenamefont {Bardakci}\ and\ \citenamefont
  {Halpern}(1968)}]{Bardakci:1969dv}%
  \BibitemOpen
  \bibfield  {author} {\bibinfo {author} {\bibfnamefont {K.}~\bibnamefont
  {Bardakci}}\ and\ \bibinfo {author} {\bibfnamefont {M.~B.}\ \bibnamefont
  {Halpern}},\ }\bibfield  {title} {\enquote {\bibinfo {title} {{Theories at
  infinite momentum}},}\ }\href {\doibase 10.1103/PhysRev.176.1686} {\bibfield
  {journal} {\bibinfo  {journal} {Phys. Rev.}\ }\textbf {\bibinfo {volume}
  {176}},\ \bibinfo {pages} {1686--1699} (\bibinfo {year} {1968})}\BibitemShut
  {NoStop}%
%%CITATION = PHRVA,176,1686;%%
\bibitem [{\citenamefont {Weinberg}(1966)}]{Weinberg:1966jm}%
  \BibitemOpen
  \bibfield  {author} {\bibinfo {author} {\bibfnamefont {Steven}\ \bibnamefont
  {Weinberg}},\ }\bibfield  {title} {\enquote {\bibinfo {title} {{Dynamics at
  infinite momentum}},}\ }\href {\doibase 10.1103/PhysRev.150.1313} {\bibfield
  {journal} {\bibinfo  {journal} {Phys. Rev.}\ }\textbf {\bibinfo {volume}
  {150}},\ \bibinfo {pages} {1313--1318} (\bibinfo {year} {1966})}\BibitemShut
  {NoStop}%
%%CITATION = PHRVA,150,1313;%%
\bibitem [{\citenamefont {Susskind}(1968)}]{Susskind:1967rg}%
  \BibitemOpen
  \bibfield  {author} {\bibinfo {author} {\bibfnamefont {Leonard}\ \bibnamefont
  {Susskind}},\ }\bibfield  {title} {\enquote {\bibinfo {title} {{Model of
  selfinduced strong interactions}},}\ }\href {\doibase
  10.1103/PhysRev.165.1535} {\bibfield  {journal} {\bibinfo  {journal} {Phys.
  Rev.}\ }\textbf {\bibinfo {volume} {165}},\ \bibinfo {pages} {1535--1546}
  (\bibinfo {year} {1968})}\BibitemShut {NoStop}%
%%CITATION = PHRVA,165,1535;%%
\bibitem [{\citenamefont {Jeon}\ and\ \citenamefont
  {Venugopalan}(2004)}]{Jeon:2004rk}%
  \BibitemOpen
  \bibfield  {author} {\bibinfo {author} {\bibfnamefont {Sangyong}\
  \bibnamefont {Jeon}}\ and\ \bibinfo {author} {\bibfnamefont {Raju}\
  \bibnamefont {Venugopalan}},\ }\bibfield  {title} {\enquote {\bibinfo {title}
  {{Random walks of partons in SU(N(c)) and classical representations of color
  charges in QCD at small x}},}\ }\href {\doibase 10.1103/PhysRevD.70.105012}
  {\bibfield  {journal} {\bibinfo  {journal} {Phys. Rev.}\ }\textbf {\bibinfo
  {volume} {D70}},\ \bibinfo {pages} {105012} (\bibinfo {year} {2004})},\
  \Eprint {http://arxiv.org/abs/hep-ph/0406169} {arXiv:hep-ph/0406169 [hep-ph]}
  \BibitemShut {NoStop}%
%%CITATION = HEP-PH/0406169;%%
\bibitem [{\citenamefont {Jalilian-Marian}\ \emph {et~al.}(1997)\citenamefont
  {Jalilian-Marian}, \citenamefont {Kovner}, \citenamefont {Leonidov},\ and\
  \citenamefont {Weigert}}]{JalilianMarian:1997jx}%
  \BibitemOpen
  \bibfield  {author} {\bibinfo {author} {\bibfnamefont {Jamal}\ \bibnamefont
  {Jalilian-Marian}}, \bibinfo {author} {\bibfnamefont {Alex}\ \bibnamefont
  {Kovner}}, \bibinfo {author} {\bibfnamefont {Andrei}\ \bibnamefont
  {Leonidov}}, \ and\ \bibinfo {author} {\bibfnamefont {Heribert}\ \bibnamefont
  {Weigert}},\ }\bibfield  {title} {\enquote {\bibinfo {title} {{The BFKL
  equation from the Wilson renormalization group}},}\ }\href {\doibase
  10.1016/S0550-3213(97)00440-9} {\bibfield  {journal} {\bibinfo  {journal}
  {Nucl. Phys.}\ }\textbf {\bibinfo {volume} {B504}},\ \bibinfo {pages}
  {415--431} (\bibinfo {year} {1997})},\ \Eprint
  {http://arxiv.org/abs/hep-ph/9701284} {arXiv:hep-ph/9701284 [hep-ph]}
  \BibitemShut {NoStop}%
%%CITATION = HEP-PH/9701284;%%
\bibitem [{\citenamefont {Jalilian-Marian}\ \emph {et~al.}(1998)\citenamefont
  {Jalilian-Marian}, \citenamefont {Kovner},\ and\ \citenamefont
  {Weigert}}]{JalilianMarian:1997dw}%
  \BibitemOpen
  \bibfield  {author} {\bibinfo {author} {\bibfnamefont {Jamal}\ \bibnamefont
  {Jalilian-Marian}}, \bibinfo {author} {\bibfnamefont {Alex}\ \bibnamefont
  {Kovner}}, \ and\ \bibinfo {author} {\bibfnamefont {Heribert}\ \bibnamefont
  {Weigert}},\ }\bibfield  {title} {\enquote {\bibinfo {title} {{The Wilson
  renormalization group for low $x$ physics: Gluon evolution at finite parton
  density}},}\ }\href {\doibase 10.1103/PhysRevD.59.014015} {\bibfield
  {journal} {\bibinfo  {journal} {Phys. Rev.}\ }\textbf {\bibinfo {volume}
  {D59}},\ \bibinfo {pages} {014015} (\bibinfo {year} {1998})},\ \Eprint
  {http://arxiv.org/abs/hep-ph/9709432} {arXiv:hep-ph/9709432 [hep-ph]}
  \BibitemShut {NoStop}%
%%CITATION = HEP-PH/9709432;%%
\bibitem [{\citenamefont {Iancu}\ \emph
  {et~al.}(2001{\natexlab{a}})\citenamefont {Iancu}, \citenamefont {Leonidov},\
  and\ \citenamefont {McLerran}}]{Iancu:2001ad}%
  \BibitemOpen
  \bibfield  {author} {\bibinfo {author} {\bibfnamefont {Edmond}\ \bibnamefont
  {Iancu}}, \bibinfo {author} {\bibfnamefont {Andrei}\ \bibnamefont
  {Leonidov}}, \ and\ \bibinfo {author} {\bibfnamefont {Larry~D.}\ \bibnamefont
  {McLerran}},\ }\bibfield  {title} {\enquote {\bibinfo {title} {{The
  Renormalization group equation for the color glass condensate}},}\ }\href
  {\doibase 10.1016/S0370-2693(01)00524-X} {\bibfield  {journal} {\bibinfo
  {journal} {Phys. Lett.}\ }\textbf {\bibinfo {volume} {B510}},\ \bibinfo
  {pages} {133--144} (\bibinfo {year} {2001}{\natexlab{a}})},\ \Eprint
  {http://arxiv.org/abs/hep-ph/0102009} {arXiv:hep-ph/0102009 [hep-ph]}
  \BibitemShut {NoStop}%
%%CITATION = HEP-PH/0102009;%%
\bibitem [{\citenamefont {Iancu}\ \emph
  {et~al.}(2001{\natexlab{b}})\citenamefont {Iancu}, \citenamefont {Leonidov},\
  and\ \citenamefont {McLerran}}]{Iancu:2000hn}%
  \BibitemOpen
  \bibfield  {author} {\bibinfo {author} {\bibfnamefont {Edmond}\ \bibnamefont
  {Iancu}}, \bibinfo {author} {\bibfnamefont {Andrei}\ \bibnamefont
  {Leonidov}}, \ and\ \bibinfo {author} {\bibfnamefont {Larry~D.}\ \bibnamefont
  {McLerran}},\ }\bibfield  {title} {\enquote {\bibinfo {title} {{Nonlinear
  gluon evolution in the color glass condensate. 1.}}}\ }\href {\doibase
  10.1016/S0375-9474(01)00642-X} {\bibfield  {journal} {\bibinfo  {journal}
  {Nucl. Phys.}\ }\textbf {\bibinfo {volume} {A692}},\ \bibinfo {pages}
  {583--645} (\bibinfo {year} {2001}{\natexlab{b}})},\ \Eprint
  {http://arxiv.org/abs/hep-ph/0011241} {arXiv:hep-ph/0011241 [hep-ph]}
  \BibitemShut {NoStop}%
%%CITATION = HEP-PH/0011241;%%
\bibitem [{\citenamefont {Ferreiro}\ \emph {et~al.}(2002)\citenamefont
  {Ferreiro}, \citenamefont {Iancu}, \citenamefont {Leonidov},\ and\
  \citenamefont {McLerran}}]{Ferreiro:2001qy}%
  \BibitemOpen
  \bibfield  {author} {\bibinfo {author} {\bibfnamefont {Elena}\ \bibnamefont
  {Ferreiro}}, \bibinfo {author} {\bibfnamefont {Edmond}\ \bibnamefont
  {Iancu}}, \bibinfo {author} {\bibfnamefont {Andrei}\ \bibnamefont
  {Leonidov}}, \ and\ \bibinfo {author} {\bibfnamefont {Larry}\ \bibnamefont
  {McLerran}},\ }\bibfield  {title} {\enquote {\bibinfo {title} {{Nonlinear
  gluon evolution in the color glass condensate. 2.}}}\ }\href {\doibase
  10.1016/S0375-9474(01)01329-X} {\bibfield  {journal} {\bibinfo  {journal}
  {Nucl. Phys.}\ }\textbf {\bibinfo {volume} {A703}},\ \bibinfo {pages}
  {489--538} (\bibinfo {year} {2002})},\ \Eprint
  {http://arxiv.org/abs/hep-ph/0109115} {arXiv:hep-ph/0109115 [hep-ph]}
  \BibitemShut {NoStop}%
%%CITATION = HEP-PH/0109115;%%
\bibitem [{\citenamefont {Weigert}(2005)}]{Weigert:2005us}%
  \BibitemOpen
  \bibfield  {author} {\bibinfo {author} {\bibfnamefont {Heribert}\
  \bibnamefont {Weigert}},\ }\bibfield  {title} {\enquote {\bibinfo {title}
  {{Evolution at small x(bj): The Color glass condensate}},}\ }\href {\doibase
  10.1016/j.ppnp.2005.01.029} {\bibfield  {journal} {\bibinfo  {journal} {Prog.
  Part. Nucl. Phys.}\ }\textbf {\bibinfo {volume} {55}},\ \bibinfo {pages}
  {461--565} (\bibinfo {year} {2005})},\ \Eprint
  {http://arxiv.org/abs/hep-ph/0501087} {arXiv:hep-ph/0501087 [hep-ph]}
  \BibitemShut {NoStop}%
%%CITATION = HEP-PH/0501087;%%
\bibitem [{\citenamefont {Balitsky}(1996)}]{Balitsky:1995ub}%
  \BibitemOpen
  \bibfield  {author} {\bibinfo {author} {\bibfnamefont {I.}~\bibnamefont
  {Balitsky}},\ }\bibfield  {title} {\enquote {\bibinfo {title} {{Operator
  expansion for high-energy scattering}},}\ }\href {\doibase
  10.1016/0550-3213(95)00638-9} {\bibfield  {journal} {\bibinfo  {journal}
  {Nucl. Phys.}\ }\textbf {\bibinfo {volume} {B463}},\ \bibinfo {pages}
  {99--160} (\bibinfo {year} {1996})},\ \Eprint
  {http://arxiv.org/abs/hep-ph/9509348} {arXiv:hep-ph/9509348 [hep-ph]}
  \BibitemShut {NoStop}%
%%CITATION = HEP-PH/9509348;%%
\bibitem [{\citenamefont {Weigert}(2002)}]{Weigert:2000gi}%
  \BibitemOpen
  \bibfield  {author} {\bibinfo {author} {\bibfnamefont {Heribert}\
  \bibnamefont {Weigert}},\ }\bibfield  {title} {\enquote {\bibinfo {title}
  {{Unitarity at small Bjorken x}},}\ }\href {\doibase
  10.1016/S0375-9474(01)01668-2} {\bibfield  {journal} {\bibinfo  {journal}
  {Nucl. Phys.}\ }\textbf {\bibinfo {volume} {A703}},\ \bibinfo {pages}
  {823--860} (\bibinfo {year} {2002})},\ \Eprint
  {http://arxiv.org/abs/hep-ph/0004044} {arXiv:hep-ph/0004044 [hep-ph]}
  \BibitemShut {NoStop}%
%%CITATION = HEP-PH/0004044;%%
\bibitem [{\citenamefont {Kovchegov}(1999)}]{Kovchegov:1999yj}%
  \BibitemOpen
  \bibfield  {author} {\bibinfo {author} {\bibfnamefont {Yuri~V.}\ \bibnamefont
  {Kovchegov}},\ }\bibfield  {title} {\enquote {\bibinfo {title} {{Small $x$
  F(2) structure function of a nucleus including multiple pomeron
  exchanges}},}\ }\href {\doibase 10.1103/PhysRevD.60.034008} {\bibfield
  {journal} {\bibinfo  {journal} {Phys. Rev.}\ }\textbf {\bibinfo {volume}
  {D60}},\ \bibinfo {pages} {034008} (\bibinfo {year} {1999})},\ \Eprint
  {http://arxiv.org/abs/hep-ph/9901281} {arXiv:hep-ph/9901281 [hep-ph]}
  \BibitemShut {NoStop}%
%%CITATION = HEP-PH/9901281;%%
\bibitem [{\citenamefont {Rummukainen}\ and\ \citenamefont
  {Weigert}(2004)}]{Rummukainen:2003ns}%
  \BibitemOpen
  \bibfield  {author} {\bibinfo {author} {\bibfnamefont {Kari}\ \bibnamefont
  {Rummukainen}}\ and\ \bibinfo {author} {\bibfnamefont {Heribert}\
  \bibnamefont {Weigert}},\ }\bibfield  {title} {\enquote {\bibinfo {title}
  {{Universal features of JIMWLK and BK evolution at small $x$}},}\ }\href
  {\doibase 10.1016/j.nuclphysa.2004.03.219} {\bibfield  {journal} {\bibinfo
  {journal} {Nucl. Phys.}\ }\textbf {\bibinfo {volume} {A739}},\ \bibinfo
  {pages} {183--226} (\bibinfo {year} {2004})},\ \Eprint
  {http://arxiv.org/abs/hep-ph/0309306} {arXiv:hep-ph/0309306 [hep-ph]}
  \BibitemShut {NoStop}%
%%CITATION = HEP-PH/0309306;%%
\bibitem [{\citenamefont {Dumitru}\ \emph {et~al.}(2011)\citenamefont
  {Dumitru}, \citenamefont {Jalilian-Marian}, \citenamefont {Lappi},
  \citenamefont {Schenke},\ and\ \citenamefont {Venugopalan}}]{Dumitru:2011vk}%
  \BibitemOpen
  \bibfield  {author} {\bibinfo {author} {\bibfnamefont {Adrian}\ \bibnamefont
  {Dumitru}}, \bibinfo {author} {\bibfnamefont {Jamal}\ \bibnamefont
  {Jalilian-Marian}}, \bibinfo {author} {\bibfnamefont {Tuomas}\ \bibnamefont
  {Lappi}}, \bibinfo {author} {\bibfnamefont {Bjoern}\ \bibnamefont {Schenke}},
  \ and\ \bibinfo {author} {\bibfnamefont {Raju}\ \bibnamefont {Venugopalan}},\
  }\bibfield  {title} {\enquote {\bibinfo {title} {{Renormalization group
  evolution of multi-gluon correlators in high energy QCD}},}\ }\href {\doibase
  10.1016/j.physletb.2011.11.002} {\bibfield  {journal} {\bibinfo  {journal}
  {Phys. Lett.}\ }\textbf {\bibinfo {volume} {B706}},\ \bibinfo {pages}
  {219--224} (\bibinfo {year} {2011})},\ \Eprint
  {http://arxiv.org/abs/1108.4764} {arXiv:1108.4764 [hep-ph]} \BibitemShut
  {NoStop}%
%%CITATION = ARXIV:1108.4764;%%
\bibitem [{\citenamefont {Benic}\ \emph {et~al.}(2017)\citenamefont {Benic},
  \citenamefont {Fukushima}, \citenamefont {Garcia-Montero},\ and\
  \citenamefont {Venugopalan}}]{Benic:2016uku}%
  \BibitemOpen
  \bibfield  {author} {\bibinfo {author} {\bibfnamefont {Sanjin}\ \bibnamefont
  {Benic}}, \bibinfo {author} {\bibfnamefont {Kenji}\ \bibnamefont
  {Fukushima}}, \bibinfo {author} {\bibfnamefont {Oscar}\ \bibnamefont
  {Garcia-Montero}}, \ and\ \bibinfo {author} {\bibfnamefont {Raju}\
  \bibnamefont {Venugopalan}},\ }\bibfield  {title} {\enquote {\bibinfo {title}
  {{Probing gluon saturation with next-to-leading order photon production at
  central rapidities in proton-nucleus collisions}},}\ }\href {\doibase
  10.1007/JHEP01(2017)115} {\bibfield  {journal} {\bibinfo  {journal} {JHEP}\
  }\textbf {\bibinfo {volume} {01}},\ \bibinfo {pages} {115} (\bibinfo {year}
  {2017})},\ \Eprint {http://arxiv.org/abs/1609.09424} {arXiv:1609.09424
  [hep-ph]} \BibitemShut {NoStop}%
%%CITATION = ARXIV:1609.09424;%%
\bibitem [{\citenamefont {Gelis}\ and\ \citenamefont
  {Jalilian-Marian}(2002)}]{Gelis:2002ki}%
  \BibitemOpen
  \bibfield  {author} {\bibinfo {author} {\bibfnamefont {Francois}\
  \bibnamefont {Gelis}}\ and\ \bibinfo {author} {\bibfnamefont {Jamal}\
  \bibnamefont {Jalilian-Marian}},\ }\bibfield  {title} {\enquote {\bibinfo
  {title} {{Photon production in high-energy proton nucleus collisions}},}\
  }\href {\doibase 10.1103/PhysRevD.66.014021} {\bibfield  {journal} {\bibinfo
  {journal} {Phys. Rev.}\ }\textbf {\bibinfo {volume} {D66}},\ \bibinfo {pages}
  {014021} (\bibinfo {year} {2002})},\ \Eprint
  {http://arxiv.org/abs/hep-ph/0205037} {arXiv:hep-ph/0205037 [hep-ph]}
  \BibitemShut {NoStop}%
%%CITATION = HEP-PH/0205037;%%
\bibitem [{\citenamefont {Dominguez}\ \emph {et~al.}(2011)\citenamefont
  {Dominguez}, \citenamefont {Marquet}, \citenamefont {Xiao},\ and\
  \citenamefont {Yuan}}]{Dominguez:2011wm}%
  \BibitemOpen
  \bibfield  {author} {\bibinfo {author} {\bibfnamefont {Fabio}\ \bibnamefont
  {Dominguez}}, \bibinfo {author} {\bibfnamefont {Cyrille}\ \bibnamefont
  {Marquet}}, \bibinfo {author} {\bibfnamefont {Bo-Wen}\ \bibnamefont {Xiao}},
  \ and\ \bibinfo {author} {\bibfnamefont {Feng}\ \bibnamefont {Yuan}},\
  }\bibfield  {title} {\enquote {\bibinfo {title} {{Universality of
  Unintegrated Gluon Distributions at small $x$}},}\ }\href {\doibase
  10.1103/PhysRevD.83.105005} {\bibfield  {journal} {\bibinfo  {journal} {Phys.
  Rev.}\ }\textbf {\bibinfo {volume} {D83}},\ \bibinfo {pages} {105005}
  (\bibinfo {year} {2011})},\ \Eprint {http://arxiv.org/abs/1101.0715}
  {arXiv:1101.0715 [hep-ph]} \BibitemShut {NoStop}%
%%CITATION = ARXIV:1101.0715;%%
\bibitem [{\citenamefont {Gribov}\ \emph {et~al.}(1966)\citenamefont {Gribov},
  \citenamefont {Ioffe},\ and\ \citenamefont {Pomeranchuk}}]{Gribov:1965hf}%
  \BibitemOpen
  \bibfield  {author} {\bibinfo {author} {\bibfnamefont {V.~N.}\ \bibnamefont
  {Gribov}}, \bibinfo {author} {\bibfnamefont {B.~L.}\ \bibnamefont {Ioffe}}, \
  and\ \bibinfo {author} {\bibfnamefont {I.~{\relax Ya}.}\ \bibnamefont
  {Pomeranchuk}},\ }\bibfield  {title} {\enquote {\bibinfo {title} {{What is
  the range of interactions at high-energies}},}\ }\href@noop {} {\bibfield
  {journal} {\bibinfo  {journal} {Sov. J. Nucl. Phys.}\ }\textbf {\bibinfo
  {volume} {2}},\ \bibinfo {pages} {549} (\bibinfo {year} {1966})},\ \bibinfo
  {note} {[Yad. Fiz.2,768(1965)]}\BibitemShut {NoStop}%
%%CITATION = SJNCA,2,549;%%
\bibitem [{\citenamefont {Ioffe}(1969)}]{Ioffe:1969kf}%
  \BibitemOpen
  \bibfield  {author} {\bibinfo {author} {\bibfnamefont {B.~L.}\ \bibnamefont
  {Ioffe}},\ }\bibfield  {title} {\enquote {\bibinfo {title} {{Space-time
  picture of photon and neutrino scattering and electroproduction cross-section
  asymptotics}},}\ }\href {\doibase 10.1016/0370-2693(69)90415-8} {\bibfield
  {journal} {\bibinfo  {journal} {Phys. Lett.}\ }\textbf {\bibinfo {volume}
  {30B}},\ \bibinfo {pages} {123--125} (\bibinfo {year} {1969})}\BibitemShut
  {NoStop}%
%%CITATION = PHLTA,30B,123;%%
\bibitem [{\citenamefont {Bjorken}\ \emph {et~al.}(1971)\citenamefont
  {Bjorken}, \citenamefont {Kogut},\ and\ \citenamefont
  {Soper}}]{Bjorken:1970ah}%
  \BibitemOpen
  \bibfield  {author} {\bibinfo {author} {\bibfnamefont {J.~D.}\ \bibnamefont
  {Bjorken}}, \bibinfo {author} {\bibfnamefont {John~B.}\ \bibnamefont
  {Kogut}}, \ and\ \bibinfo {author} {\bibfnamefont {Davison~E.}\ \bibnamefont
  {Soper}},\ }\bibfield  {title} {\enquote {\bibinfo {title} {{Quantum
  Electrodynamics at Infinite Momentum: Scattering from an External Field}},}\
  }\href {\doibase 10.1103/PhysRevD.3.1382} {\bibfield  {journal} {\bibinfo
  {journal} {Phys. Rev.}\ }\textbf {\bibinfo {volume} {D3}},\ \bibinfo {pages}
  {1382} (\bibinfo {year} {1971})}\BibitemShut {NoStop}%
%%CITATION = PHRVA,D3,1382;%%
\bibitem [{\citenamefont {Nikolaev}\ and\ \citenamefont
  {Zakharov}(1991)}]{Nikolaev:1990ja}%
  \BibitemOpen
  \bibfield  {author} {\bibinfo {author} {\bibfnamefont {Nikolai~N.}\
  \bibnamefont {Nikolaev}}\ and\ \bibinfo {author} {\bibfnamefont {B.~G.}\
  \bibnamefont {Zakharov}},\ }\bibfield  {title} {\enquote {\bibinfo {title}
  {{Color transparency and scaling properties of nuclear shadowing in deep
  inelastic scattering}},}\ }\href {\doibase 10.1007/BF01483577} {\bibfield
  {journal} {\bibinfo  {journal} {Z. Phys.}\ }\textbf {\bibinfo {volume}
  {C49}},\ \bibinfo {pages} {607--618} (\bibinfo {year} {1991})}\BibitemShut
  {NoStop}%
%%CITATION = ZEPYA,C49,607;%%
\bibitem [{\citenamefont {Benic}\ and\ \citenamefont
  {Fukushima}(2017)}]{Benic:2016yqt}%
  \BibitemOpen
  \bibfield  {author} {\bibinfo {author} {\bibfnamefont {Sanjin}\ \bibnamefont
  {Benic}}\ and\ \bibinfo {author} {\bibfnamefont {Kenji}\ \bibnamefont
  {Fukushima}},\ }\bibfield  {title} {\enquote {\bibinfo {title} {{Photon from
  the annihilation process with CGC in the $pA$ collision}},}\ }\href {\doibase
  10.1016/j.nuclphysa.2016.11.003} {\bibfield  {journal} {\bibinfo  {journal}
  {Nucl. Phys.}\ }\textbf {\bibinfo {volume} {A958}},\ \bibinfo {pages} {1--24}
  (\bibinfo {year} {2017})},\ \Eprint {http://arxiv.org/abs/1602.01989}
  {arXiv:1602.01989 [hep-ph]} \BibitemShut {NoStop}%
%%CITATION = ARXIV:1602.01989;%%
\bibitem [{\citenamefont {Altinoluk}\ \emph {et~al.}(2018)\citenamefont
  {Altinoluk}, \citenamefont {Armesto}, \citenamefont {Kovner}, \citenamefont
  {Lublinsky},\ and\ \citenamefont {Petreska}}]{Altinoluk:2018uax}%
  \BibitemOpen
  \bibfield  {author} {\bibinfo {author} {\bibfnamefont {Tolga}\ \bibnamefont
  {Altinoluk}}, \bibinfo {author} {\bibfnamefont {N\'{e}stor}\ \bibnamefont
  {Armesto}}, \bibinfo {author} {\bibfnamefont {Alex}\ \bibnamefont {Kovner}},
  \bibinfo {author} {\bibfnamefont {Michael}\ \bibnamefont {Lublinsky}}, \ and\
  \bibinfo {author} {\bibfnamefont {Elena}\ \bibnamefont {Petreska}},\
  }\bibfield  {title} {\enquote {\bibinfo {title} {{Soft photon and two hard
  jets forward production in proton-nucleus collisions}},}\ }\href@noop {} {\
  (\bibinfo {year} {2018})},\ \Eprint {http://arxiv.org/abs/1802.01398}
  {arXiv:1802.01398 [hep-ph]} \BibitemShut {NoStop}%
%%CITATION = ARXIV:1802.01398;%%
\bibitem [{\citenamefont {Dokshitzer}(1977)}]{Dokshitzer:1977sg}%
  \BibitemOpen
  \bibfield  {author} {\bibinfo {author} {\bibfnamefont {Yuri~L.}\ \bibnamefont
  {Dokshitzer}},\ }\bibfield  {title} {\enquote {\bibinfo {title} {{Calculation
  of the Structure Functions for Deep Inelastic Scattering and e+ e-
  Annihilation by Perturbation Theory in Quantum Chromodynamics.}}}\
  }\href@noop {} {\bibfield  {journal} {\bibinfo  {journal} {Sov. Phys. JETP}\
  }\textbf {\bibinfo {volume} {46}},\ \bibinfo {pages} {641--653} (\bibinfo
  {year} {1977})},\ \bibinfo {note} {[Zh. Eksp. Teor.
  Fiz.73,1216(1977)]}\BibitemShut {NoStop}%
%%CITATION = SPHJA,46,641;%%
\bibitem [{\citenamefont {Gribov}\ and\ \citenamefont
  {Lipatov}(1972)}]{Gribov:1972ri}%
  \BibitemOpen
  \bibfield  {author} {\bibinfo {author} {\bibfnamefont {V.~N.}\ \bibnamefont
  {Gribov}}\ and\ \bibinfo {author} {\bibfnamefont {L.~N.}\ \bibnamefont
  {Lipatov}},\ }\bibfield  {title} {\enquote {\bibinfo {title} {{Deep inelastic
  e p scattering in perturbation theory}},}\ }\href@noop {} {\bibfield
  {journal} {\bibinfo  {journal} {Sov. J. Nucl. Phys.}\ }\textbf {\bibinfo
  {volume} {15}},\ \bibinfo {pages} {438--450} (\bibinfo {year} {1972})},\
  \bibinfo {note} {[Yad. Fiz.15,781(1972)]}\BibitemShut {NoStop}%
%%CITATION = SJNCA,15,438;%%
\bibitem [{\citenamefont {Altarelli}\ and\ \citenamefont
  {Parisi}(1977)}]{Altarelli:1977zs}%
  \BibitemOpen
  \bibfield  {author} {\bibinfo {author} {\bibfnamefont {Guido}\ \bibnamefont
  {Altarelli}}\ and\ \bibinfo {author} {\bibfnamefont {G.}~\bibnamefont
  {Parisi}},\ }\bibfield  {title} {\enquote {\bibinfo {title} {{Asymptotic
  Freedom in Parton Language}},}\ }\href {\doibase
  10.1016/0550-3213(77)90384-4} {\bibfield  {journal} {\bibinfo  {journal}
  {Nucl. Phys.}\ }\textbf {\bibinfo {volume} {B126}},\ \bibinfo {pages}
  {298--318} (\bibinfo {year} {1977})}\BibitemShut {NoStop}%
%%CITATION = NUPHA,B126,298;%%
\bibitem [{\citenamefont {Abe}\ \emph {et~al.}(1993)\citenamefont {Abe} \emph
  {et~al.}}]{Abt:1993cb}%
  \BibitemOpen
  \bibfield  {author} {\bibinfo {author} {\bibfnamefont {I.}~\bibnamefont
  {Abe}} \emph {et~al.} (\bibinfo {collaboration} {H1}),\ }\bibfield  {title}
  {\enquote {\bibinfo {title} {{Measurement of the proton structure function F2
  $(x, Q^{2})$ in the low $x$ region at HERA}},}\ }\href {\doibase
  10.1016/0550-3213(93)90090-C} {\bibfield  {journal} {\bibinfo  {journal}
  {Nucl. Phys.}\ }\textbf {\bibinfo {volume} {B407}},\ \bibinfo {pages}
  {515--538} (\bibinfo {year} {1993})}\BibitemShut {NoStop}%
%%CITATION = NUPHA,B407,515;%%
\bibitem [{\citenamefont {Ahmed}\ \emph {et~al.}(1995)\citenamefont {Ahmed}
  \emph {et~al.}}]{Ahmed:1995fd}%
  \BibitemOpen
  \bibfield  {author} {\bibinfo {author} {\bibfnamefont {T.}~\bibnamefont
  {Ahmed}} \emph {et~al.} (\bibinfo {collaboration} {H1}),\ }\bibfield  {title}
  {\enquote {\bibinfo {title} {{A Measurement of the proton structure function
  $f2 (x, Q^{2})$}},}\ }\href {\doibase 10.1016/0550-3213(95)98236-U}
  {\bibfield  {journal} {\bibinfo  {journal} {Nucl. Phys.}\ }\textbf {\bibinfo
  {volume} {B439}},\ \bibinfo {pages} {471--502} (\bibinfo {year} {1995})},\
  \Eprint {http://arxiv.org/abs/hep-ex/9503001} {arXiv:hep-ex/9503001 [hep-ex]}
  \BibitemShut {NoStop}%
%%CITATION = HEP-EX/9503001;%%
\bibitem [{\citenamefont {Derrick}\ \emph {et~al.}(1993)\citenamefont {Derrick}
  \emph {et~al.}}]{Derrick:1993fta}%
  \BibitemOpen
  \bibfield  {author} {\bibinfo {author} {\bibfnamefont {M.}~\bibnamefont
  {Derrick}} \emph {et~al.} (\bibinfo {collaboration} {ZEUS}),\ }\bibfield
  {title} {\enquote {\bibinfo {title} {{Measurement of the proton structure
  function F2 in ep scattering at HERA}},}\ }\href {\doibase
  10.1016/0370-2693(93)90347-K} {\bibfield  {journal} {\bibinfo  {journal}
  {Phys. Lett.}\ }\textbf {\bibinfo {volume} {B316}},\ \bibinfo {pages}
  {412--426} (\bibinfo {year} {1993})}\BibitemShut {NoStop}%
%%CITATION = PHLTA,B316,412;%%
\bibitem [{\citenamefont {Derrick}\ \emph {et~al.}(1995)\citenamefont {Derrick}
  \emph {et~al.}}]{Derrick:1994sz}%
  \BibitemOpen
  \bibfield  {author} {\bibinfo {author} {\bibfnamefont {M.}~\bibnamefont
  {Derrick}} \emph {et~al.} (\bibinfo {collaboration} {ZEUS}),\ }\bibfield
  {title} {\enquote {\bibinfo {title} {{Measurement of the proton structure
  function F2 from the 1993 HERA data}},}\ }\href {\doibase 10.1007/BF01556128}
  {\bibfield  {journal} {\bibinfo  {journal} {Z. Phys.}\ }\textbf {\bibinfo
  {volume} {C65}},\ \bibinfo {pages} {379--398} (\bibinfo {year}
  {1995})}\BibitemShut {NoStop}%
%%CITATION = ZEPYA,C65,379;%%
\bibitem [{\citenamefont {Martin}\ \emph {et~al.}(1994)\citenamefont {Martin},
  \citenamefont {Stirling},\ and\ \citenamefont {Roberts}}]{Martin:1994kn}%
  \BibitemOpen
  \bibfield  {author} {\bibinfo {author} {\bibfnamefont {Alan~D.}\ \bibnamefont
  {Martin}}, \bibinfo {author} {\bibfnamefont {W.~James}\ \bibnamefont
  {Stirling}}, \ and\ \bibinfo {author} {\bibfnamefont {R.~G.}\ \bibnamefont
  {Roberts}},\ }\bibfield  {title} {\enquote {\bibinfo {title} {{Parton
  distributions of the proton}},}\ }\href {\doibase 10.1103/PhysRevD.50.6734}
  {\bibfield  {journal} {\bibinfo  {journal} {Phys. Rev.}\ }\textbf {\bibinfo
  {volume} {D50}},\ \bibinfo {pages} {6734--6752} (\bibinfo {year} {1994})},\
  \Eprint {http://arxiv.org/abs/hep-ph/9406315} {arXiv:hep-ph/9406315 [hep-ph]}
  \BibitemShut {NoStop}%
%%CITATION = HEP-PH/9406315;%%
\bibitem [{\citenamefont {Lai}\ \emph {et~al.}(1995)\citenamefont {Lai},
  \citenamefont {Botts}, \citenamefont {Huston}, \citenamefont {Morfin},
  \citenamefont {Owens}, \citenamefont {Qiu}, \citenamefont {Tung},\ and\
  \citenamefont {Weerts}}]{Lai:1994bb}%
  \BibitemOpen
  \bibfield  {author} {\bibinfo {author} {\bibfnamefont {H.~L.}\ \bibnamefont
  {Lai}}, \bibinfo {author} {\bibfnamefont {J.}~\bibnamefont {Botts}}, \bibinfo
  {author} {\bibfnamefont {J.}~\bibnamefont {Huston}}, \bibinfo {author}
  {\bibfnamefont {J.~G.}\ \bibnamefont {Morfin}}, \bibinfo {author}
  {\bibfnamefont {J.~F.}\ \bibnamefont {Owens}}, \bibinfo {author}
  {\bibfnamefont {Jian-wei}\ \bibnamefont {Qiu}}, \bibinfo {author}
  {\bibfnamefont {W.~K.}\ \bibnamefont {Tung}}, \ and\ \bibinfo {author}
  {\bibfnamefont {H.}~\bibnamefont {Weerts}},\ }\bibfield  {title} {\enquote
  {\bibinfo {title} {{Global QCD analysis and the CTEQ parton
  distributions}},}\ }\href {\doibase 10.1103/PhysRevD.51.4763} {\bibfield
  {journal} {\bibinfo  {journal} {Phys. Rev.}\ }\textbf {\bibinfo {volume}
  {D51}},\ \bibinfo {pages} {4763--4782} (\bibinfo {year} {1995})},\ \Eprint
  {http://arxiv.org/abs/hep-ph/9410404} {arXiv:hep-ph/9410404 [hep-ph]}
  \BibitemShut {NoStop}%
%%CITATION = HEP-PH/9410404;%%
\bibitem [{\citenamefont {Gelis}\ and\ \citenamefont
  {Venugopalan}(2006{\natexlab{a}})}]{Gelis:2006yv}%
  \BibitemOpen
  \bibfield  {author} {\bibinfo {author} {\bibfnamefont {Francois}\
  \bibnamefont {Gelis}}\ and\ \bibinfo {author} {\bibfnamefont {Raju}\
  \bibnamefont {Venugopalan}},\ }\bibfield  {title} {\enquote {\bibinfo {title}
  {{Particle production in field theories coupled to strong external
  sources}},}\ }\href {\doibase 10.1016/j.nuclphysa.2006.07.020} {\bibfield
  {journal} {\bibinfo  {journal} {Nucl. Phys.}\ }\textbf {\bibinfo {volume}
  {A776}},\ \bibinfo {pages} {135--171} (\bibinfo {year}
  {2006}{\natexlab{a}})},\ \Eprint {http://arxiv.org/abs/hep-ph/0601209}
  {arXiv:hep-ph/0601209 [hep-ph]} \BibitemShut {NoStop}%
%%CITATION = HEP-PH/0601209;%%
\bibitem [{\citenamefont {Gelis}\ and\ \citenamefont
  {Venugopalan}(2006{\natexlab{b}})}]{Gelis:2006cr}%
  \BibitemOpen
  \bibfield  {author} {\bibinfo {author} {\bibfnamefont {Francois}\
  \bibnamefont {Gelis}}\ and\ \bibinfo {author} {\bibfnamefont {Raju}\
  \bibnamefont {Venugopalan}},\ }\bibfield  {title} {\enquote {\bibinfo {title}
  {{Particle production in field theories coupled to strong external sources.
  II. Generating functions}},}\ }\href {\doibase
  10.1016/j.nuclphysa.2006.08.015} {\bibfield  {journal} {\bibinfo  {journal}
  {Nucl. Phys.}\ }\textbf {\bibinfo {volume} {A779}},\ \bibinfo {pages}
  {177--196} (\bibinfo {year} {2006}{\natexlab{b}})},\ \Eprint
  {http://arxiv.org/abs/hep-ph/0605246} {arXiv:hep-ph/0605246 [hep-ph]}
  \BibitemShut {NoStop}%
%%CITATION = HEP-PH/0605246;%%
\bibitem [{\citenamefont {Gelis}\ \emph {et~al.}(2007)\citenamefont {Gelis},
  \citenamefont {Lappi},\ and\ \citenamefont {Venugopalan}}]{Gelis:2007kn}%
  \BibitemOpen
  \bibfield  {author} {\bibinfo {author} {\bibfnamefont {Francois}\
  \bibnamefont {Gelis}}, \bibinfo {author} {\bibfnamefont {Tuomas}\
  \bibnamefont {Lappi}}, \ and\ \bibinfo {author} {\bibfnamefont {Raju}\
  \bibnamefont {Venugopalan}},\ }\bibfield  {title} {\enquote {\bibinfo {title}
  {{High energy scattering in Quantum Chromodynamics}},}\ }\bibfield
  {booktitle} {\emph {\bibinfo {booktitle} {{Hadron physics. Proceedings, 10th
  International Workshop, Florianopolis, Brazil, April 26-31, 2007}}},\ }\href
  {\doibase 10.1142/S0218301307008331} {\bibfield  {journal} {\bibinfo
  {journal} {Int. J. Mod. Phys.}\ }\textbf {\bibinfo {volume} {E16}},\ \bibinfo
  {pages} {2595--2637} (\bibinfo {year} {2007})},\ \Eprint
  {http://arxiv.org/abs/0708.0047} {arXiv:0708.0047 [hep-ph]} \BibitemShut
  {NoStop}%
%%CITATION = ARXIV:0708.0047;%%
\bibitem [{\citenamefont {Gelis}\ \emph
  {et~al.}(2008{\natexlab{a}})\citenamefont {Gelis}, \citenamefont {Lappi},\
  and\ \citenamefont {Venugopalan}}]{Gelis:2008rw}%
  \BibitemOpen
  \bibfield  {author} {\bibinfo {author} {\bibfnamefont {Francois}\
  \bibnamefont {Gelis}}, \bibinfo {author} {\bibfnamefont {Tuomas}\
  \bibnamefont {Lappi}}, \ and\ \bibinfo {author} {\bibfnamefont {Raju}\
  \bibnamefont {Venugopalan}},\ }\bibfield  {title} {\enquote {\bibinfo {title}
  {{High energy factorization in nucleus-nucleus collisions}},}\ }\href
  {\doibase 10.1103/PhysRevD.78.054019} {\bibfield  {journal} {\bibinfo
  {journal} {Phys. Rev.}\ }\textbf {\bibinfo {volume} {D78}},\ \bibinfo {pages}
  {054019} (\bibinfo {year} {2008}{\natexlab{a}})},\ \Eprint
  {http://arxiv.org/abs/0804.2630} {arXiv:0804.2630 [hep-ph]} \BibitemShut
  {NoStop}%
%%CITATION = ARXIV:0804.2630;%%
\bibitem [{\citenamefont {Gelis}\ \emph
  {et~al.}(2008{\natexlab{b}})\citenamefont {Gelis}, \citenamefont {Lappi},\
  and\ \citenamefont {Venugopalan}}]{Gelis:2008ad}%
  \BibitemOpen
  \bibfield  {author} {\bibinfo {author} {\bibfnamefont {Francois}\
  \bibnamefont {Gelis}}, \bibinfo {author} {\bibfnamefont {Tuomas}\
  \bibnamefont {Lappi}}, \ and\ \bibinfo {author} {\bibfnamefont {Raju}\
  \bibnamefont {Venugopalan}},\ }\bibfield  {title} {\enquote {\bibinfo {title}
  {{High energy factorization in nucleus-nucleus collisions. II. Multigluon
  correlations}},}\ }\href {\doibase 10.1103/PhysRevD.78.054020} {\bibfield
  {journal} {\bibinfo  {journal} {Phys. Rev.}\ }\textbf {\bibinfo {volume}
  {D78}},\ \bibinfo {pages} {054020} (\bibinfo {year} {2008}{\natexlab{b}})},\
  \Eprint {http://arxiv.org/abs/0807.1306} {arXiv:0807.1306 [hep-ph]}
  \BibitemShut {NoStop}%
%%CITATION = ARXIV:0807.1306;%%
\bibitem [{\citenamefont {Balitsky}\ and\ \citenamefont
  {Chirilli}(2013)}]{Balitsky:2013fea}%
  \BibitemOpen
  \bibfield  {author} {\bibinfo {author} {\bibfnamefont {Ian}\ \bibnamefont
  {Balitsky}}\ and\ \bibinfo {author} {\bibfnamefont {Giovanni~A.}\
  \bibnamefont {Chirilli}},\ }\bibfield  {title} {\enquote {\bibinfo {title}
  {{Rapidity evolution of Wilson lines at the next-to-leading order}},}\ }\href
  {\doibase 10.1103/PhysRevD.88.111501} {\bibfield  {journal} {\bibinfo
  {journal} {Phys. Rev.}\ }\textbf {\bibinfo {volume} {D88}},\ \bibinfo {pages}
  {111501} (\bibinfo {year} {2013})},\ \Eprint {http://arxiv.org/abs/1309.7644}
  {arXiv:1309.7644 [hep-ph]} \BibitemShut {NoStop}%
%%CITATION = ARXIV:1309.7644;%%
\bibitem [{\citenamefont {Grabovsky}(2013)}]{Grabovsky:2013mba}%
  \BibitemOpen
  \bibfield  {author} {\bibinfo {author} {\bibfnamefont {A.~V.}\ \bibnamefont
  {Grabovsky}},\ }\bibfield  {title} {\enquote {\bibinfo {title} {{Connected
  contribution to the kernel of the evolution equation for 3-quark Wilson loop
  operator}},}\ }\href {\doibase 10.1007/JHEP09(2013)141} {\bibfield  {journal}
  {\bibinfo  {journal} {JHEP}\ }\textbf {\bibinfo {volume} {09}},\ \bibinfo
  {pages} {141} (\bibinfo {year} {2013})},\ \Eprint
  {http://arxiv.org/abs/1307.5414} {arXiv:1307.5414 [hep-ph]} \BibitemShut
  {NoStop}%
%%CITATION = ARXIV:1307.5414;%%
\bibitem [{\citenamefont {Kovner}\ \emph {et~al.}(2014)\citenamefont {Kovner},
  \citenamefont {Lublinsky},\ and\ \citenamefont {Mulian}}]{Kovner:2013ona}%
  \BibitemOpen
  \bibfield  {author} {\bibinfo {author} {\bibfnamefont {Alex}\ \bibnamefont
  {Kovner}}, \bibinfo {author} {\bibfnamefont {Michael}\ \bibnamefont
  {Lublinsky}}, \ and\ \bibinfo {author} {\bibfnamefont {Yair}\ \bibnamefont
  {Mulian}},\ }\bibfield  {title} {\enquote {\bibinfo {title}
  {{Jalilian-Marian, Iancu, McLerran, Weigert, Leonidov, Kovner evolution at
  next to leading order}},}\ }\href {\doibase 10.1103/PhysRevD.89.061704}
  {\bibfield  {journal} {\bibinfo  {journal} {Phys. Rev.}\ }\textbf {\bibinfo
  {volume} {D89}},\ \bibinfo {pages} {061704} (\bibinfo {year} {2014})},\
  \Eprint {http://arxiv.org/abs/1310.0378} {arXiv:1310.0378 [hep-ph]}
  \BibitemShut {NoStop}%
%%CITATION = ARXIV:1310.0378;%%
\bibitem [{\citenamefont {Kovchegov}\ and\ \citenamefont
  {Weigert}(2007)}]{Kovchegov:2006vj}%
  \BibitemOpen
  \bibfield  {author} {\bibinfo {author} {\bibfnamefont {Yuri~V.}\ \bibnamefont
  {Kovchegov}}\ and\ \bibinfo {author} {\bibfnamefont {Heribert}\ \bibnamefont
  {Weigert}},\ }\bibfield  {title} {\enquote {\bibinfo {title} {{Triumvirate of
  Running Couplings in Small-x Evolution}},}\ }\href {\doibase
  10.1016/j.nuclphysa.2006.10.075} {\bibfield  {journal} {\bibinfo  {journal}
  {Nucl. Phys.}\ }\textbf {\bibinfo {volume} {A784}},\ \bibinfo {pages}
  {188--226} (\bibinfo {year} {2007})},\ \Eprint
  {http://arxiv.org/abs/hep-ph/0609090} {arXiv:hep-ph/0609090 [hep-ph]}
  \BibitemShut {NoStop}%
%%CITATION = HEP-PH/0609090;%%
\bibitem [{\citenamefont {Balitsky}\ and\ \citenamefont
  {Chirilli}(2008)}]{Balitsky:2008zza}%
  \BibitemOpen
  \bibfield  {author} {\bibinfo {author} {\bibfnamefont {Ian}\ \bibnamefont
  {Balitsky}}\ and\ \bibinfo {author} {\bibfnamefont {Giovanni~A.}\
  \bibnamefont {Chirilli}},\ }\bibfield  {title} {\enquote {\bibinfo {title}
  {{Next-to-leading order evolution of color dipoles}},}\ }\href {\doibase
  10.1103/PhysRevD.77.014019} {\bibfield  {journal} {\bibinfo  {journal} {Phys.
  Rev.}\ }\textbf {\bibinfo {volume} {D77}},\ \bibinfo {pages} {014019}
  (\bibinfo {year} {2008})},\ \Eprint {http://arxiv.org/abs/0710.4330}
  {arXiv:0710.4330 [hep-ph]} \BibitemShut {NoStop}%
%%CITATION = ARXIV:0710.4330;%%
\bibitem [{\citenamefont {Balitsky}(2007)}]{Balitsky:2006wa}%
  \BibitemOpen
  \bibfield  {author} {\bibinfo {author} {\bibfnamefont {Ian}\ \bibnamefont
  {Balitsky}},\ }\bibfield  {title} {\enquote {\bibinfo {title} {{Quark
  contribution to the small-x evolution of color dipole}},}\ }\href {\doibase
  10.1103/PhysRevD.75.014001} {\bibfield  {journal} {\bibinfo  {journal} {Phys.
  Rev.}\ }\textbf {\bibinfo {volume} {D75}},\ \bibinfo {pages} {014001}
  (\bibinfo {year} {2007})},\ \Eprint {http://arxiv.org/abs/hep-ph/0609105}
  {arXiv:hep-ph/0609105 [hep-ph]} \BibitemShut {NoStop}%
%%CITATION = HEP-PH/0609105;%%
\bibitem [{\citenamefont {Low}(1958)}]{Low:1958sn}%
  \BibitemOpen
  \bibfield  {author} {\bibinfo {author} {\bibfnamefont {F.~E.}\ \bibnamefont
  {Low}},\ }\bibfield  {title} {\enquote {\bibinfo {title} {{Bremsstrahlung of
  very low-energy quanta in elementary particle collisions}},}\ }\href
  {\doibase 10.1103/PhysRev.110.974} {\bibfield  {journal} {\bibinfo  {journal}
  {Phys. Rev.}\ }\textbf {\bibinfo {volume} {110}},\ \bibinfo {pages}
  {974--977} (\bibinfo {year} {1958})}\BibitemShut {NoStop}%
%%CITATION = PHRVA,110,974;%%
\bibitem [{\citenamefont {Burnett}\ and\ \citenamefont
  {Kroll}(1968)}]{Burnett:1967km}%
  \BibitemOpen
  \bibfield  {author} {\bibinfo {author} {\bibfnamefont {T.~H.}\ \bibnamefont
  {Burnett}}\ and\ \bibinfo {author} {\bibfnamefont {Norman~M.}\ \bibnamefont
  {Kroll}},\ }\bibfield  {title} {\enquote {\bibinfo {title} {{Extension of the
  low soft photon theorem}},}\ }\href {\doibase 10.1103/PhysRevLett.20.86}
  {\bibfield  {journal} {\bibinfo  {journal} {Phys. Rev. Lett.}\ }\textbf
  {\bibinfo {volume} {20}},\ \bibinfo {pages} {86} (\bibinfo {year}
  {1968})}\BibitemShut {NoStop}%
%%CITATION = PRLTA,20,86;%%
\bibitem [{\citenamefont {Bell}\ and\ \citenamefont
  {Van~Royen}(1969)}]{Bell:1969yw}%
  \BibitemOpen
  \bibfield  {author} {\bibinfo {author} {\bibfnamefont {J.~S.}\ \bibnamefont
  {Bell}}\ and\ \bibinfo {author} {\bibfnamefont {R.}~\bibnamefont
  {Van~Royen}},\ }\bibfield  {title} {\enquote {\bibinfo {title} {{On the
  low-burnett-kroll theorem for soft-photon emission}},}\ }\href {\doibase
  10.1007/BF02823297} {\bibfield  {journal} {\bibinfo  {journal} {Nuovo Cim.}\
  }\textbf {\bibinfo {volume} {A60}},\ \bibinfo {pages} {62--68} (\bibinfo
  {year} {1969})}\BibitemShut {NoStop}%
%%CITATION = NUCIA,A60,62;%%
\bibitem [{\citenamefont {McLerran}\ and\ \citenamefont
  {Venugopalan}(1999)}]{McLerran:1998nk}%
  \BibitemOpen
  \bibfield  {author} {\bibinfo {author} {\bibfnamefont {Larry~D.}\
  \bibnamefont {McLerran}}\ and\ \bibinfo {author} {\bibfnamefont {Raju}\
  \bibnamefont {Venugopalan}},\ }\bibfield  {title} {\enquote {\bibinfo {title}
  {{Fock space distributions, structure functions, higher twists and small
  $x$}},}\ }\href {\doibase 10.1103/PhysRevD.59.094002} {\bibfield  {journal}
  {\bibinfo  {journal} {Phys. Rev.}\ }\textbf {\bibinfo {volume} {D59}},\
  \bibinfo {pages} {094002} (\bibinfo {year} {1999})},\ \Eprint
  {http://arxiv.org/abs/hep-ph/9809427} {arXiv:hep-ph/9809427 [hep-ph]}
  \BibitemShut {NoStop}%
%%CITATION = HEP-PH/9809427;%%
\bibitem [{\citenamefont {Peskin}\ and\ \citenamefont
  {Schroeder}(1995)}]{Peskin:1995ev}%
  \BibitemOpen
  \bibfield  {author} {\bibinfo {author} {\bibfnamefont {Michael~E.}\
  \bibnamefont {Peskin}}\ and\ \bibinfo {author} {\bibfnamefont {Daniel~V.}\
  \bibnamefont {Schroeder}},\ }\href
  {http://www.slac.stanford.edu/~mpeskin/QFT.html} {\emph {\bibinfo {title}
  {{An Introduction to quantum field theory}}}}\ (\bibinfo  {publisher}
  {Addison-Wesley},\ \bibinfo {address} {Reading, USA},\ \bibinfo {year}
  {1995})\BibitemShut {NoStop}%
%%CITATION = INSPIRE-407703;%%
\bibitem [{\citenamefont {Blaizot}\ \emph
  {et~al.}(2004{\natexlab{a}})\citenamefont {Blaizot}, \citenamefont {Gelis},\
  and\ \citenamefont {Venugopalan}}]{Blaizot:2004wu}%
  \BibitemOpen
  \bibfield  {author} {\bibinfo {author} {\bibfnamefont {Jean~Paul}\
  \bibnamefont {Blaizot}}, \bibinfo {author} {\bibfnamefont {Francois}\
  \bibnamefont {Gelis}}, \ and\ \bibinfo {author} {\bibfnamefont {Raju}\
  \bibnamefont {Venugopalan}},\ }\bibfield  {title} {\enquote {\bibinfo {title}
  {{High-energy pA collisions in the color glass condensate approach. 1. Gluon
  production and the Cronin effect}},}\ }\href {\doibase
  10.1016/j.nuclphysa.2004.07.005} {\bibfield  {journal} {\bibinfo  {journal}
  {Nucl. Phys.}\ }\textbf {\bibinfo {volume} {A743}},\ \bibinfo {pages}
  {13--56} (\bibinfo {year} {2004}{\natexlab{a}})},\ \Eprint
  {http://arxiv.org/abs/hep-ph/0402256} {arXiv:hep-ph/0402256 [hep-ph]}
  \BibitemShut {NoStop}%
%%CITATION = HEP-PH/0402256;%%
\bibitem [{\citenamefont {Jalilian-Marian}\ and\ \citenamefont
  {Kovchegov}(2006)}]{JalilianMarian:2005jf}%
  \BibitemOpen
  \bibfield  {author} {\bibinfo {author} {\bibfnamefont {Jamal}\ \bibnamefont
  {Jalilian-Marian}}\ and\ \bibinfo {author} {\bibfnamefont {Yuri~V.}\
  \bibnamefont {Kovchegov}},\ }\bibfield  {title} {\enquote {\bibinfo {title}
  {{Saturation physics and deuteron-Gold collisions at RHIC}},}\ }\href
  {\doibase 10.1016/j.ppnp.2005.07.002} {\bibfield  {journal} {\bibinfo
  {journal} {Prog. Part. Nucl. Phys.}\ }\textbf {\bibinfo {volume} {56}},\
  \bibinfo {pages} {104--231} (\bibinfo {year} {2006})},\ \Eprint
  {http://arxiv.org/abs/hep-ph/0505052} {arXiv:hep-ph/0505052 [hep-ph]}
  \BibitemShut {NoStop}%
%%CITATION = HEP-PH/0505052;%%
\bibitem [{\citenamefont {Fujii}\ \emph {et~al.}(2006)\citenamefont {Fujii},
  \citenamefont {Gelis},\ and\ \citenamefont {Venugopalan}}]{Fujii:2006ab}%
  \BibitemOpen
  \bibfield  {author} {\bibinfo {author} {\bibfnamefont {Hirotsugu}\
  \bibnamefont {Fujii}}, \bibinfo {author} {\bibfnamefont {Francois}\
  \bibnamefont {Gelis}}, \ and\ \bibinfo {author} {\bibfnamefont {Raju}\
  \bibnamefont {Venugopalan}},\ }\bibfield  {title} {\enquote {\bibinfo {title}
  {{Quark pair production in high energy pA collisions: General features}},}\
  }\href {\doibase 10.1016/j.nuclphysa.2006.09.012} {\bibfield  {journal}
  {\bibinfo  {journal} {Nucl. Phys.}\ }\textbf {\bibinfo {volume} {A780}},\
  \bibinfo {pages} {146--174} (\bibinfo {year} {2006})},\ \Eprint
  {http://arxiv.org/abs/hep-ph/0603099} {arXiv:hep-ph/0603099 [hep-ph]}
  \BibitemShut {NoStop}%
%%CITATION = HEP-PH/0603099;%%
\bibitem [{\citenamefont {Kuraev}\ \emph {et~al.}(1977)\citenamefont {Kuraev},
  \citenamefont {Lipatov},\ and\ \citenamefont {Fadin}}]{Kuraev:1977fs}%
  \BibitemOpen
  \bibfield  {author} {\bibinfo {author} {\bibfnamefont {E.~A.}\ \bibnamefont
  {Kuraev}}, \bibinfo {author} {\bibfnamefont {L.~N.}\ \bibnamefont {Lipatov}},
  \ and\ \bibinfo {author} {\bibfnamefont {Victor~S.}\ \bibnamefont {Fadin}},\
  }\bibfield  {title} {\enquote {\bibinfo {title} {{The Pomeranchuk Singularity
  in Nonabelian Gauge Theories}},}\ }\href@noop {} {\bibfield  {journal}
  {\bibinfo  {journal} {Sov. Phys. JETP}\ }\textbf {\bibinfo {volume} {45}},\
  \bibinfo {pages} {199--204} (\bibinfo {year} {1977})},\ \bibinfo {note} {[Zh.
  Eksp. Teor. Fiz.72,377(1977)]}\BibitemShut {NoStop}%
%%CITATION = SPHJA,45,199;%%
\bibitem [{\citenamefont {Balitsky}\ and\ \citenamefont
  {Lipatov}(1978)}]{Balitsky:1978ic}%
  \BibitemOpen
  \bibfield  {author} {\bibinfo {author} {\bibfnamefont {I.~I.}\ \bibnamefont
  {Balitsky}}\ and\ \bibinfo {author} {\bibfnamefont {L.~N.}\ \bibnamefont
  {Lipatov}},\ }\bibfield  {title} {\enquote {\bibinfo {title} {{The
  Pomeranchuk Singularity in Quantum Chromodynamics}},}\ }\href@noop {}
  {\bibfield  {journal} {\bibinfo  {journal} {Sov. J. Nucl. Phys.}\ }\textbf
  {\bibinfo {volume} {28}},\ \bibinfo {pages} {822--829} (\bibinfo {year}
  {1978})},\ \bibinfo {note} {[Yad. Fiz.28,1597(1978)]}\BibitemShut {NoStop}%
%%CITATION = SJNCA,28,822;%%
\bibitem [{\citenamefont {Gelis}\ and\ \citenamefont
  {Venugopalan}(2004)}]{Gelis:2003vh}%
  \BibitemOpen
  \bibfield  {author} {\bibinfo {author} {\bibfnamefont {Francois}\
  \bibnamefont {Gelis}}\ and\ \bibinfo {author} {\bibfnamefont {Raju}\
  \bibnamefont {Venugopalan}},\ }\bibfield  {title} {\enquote {\bibinfo {title}
  {{Large mass q$\bar{q}$ production from the color glass condensate}},}\
  }\href {\doibase 10.1103/PhysRevD.69.014019} {\bibfield  {journal} {\bibinfo
  {journal} {Phys. Rev.}\ }\textbf {\bibinfo {volume} {D69}},\ \bibinfo {pages}
  {014019} (\bibinfo {year} {2004})},\ \Eprint
  {http://arxiv.org/abs/hep-ph/0310090} {arXiv:hep-ph/0310090 [hep-ph]}
  \BibitemShut {NoStop}%
%%CITATION = HEP-PH/0310090;%%
\bibitem [{\citenamefont {Aurenche}\ \emph {et~al.}(1984)\citenamefont
  {Aurenche}, \citenamefont {Douiri}, \citenamefont {Baier}, \citenamefont
  {Fontannaz},\ and\ \citenamefont {Schiff}}]{Aurenche:1983hc}%
  \BibitemOpen
  \bibfield  {author} {\bibinfo {author} {\bibfnamefont {P.}~\bibnamefont
  {Aurenche}}, \bibinfo {author} {\bibfnamefont {A.}~\bibnamefont {Douiri}},
  \bibinfo {author} {\bibfnamefont {R.}~\bibnamefont {Baier}}, \bibinfo
  {author} {\bibfnamefont {M.}~\bibnamefont {Fontannaz}}, \ and\ \bibinfo
  {author} {\bibfnamefont {D.}~\bibnamefont {Schiff}},\ }\bibfield  {title}
  {\enquote {\bibinfo {title} {{The Deep Compton Scattering: Single Photon
  Spectrum and Photon - Hadron Correlations Beyond Leading Logarithms}},}\
  }\href {\doibase 10.1007/BF01410370} {\bibfield  {journal} {\bibinfo
  {journal} {Z. Phys.}\ }\textbf {\bibinfo {volume} {C24}},\ \bibinfo {pages}
  {309} (\bibinfo {year} {1984})}\BibitemShut {NoStop}%
%%CITATION = ZEPYA,C24,309;%%
\bibitem [{\citenamefont {Gehrmann-De~Ridder}\ \emph
  {et~al.}(2000)\citenamefont {Gehrmann-De~Ridder}, \citenamefont {Kramer},\
  and\ \citenamefont {Spiesberger}}]{GehrmannDeRidder:2000ce}%
  \BibitemOpen
  \bibfield  {author} {\bibinfo {author} {\bibfnamefont {A.}~\bibnamefont
  {Gehrmann-De~Ridder}}, \bibinfo {author} {\bibfnamefont {G.}~\bibnamefont
  {Kramer}}, \ and\ \bibinfo {author} {\bibfnamefont {H.}~\bibnamefont
  {Spiesberger}},\ }\bibfield  {title} {\enquote {\bibinfo {title} {{Photon
  plus jet-cross sections in deep inelastic e p collisions at order O(alpha**2
  alpha(s))}},}\ }\href {\doibase 10.1016/S0550-3213(00)00228-5} {\bibfield
  {journal} {\bibinfo  {journal} {Nucl. Phys.}\ }\textbf {\bibinfo {volume}
  {B578}},\ \bibinfo {pages} {326--350} (\bibinfo {year} {2000})},\ \Eprint
  {http://arxiv.org/abs/hep-ph/0003082} {arXiv:hep-ph/0003082 [hep-ph]}
  \BibitemShut {NoStop}%
%%CITATION = HEP-PH/0003082;%%
\bibitem [{\citenamefont {Kramer}\ \emph {et~al.}(1998)\citenamefont {Kramer},
  \citenamefont {Michelsen},\ and\ \citenamefont
  {Spiesberger}}]{Kramer:1997nb}%
  \BibitemOpen
  \bibfield  {author} {\bibinfo {author} {\bibfnamefont {G.}~\bibnamefont
  {Kramer}}, \bibinfo {author} {\bibfnamefont {D.}~\bibnamefont {Michelsen}}, \
  and\ \bibinfo {author} {\bibfnamefont {H.}~\bibnamefont {Spiesberger}},\
  }\bibfield  {title} {\enquote {\bibinfo {title} {{Production of hard photons
  and jets in deep inelastic lepton proton scattering at order O (alpha-s)}},}\
  }\href {\doibase 10.1007/s100520050272, 10.1007/s100529800836} {\bibfield
  {journal} {\bibinfo  {journal} {Eur. Phys. J.}\ }\textbf {\bibinfo {volume}
  {C5}},\ \bibinfo {pages} {293--301} (\bibinfo {year} {1998})},\ \Eprint
  {http://arxiv.org/abs/hep-ph/9712309} {arXiv:hep-ph/9712309 [hep-ph]}
  \BibitemShut {NoStop}%
%%CITATION = HEP-PH/9712309;%%
\bibitem [{\citenamefont {Michelsen}(1995)}]{Michelsen:1995ag}%
  \BibitemOpen
  \bibfield  {author} {\bibinfo {author} {\bibfnamefont {Dirk}\ \bibnamefont
  {Michelsen}},\ }\emph {\bibinfo {title} {{Radiative corrections of order
  alpha-s for production of hard photons by HERA}}},\ \href@noop {} {Ph.D.
  thesis},\ \bibinfo  {school} {Hamburg U.} (\bibinfo {year}
  {1995})\BibitemShut {NoStop}%
%%CITATION = DESY-95-146;%%
\bibitem [{\citenamefont {Gelis}\ and\ \citenamefont
  {Jalilian-Marian}(2003)}]{Gelis:2002nn}%
  \BibitemOpen
  \bibfield  {author} {\bibinfo {author} {\bibfnamefont {Francois}\
  \bibnamefont {Gelis}}\ and\ \bibinfo {author} {\bibfnamefont {Jamal}\
  \bibnamefont {Jalilian-Marian}},\ }\bibfield  {title} {\enquote {\bibinfo
  {title} {{From DIS to proton nucleus collisions in the color glass condensate
  model}},}\ }\href {\doibase 10.1103/PhysRevD.67.074019} {\bibfield  {journal}
  {\bibinfo  {journal} {Phys. Rev.}\ }\textbf {\bibinfo {volume} {D67}},\
  \bibinfo {pages} {074019} (\bibinfo {year} {2003})},\ \Eprint
  {http://arxiv.org/abs/hep-ph/0211363} {arXiv:hep-ph/0211363 [hep-ph]}
  \BibitemShut {NoStop}%
%%CITATION = HEP-PH/0211363;%%
\bibitem [{\citenamefont {Dominguez}\ \emph {et~al.}(2012)\citenamefont
  {Dominguez}, \citenamefont {Qiu}, \citenamefont {Xiao},\ and\ \citenamefont
  {Yuan}}]{Dominguez:2011br}%
  \BibitemOpen
  \bibfield  {author} {\bibinfo {author} {\bibfnamefont {Fabio}\ \bibnamefont
  {Dominguez}}, \bibinfo {author} {\bibfnamefont {Jian-Wei}\ \bibnamefont
  {Qiu}}, \bibinfo {author} {\bibfnamefont {Bo-Wen}\ \bibnamefont {Xiao}}, \
  and\ \bibinfo {author} {\bibfnamefont {Feng}\ \bibnamefont {Yuan}},\
  }\bibfield  {title} {\enquote {\bibinfo {title} {{On the linearly polarized
  gluon distributions in the color dipole model}},}\ }\href {\doibase
  10.1103/PhysRevD.85.045003} {\bibfield  {journal} {\bibinfo  {journal} {Phys.
  Rev.}\ }\textbf {\bibinfo {volume} {D85}},\ \bibinfo {pages} {045003}
  (\bibinfo {year} {2012})},\ \Eprint {http://arxiv.org/abs/1109.6293}
  {arXiv:1109.6293 [hep-ph]} \BibitemShut {NoStop}%
%%CITATION = ARXIV:1109.6293;%%
\bibitem [{\citenamefont {Kovchegov}\ and\ \citenamefont
  {Mueller}(1998)}]{Kovchegov:1998bi}%
  \BibitemOpen
  \bibfield  {author} {\bibinfo {author} {\bibfnamefont {Yuri~V.}\ \bibnamefont
  {Kovchegov}}\ and\ \bibinfo {author} {\bibfnamefont {Alfred~H.}\ \bibnamefont
  {Mueller}},\ }\bibfield  {title} {\enquote {\bibinfo {title} {{Gluon
  production in current nucleus and nucleon - nucleus collisions in a
  quasiclassical approximation}},}\ }\href {\doibase
  10.1016/S0550-3213(98)00384-8} {\bibfield  {journal} {\bibinfo  {journal}
  {Nucl. Phys.}\ }\textbf {\bibinfo {volume} {B529}},\ \bibinfo {pages}
  {451--479} (\bibinfo {year} {1998})},\ \Eprint
  {http://arxiv.org/abs/hep-ph/9802440} {arXiv:hep-ph/9802440 [hep-ph]}
  \BibitemShut {NoStop}%
%%CITATION = HEP-PH/9802440;%%
\bibitem [{\citenamefont {Diehl}(2003)}]{Diehl:2003ny}%
  \BibitemOpen
  \bibfield  {author} {\bibinfo {author} {\bibfnamefont {M.}~\bibnamefont
  {Diehl}},\ }\bibfield  {title} {\enquote {\bibinfo {title} {{Generalized
  parton distributions}},}\ }\href {\doibase 10.1016/j.physrep.2003.08.002,
  10.3204/DESY-THESIS-2003-018} {\bibfield  {journal} {\bibinfo  {journal}
  {Phys. Rept.}\ }\textbf {\bibinfo {volume} {388}},\ \bibinfo {pages}
  {41--277} (\bibinfo {year} {2003})},\ \Eprint
  {http://arxiv.org/abs/hep-ph/0307382} {arXiv:hep-ph/0307382 [hep-ph]}
  \BibitemShut {NoStop}%
%%CITATION = HEP-PH/0307382;%%
\bibitem [{\citenamefont {Kogut}\ and\ \citenamefont
  {Soper}(1970)}]{Kogut:1969xa}%
  \BibitemOpen
  \bibfield  {author} {\bibinfo {author} {\bibfnamefont {John~B.}\ \bibnamefont
  {Kogut}}\ and\ \bibinfo {author} {\bibfnamefont {Davison~E.}\ \bibnamefont
  {Soper}},\ }\bibfield  {title} {\enquote {\bibinfo {title} {{Quantum
  Electrodynamics in the Infinite Momentum Frame}},}\ }\href {\doibase
  10.1103/PhysRevD.1.2901} {\bibfield  {journal} {\bibinfo  {journal} {Phys.
  Rev.}\ }\textbf {\bibinfo {volume} {D1}},\ \bibinfo {pages} {2901--2913}
  (\bibinfo {year} {1970})}\BibitemShut {NoStop}%
%%CITATION = PHRVA,D1,2901;%%
\bibitem [{\citenamefont {Brodsky}\ and\ \citenamefont
  {Robertson}(1995)}]{Brodsky:1995rn}%
  \BibitemOpen
  \bibfield  {author} {\bibinfo {author} {\bibfnamefont {Stanley~J.}\
  \bibnamefont {Brodsky}}\ and\ \bibinfo {author} {\bibfnamefont {David~G.}\
  \bibnamefont {Robertson}},\ }\bibfield  {title} {\enquote {\bibinfo {title}
  {{Light cone quantization and QCD phenomenology}},}\ }in\ \href
  {http://www-public.slac.stanford.edu/sciDoc/docMeta.aspx?slacPubNumber=SLAC-PUB-7056}
  {\emph {\bibinfo {booktitle} {{Confinement physics. Proceedings, 1st ELFE
  Summer School, Cambridge, UK, July 22-28, 1995}}}}\ (\bibinfo {year} {1995})\
  pp.\ \bibinfo {pages} {71--110},\ \Eprint
  {http://arxiv.org/abs/hep-ph/9511374} {arXiv:hep-ph/9511374 [hep-ph]}
  \BibitemShut {NoStop}%
%%CITATION = HEP-PH/9511374;%%
\bibitem [{\citenamefont {Brodsky}\ \emph {et~al.}(1998)\citenamefont
  {Brodsky}, \citenamefont {Pauli},\ and\ \citenamefont
  {Pinsky}}]{Brodsky:1997de}%
  \BibitemOpen
  \bibfield  {author} {\bibinfo {author} {\bibfnamefont {Stanley~J.}\
  \bibnamefont {Brodsky}}, \bibinfo {author} {\bibfnamefont {Hans-Christian}\
  \bibnamefont {Pauli}}, \ and\ \bibinfo {author} {\bibfnamefont {Stephen~S.}\
  \bibnamefont {Pinsky}},\ }\bibfield  {title} {\enquote {\bibinfo {title}
  {{Quantum chromodynamics and other field theories on the light cone}},}\
  }\href {\doibase 10.1016/S0370-1573(97)00089-6} {\bibfield  {journal}
  {\bibinfo  {journal} {Phys. Rept.}\ }\textbf {\bibinfo {volume} {301}},\
  \bibinfo {pages} {299--486} (\bibinfo {year} {1998})},\ \Eprint
  {http://arxiv.org/abs/hep-ph/9705477} {arXiv:hep-ph/9705477 [hep-ph]}
  \BibitemShut {NoStop}%
%%CITATION = HEP-PH/9705477;%%
\bibitem [{\citenamefont {Venugopalan}(1998)}]{Venugopalan:1998zd}%
  \BibitemOpen
  \bibfield  {author} {\bibinfo {author} {\bibfnamefont {Raju}\ \bibnamefont
  {Venugopalan}},\ }\bibfield  {title} {\enquote {\bibinfo {title}
  {{Introduction to light cone field theory and high-energy scattering}},}\
  }\bibfield  {booktitle} {\emph {\bibinfo {booktitle} {{Hadrons in dense
  matter and hadrosynthesis. Proceedings, 11th Chris Engelbrecht Summer School,
  Cape Town, South Africa, February 4-13, 1998}}},\ }\href {\doibase
  10.1007/BFb0107312} {\  (\bibinfo {year} {1998}),\ 10.1007/BFb0107312},\
  \bibinfo {note} {[Lect. Notes Phys.516,89(1999)]},\ \Eprint
  {http://arxiv.org/abs/nucl-th/9808023} {arXiv:nucl-th/9808023 [nucl-th]}
  \BibitemShut {NoStop}%
%%CITATION = NUCL-TH/9808023;%%
\bibitem [{\citenamefont {Venugopalan}(1999)}]{Venugopalan:1999wu}%
  \BibitemOpen
  \bibfield  {author} {\bibinfo {author} {\bibfnamefont {Raju}\ \bibnamefont
  {Venugopalan}},\ }\bibfield  {title} {\enquote {\bibinfo {title} {{Classical
  methods in DIS and nuclear scattering at small $x$}},}\ }\bibfield
  {booktitle} {\emph {\bibinfo {booktitle} {{Strong interactions at low and
  high-energies. Proceedings, 39th Cracow School of Theoretical Physics,
  Zakopane, Poland, May 29-June 8, 1999}}},\ }\href@noop {} {\bibfield
  {journal} {\bibinfo  {journal} {Acta Phys. Polon.}\ }\textbf {\bibinfo
  {volume} {B30}},\ \bibinfo {pages} {3731--3761} (\bibinfo {year} {1999})},\
  \Eprint {http://arxiv.org/abs/hep-ph/9911371} {arXiv:hep-ph/9911371 [hep-ph]}
  \BibitemShut {NoStop}%
%%CITATION = HEP-PH/9911371;%%
\bibitem [{\citenamefont {Ayala}\ \emph {et~al.}(1995)\citenamefont {Ayala},
  \citenamefont {Jalilian-Marian}, \citenamefont {McLerran},\ and\
  \citenamefont {Venugopalan}}]{Ayala:1995kg}%
  \BibitemOpen
  \bibfield  {author} {\bibinfo {author} {\bibfnamefont {Alejandro}\
  \bibnamefont {Ayala}}, \bibinfo {author} {\bibfnamefont {Jamal}\ \bibnamefont
  {Jalilian-Marian}}, \bibinfo {author} {\bibfnamefont {Larry~D.}\ \bibnamefont
  {McLerran}}, \ and\ \bibinfo {author} {\bibfnamefont {Raju}\ \bibnamefont
  {Venugopalan}},\ }\bibfield  {title} {\enquote {\bibinfo {title} {{The Gluon
  propagator in nonAbelian Weizs\"{a}cker-Williams fields}},}\ }\href {\doibase
  10.1103/PhysRevD.52.2935} {\bibfield  {journal} {\bibinfo  {journal} {Phys.
  Rev.}\ }\textbf {\bibinfo {volume} {D52}},\ \bibinfo {pages} {2935--2943}
  (\bibinfo {year} {1995})},\ \Eprint {http://arxiv.org/abs/hep-ph/9501324}
  {arXiv:hep-ph/9501324 [hep-ph]} \BibitemShut {NoStop}%
%%CITATION = HEP-PH/9501324;%%
\bibitem [{\citenamefont {Ayala}\ \emph {et~al.}(1996)\citenamefont {Ayala},
  \citenamefont {Jalilian-Marian}, \citenamefont {McLerran},\ and\
  \citenamefont {Venugopalan}}]{Ayala:1995hx}%
  \BibitemOpen
  \bibfield  {author} {\bibinfo {author} {\bibfnamefont {Alejandro}\
  \bibnamefont {Ayala}}, \bibinfo {author} {\bibfnamefont {Jamal}\ \bibnamefont
  {Jalilian-Marian}}, \bibinfo {author} {\bibfnamefont {Larry~D.}\ \bibnamefont
  {McLerran}}, \ and\ \bibinfo {author} {\bibfnamefont {Raju}\ \bibnamefont
  {Venugopalan}},\ }\bibfield  {title} {\enquote {\bibinfo {title} {{Quantum
  corrections to the Weizsacker-Williams gluon distribution function at small
  x}},}\ }\href {\doibase 10.1103/PhysRevD.53.458} {\bibfield  {journal}
  {\bibinfo  {journal} {Phys. Rev.}\ }\textbf {\bibinfo {volume} {D53}},\
  \bibinfo {pages} {458--475} (\bibinfo {year} {1996})},\ \Eprint
  {http://arxiv.org/abs/hep-ph/9508302} {arXiv:hep-ph/9508302 [hep-ph]}
  \BibitemShut {NoStop}%
%%CITATION = HEP-PH/9508302;%%
\bibitem [{\citenamefont {Gelis}\ and\ \citenamefont
  {Mehtar-Tani}(2006)}]{Gelis:2005pt}%
  \BibitemOpen
  \bibfield  {author} {\bibinfo {author} {\bibfnamefont {Francois}\
  \bibnamefont {Gelis}}\ and\ \bibinfo {author} {\bibfnamefont {Yacine}\
  \bibnamefont {Mehtar-Tani}},\ }\bibfield  {title} {\enquote {\bibinfo {title}
  {{Gluon propagation inside a high-energy nucleus}},}\ }\href {\doibase
  10.1103/PhysRevD.73.034019} {\bibfield  {journal} {\bibinfo  {journal} {Phys.
  Rev.}\ }\textbf {\bibinfo {volume} {D73}},\ \bibinfo {pages} {034019}
  (\bibinfo {year} {2006})},\ \Eprint {http://arxiv.org/abs/hep-ph/0512079}
  {arXiv:hep-ph/0512079 [hep-ph]} \BibitemShut {NoStop}%
%%CITATION = HEP-PH/0512079;%%
\bibitem [{\citenamefont {Blaizot}\ \emph
  {et~al.}(2004{\natexlab{b}})\citenamefont {Blaizot}, \citenamefont {Gelis},\
  and\ \citenamefont {Venugopalan}}]{Blaizot:2004wv}%
  \BibitemOpen
  \bibfield  {author} {\bibinfo {author} {\bibfnamefont {Jean~Paul}\
  \bibnamefont {Blaizot}}, \bibinfo {author} {\bibfnamefont {Francois}\
  \bibnamefont {Gelis}}, \ and\ \bibinfo {author} {\bibfnamefont {Raju}\
  \bibnamefont {Venugopalan}},\ }\bibfield  {title} {\enquote {\bibinfo {title}
  {{High-energy pA collisions in the color glass condensate approach. 2. Quark
  production}},}\ }\href {\doibase 10.1016/j.nuclphysa.2004.07.006} {\bibfield
  {journal} {\bibinfo  {journal} {Nucl. Phys.}\ }\textbf {\bibinfo {volume}
  {A743}},\ \bibinfo {pages} {57--91} (\bibinfo {year} {2004}{\natexlab{b}})},\
  \Eprint {http://arxiv.org/abs/hep-ph/0402257} {arXiv:hep-ph/0402257 [hep-ph]}
  \BibitemShut {NoStop}%
%%CITATION = HEP-PH/0402257;%%
\bibitem [{\citenamefont {Balitsky}\ and\ \citenamefont
  {Belitsky}(2002)}]{Balitsky:2001mr}%
  \BibitemOpen
  \bibfield  {author} {\bibinfo {author} {\bibfnamefont {I.~I.}\ \bibnamefont
  {Balitsky}}\ and\ \bibinfo {author} {\bibfnamefont {Andrei~V.}\ \bibnamefont
  {Belitsky}},\ }\bibfield  {title} {\enquote {\bibinfo {title} {{Nonlinear
  evolution in high density QCD}},}\ }\href {\doibase
  10.1016/S0550-3213(02)00149-9} {\bibfield  {journal} {\bibinfo  {journal}
  {Nucl. Phys.}\ }\textbf {\bibinfo {volume} {B629}},\ \bibinfo {pages}
  {290--322} (\bibinfo {year} {2002})},\ \Eprint
  {http://arxiv.org/abs/hep-ph/0110158} {arXiv:hep-ph/0110158 [hep-ph]}
  \BibitemShut {NoStop}%
%%CITATION = HEP-PH/0110158;%%
\bibitem [{\citenamefont {Hebecker}\ and\ \citenamefont
  {Weigert}(1998)}]{Hebecker:1998kv}%
  \BibitemOpen
  \bibfield  {author} {\bibinfo {author} {\bibfnamefont {Arthur}\ \bibnamefont
  {Hebecker}}\ and\ \bibinfo {author} {\bibfnamefont {H.}~\bibnamefont
  {Weigert}},\ }\bibfield  {title} {\enquote {\bibinfo {title} {{Small x parton
  distributions of large hadronic targets}},}\ }\href {\doibase
  10.1016/S0370-2693(98)00623-6} {\bibfield  {journal} {\bibinfo  {journal}
  {Phys. Lett.}\ }\textbf {\bibinfo {volume} {B432}},\ \bibinfo {pages}
  {215--221} (\bibinfo {year} {1998})},\ \Eprint
  {http://arxiv.org/abs/hep-ph/9804217} {arXiv:hep-ph/9804217 [hep-ph]}
  \BibitemShut {NoStop}%
%%CITATION = HEP-PH/9804217;%%
\bibitem [{\citenamefont {Kovchegov}\ and\ \citenamefont
  {Tuchin}(2002)}]{Kovchegov:2001sc}%
  \BibitemOpen
  \bibfield  {author} {\bibinfo {author} {\bibfnamefont {Yuri~V.}\ \bibnamefont
  {Kovchegov}}\ and\ \bibinfo {author} {\bibfnamefont {Kirill}\ \bibnamefont
  {Tuchin}},\ }\bibfield  {title} {\enquote {\bibinfo {title} {{Inclusive gluon
  production in DIS at high parton density}},}\ }\href {\doibase
  10.1103/PhysRevD.65.074026} {\bibfield  {journal} {\bibinfo  {journal} {Phys.
  Rev.}\ }\textbf {\bibinfo {volume} {D65}},\ \bibinfo {pages} {074026}
  (\bibinfo {year} {2002})},\ \Eprint {http://arxiv.org/abs/hep-ph/0111362}
  {arXiv:hep-ph/0111362 [hep-ph]} \BibitemShut {NoStop}%
%%CITATION = HEP-PH/0111362;%%
\bibitem [{\citenamefont {Roy}\ and\ \citenamefont
  {Venugopalan}()}]{Roy-Venugopalan-2}%
  \BibitemOpen
  \bibfield  {author} {\bibinfo {author} {\bibfnamefont {Kaushik}\ \bibnamefont
  {Roy}}\ and\ \bibinfo {author} {\bibfnamefont {Raju}\ \bibnamefont
  {Venugopalan}},\ }\href@noop {} {\bibinfo  {journal} {in preparation}\
  }\BibitemShut {NoStop}%
\bibitem [{\citenamefont {Bartels}\ \emph {et~al.}(2001)\citenamefont
  {Bartels}, \citenamefont {Gieseke},\ and\ \citenamefont
  {Qiao}}]{Bartels:2000gt}%
  \BibitemOpen
\bibfield  {journal} {  }\bibfield  {author} {\bibinfo {author} {\bibfnamefont
  {Jochen}\ \bibnamefont {Bartels}}, \bibinfo {author} {\bibfnamefont
  {S.}~\bibnamefont {Gieseke}}, \ and\ \bibinfo {author} {\bibfnamefont
  {C.~F.}\ \bibnamefont {Qiao}},\ }\bibfield  {title} {\enquote {\bibinfo
  {title} {{The $\gamma^{*} \rightarrow q \bar{q}$ Reggeon vertex in
  next-to-leading order QCD}},}\ }\href {\doibase 10.1103/PhysRevD.63.056014,
  10.1103/PhysRevD.65.079902} {\bibfield  {journal} {\bibinfo  {journal} {Phys.
  Rev.}\ }\textbf {\bibinfo {volume} {D63}},\ \bibinfo {pages} {056014}
  (\bibinfo {year} {2001})},\ \bibinfo {note} {[Erratum: Phys.
  Rev.D65,079902(2002)]},\ \Eprint {http://arxiv.org/abs/hep-ph/0009102}
  {arXiv:hep-ph/0009102 [hep-ph]} \BibitemShut {NoStop}%
%%CITATION = HEP-PH/0009102;%%
\bibitem [{\citenamefont {Bartels}\ \emph
  {et~al.}(2002{\natexlab{a}})\citenamefont {Bartels}, \citenamefont
  {Colferai}, \citenamefont {Gieseke},\ and\ \citenamefont
  {Kyrieleis}}]{Bartels:2002uz}%
  \BibitemOpen
  \bibfield  {author} {\bibinfo {author} {\bibfnamefont {J.}~\bibnamefont
  {Bartels}}, \bibinfo {author} {\bibfnamefont {D.}~\bibnamefont {Colferai}},
  \bibinfo {author} {\bibfnamefont {S.}~\bibnamefont {Gieseke}}, \ and\
  \bibinfo {author} {\bibfnamefont {A.}~\bibnamefont {Kyrieleis}},\ }\bibfield
  {title} {\enquote {\bibinfo {title} {{NLO corrections to the photon impact
  factor: Combining real and virtual corrections}},}\ }\href {\doibase
  10.1103/PhysRevD.66.094017} {\bibfield  {journal} {\bibinfo  {journal} {Phys.
  Rev.}\ }\textbf {\bibinfo {volume} {D66}},\ \bibinfo {pages} {094017}
  (\bibinfo {year} {2002}{\natexlab{a}})},\ \Eprint
  {http://arxiv.org/abs/hep-ph/0208130} {arXiv:hep-ph/0208130 [hep-ph]}
  \BibitemShut {NoStop}%
%%CITATION = HEP-PH/0208130;%%
\bibitem [{\citenamefont {Bartels}\ \emph
  {et~al.}(2002{\natexlab{b}})\citenamefont {Bartels}, \citenamefont
  {Gieseke},\ and\ \citenamefont {Kyrieleis}}]{Bartels:2001mv}%
  \BibitemOpen
  \bibfield  {author} {\bibinfo {author} {\bibfnamefont {Jochen}\ \bibnamefont
  {Bartels}}, \bibinfo {author} {\bibfnamefont {S.}~\bibnamefont {Gieseke}}, \
  and\ \bibinfo {author} {\bibfnamefont {A.}~\bibnamefont {Kyrieleis}},\
  }\bibfield  {title} {\enquote {\bibinfo {title} {{The Process $\gamma^{*L} +
  q \rightarrow q \bar{q} g + q$: Real corrections to the virtual photon impact
  factor}},}\ }\href {\doibase 10.1103/PhysRevD.65.014006} {\bibfield
  {journal} {\bibinfo  {journal} {Phys. Rev.}\ }\textbf {\bibinfo {volume}
  {D65}},\ \bibinfo {pages} {014006} (\bibinfo {year} {2002}{\natexlab{b}})},\
  \Eprint {http://arxiv.org/abs/hep-ph/0107152} {arXiv:hep-ph/0107152 [hep-ph]}
  \BibitemShut {NoStop}%
%%CITATION = HEP-PH/0107152;%%
\bibitem [{\citenamefont {Balitsky}\ and\ \citenamefont
  {Chirilli}(2011)}]{Balitsky:2010ze}%
  \BibitemOpen
  \bibfield  {author} {\bibinfo {author} {\bibfnamefont {Ian}\ \bibnamefont
  {Balitsky}}\ and\ \bibinfo {author} {\bibfnamefont {Giovanni~A.}\
  \bibnamefont {Chirilli}},\ }\bibfield  {title} {\enquote {\bibinfo {title}
  {{Photon impact factor in the next-to-leading order}},}\ }\href {\doibase
  10.1103/PhysRevD.83.031502} {\bibfield  {journal} {\bibinfo  {journal} {Phys.
  Rev.}\ }\textbf {\bibinfo {volume} {D83}},\ \bibinfo {pages} {031502}
  (\bibinfo {year} {2011})},\ \Eprint {http://arxiv.org/abs/1009.4729}
  {arXiv:1009.4729 [hep-ph]} \BibitemShut {NoStop}%
%%CITATION = ARXIV:1009.4729;%%
\bibitem [{\citenamefont {Beuf}(2012)}]{Beuf:2011xd}%
  \BibitemOpen
  \bibfield  {author} {\bibinfo {author} {\bibfnamefont {Guillaume}\
  \bibnamefont {Beuf}},\ }\bibfield  {title} {\enquote {\bibinfo {title} {{NLO
  corrections for the dipole factorization of DIS structure functions at low
  $x$}},}\ }\href {\doibase 10.1103/PhysRevD.85.034039} {\bibfield  {journal}
  {\bibinfo  {journal} {Phys. Rev.}\ }\textbf {\bibinfo {volume} {D85}},\
  \bibinfo {pages} {034039} (\bibinfo {year} {2012})},\ \Eprint
  {http://arxiv.org/abs/1112.4501} {arXiv:1112.4501 [hep-ph]} \BibitemShut
  {NoStop}%
%%CITATION = ARXIV:1112.4501;%%
\bibitem [{\citenamefont {Beuf}(2016)}]{Beuf:2016wdz}%
  \BibitemOpen
  \bibfield  {author} {\bibinfo {author} {\bibfnamefont {Guillaume}\
  \bibnamefont {Beuf}},\ }\bibfield  {title} {\enquote {\bibinfo {title}
  {{Dipole factorization for DIS at NLO: Loop correction to the
  $\gamma^*_{T,L}\to q\overline q$ light-front wave functions}},}\ }\href
  {\doibase 10.1103/PhysRevD.94.054016} {\bibfield  {journal} {\bibinfo
  {journal} {Phys. Rev.}\ }\textbf {\bibinfo {volume} {D94}},\ \bibinfo {pages}
  {054016} (\bibinfo {year} {2016})},\ \Eprint
  {http://arxiv.org/abs/1606.00777} {arXiv:1606.00777 [hep-ph]} \BibitemShut
  {NoStop}%
%%CITATION = ARXIV:1606.00777;%%
\bibitem [{\citenamefont {Beuf}(2017)}]{Beuf:2017bpd}%
  \BibitemOpen
  \bibfield  {author} {\bibinfo {author} {\bibfnamefont {Guillaume}\
  \bibnamefont {Beuf}},\ }\bibfield  {title} {\enquote {\bibinfo {title}
  {{Dipole factorization for DIS at NLO: Combining the $q\bar{q}$ and
  $q\bar{q}g$ contributions}},}\ }\href {\doibase 10.1103/PhysRevD.96.074033}
  {\bibfield  {journal} {\bibinfo  {journal} {Phys. Rev.}\ }\textbf {\bibinfo
  {volume} {D96}},\ \bibinfo {pages} {074033} (\bibinfo {year} {2017})},\
  \Eprint {http://arxiv.org/abs/1708.06557} {arXiv:1708.06557 [hep-ph]}
  \BibitemShut {NoStop}%
%%CITATION = ARXIV:1708.06557;%%
\bibitem [{\citenamefont {Hanninen}\ \emph {et~al.}(2017)\citenamefont
  {Hanninen}, \citenamefont {Lappi},\ and\ \citenamefont
  {Paatelainen}}]{Hanninen:2017ddy}%
  \BibitemOpen
  \bibfield  {author} {\bibinfo {author} {\bibfnamefont {H.}~\bibnamefont
  {Hanninen}}, \bibinfo {author} {\bibfnamefont {T.}~\bibnamefont {Lappi}}, \
  and\ \bibinfo {author} {\bibfnamefont {R.}~\bibnamefont {Paatelainen}},\
  }\bibfield  {title} {\enquote {\bibinfo {title} {{One-loop corrections to
  light cone wave functions: the dipole picture DIS cross section}},}\
  }\href@noop {} {\  (\bibinfo {year} {2017})},\ \Eprint
  {http://arxiv.org/abs/1711.08207} {arXiv:1711.08207 [hep-ph]} \BibitemShut
  {NoStop}%
%%CITATION = ARXIV:1711.08207;%%
\bibitem [{\citenamefont {Ducloue}\ \emph
  {et~al.}(2017{\natexlab{a}})\citenamefont {Ducloue}, \citenamefont
  {Hanninen}, \citenamefont {Lappi},\ and\ \citenamefont
  {Zhu}}]{Ducloue:2017ftk}%
  \BibitemOpen
  \bibfield  {author} {\bibinfo {author} {\bibfnamefont {B.}~\bibnamefont
  {Ducloue}}, \bibinfo {author} {\bibfnamefont {H.}~\bibnamefont {Hanninen}},
  \bibinfo {author} {\bibfnamefont {T.}~\bibnamefont {Lappi}}, \ and\ \bibinfo
  {author} {\bibfnamefont {Y.}~\bibnamefont {Zhu}},\ }\bibfield  {title}
  {\enquote {\bibinfo {title} {{Deep inelastic scattering in the dipole picture
  at next-to-leading order}},}\ }\href {\doibase 10.1103/PhysRevD.96.094017}
  {\bibfield  {journal} {\bibinfo  {journal} {Phys. Rev.}\ }\textbf {\bibinfo
  {volume} {D96}},\ \bibinfo {pages} {094017} (\bibinfo {year}
  {2017}{\natexlab{a}})},\ \Eprint {http://arxiv.org/abs/1708.07328}
  {arXiv:1708.07328 [hep-ph]} \BibitemShut {NoStop}%
%%CITATION = ARXIV:1708.07328;%%
\bibitem [{\citenamefont {Kovchegov}\ and\ \citenamefont
  {Weigert}(2008)}]{Kovchegov:2007vf}%
  \BibitemOpen
  \bibfield  {author} {\bibinfo {author} {\bibfnamefont {Yuri~V.}\ \bibnamefont
  {Kovchegov}}\ and\ \bibinfo {author} {\bibfnamefont {Heribert}\ \bibnamefont
  {Weigert}},\ }\bibfield  {title} {\enquote {\bibinfo {title} {{Collinear
  Singularities and Running Coupling Corrections to Gluon Production in
  CGC}},}\ }\href {\doibase 10.1016/j.nuclphysa.2008.04.008} {\bibfield
  {journal} {\bibinfo  {journal} {Nucl. Phys.}\ }\textbf {\bibinfo {volume}
  {A807}},\ \bibinfo {pages} {158--189} (\bibinfo {year} {2008})},\ \Eprint
  {http://arxiv.org/abs/0712.3732} {arXiv:0712.3732 [hep-ph]} \BibitemShut
  {NoStop}%
%%CITATION = ARXIV:0712.3732;%%
\bibitem [{\citenamefont {Ducloue}\ \emph
  {et~al.}(2017{\natexlab{b}})\citenamefont {Ducloue}, \citenamefont {Iancu},
  \citenamefont {Lappi}, \citenamefont {Mueller}, \citenamefont {Soyez},
  \citenamefont {Triantafyllopoulos},\ and\ \citenamefont
  {Zhu}}]{Ducloue:2017dit}%
  \BibitemOpen
  \bibfield  {author} {\bibinfo {author} {\bibfnamefont {B.}~\bibnamefont
  {Ducloue}}, \bibinfo {author} {\bibfnamefont {E.}~\bibnamefont {Iancu}},
  \bibinfo {author} {\bibfnamefont {T.}~\bibnamefont {Lappi}}, \bibinfo
  {author} {\bibfnamefont {A.~H.}\ \bibnamefont {Mueller}}, \bibinfo {author}
  {\bibfnamefont {G.}~\bibnamefont {Soyez}}, \bibinfo {author} {\bibfnamefont
  {D.~N.}\ \bibnamefont {Triantafyllopoulos}}, \ and\ \bibinfo {author}
  {\bibfnamefont {Y.}~\bibnamefont {Zhu}},\ }\bibfield  {title} {\enquote
  {\bibinfo {title} {{On the use of a running coupling in the NLO calculation
  of forward hadron production}},}\ }\href@noop {} {\  (\bibinfo {year}
  {2017}{\natexlab{b}})},\ \Eprint {http://arxiv.org/abs/1712.07480}
  {arXiv:1712.07480 [hep-ph]} \BibitemShut {NoStop}%
%%CITATION = ARXIV:1712.07480;%%
\bibitem [{\citenamefont {Accardi}\ \emph {et~al.}(2016)\citenamefont {Accardi}
  \emph {et~al.}}]{Accardi:2012qut}%
  \BibitemOpen
  \bibfield  {author} {\bibinfo {author} {\bibfnamefont {A.}~\bibnamefont
  {Accardi}} \emph {et~al.},\ }\bibfield  {title} {\enquote {\bibinfo {title}
  {{Electron Ion Collider: The Next QCD Frontier}},}\ }\href {\doibase
  10.1140/epja/i2016-16268-9} {\bibfield  {journal} {\bibinfo  {journal} {Eur.
  Phys. J.}\ }\textbf {\bibinfo {volume} {A52}},\ \bibinfo {pages} {268}
  (\bibinfo {year} {2016})},\ \Eprint {http://arxiv.org/abs/1212.1701}
  {arXiv:1212.1701 [nucl-ex]} \BibitemShut {NoStop}%
%%CITATION = ARXIV:1212.1701;%%
\bibitem [{\citenamefont {Aschenauer}\ \emph {et~al.}(2017)\citenamefont
  {Aschenauer}, \citenamefont {Fazio}, \citenamefont {Lee}, \citenamefont
  {Mantysaari}, \citenamefont {Page}, \citenamefont {Schenke}, \citenamefont
  {Ullrich}, \citenamefont {Venugopalan},\ and\ \citenamefont
  {Zurita}}]{Aschenauer:2017jsk}%
  \BibitemOpen
  \bibfield  {author} {\bibinfo {author} {\bibfnamefont {E.~C.}\ \bibnamefont
  {Aschenauer}}, \bibinfo {author} {\bibfnamefont {S.}~\bibnamefont {Fazio}},
  \bibinfo {author} {\bibfnamefont {J.~H.}\ \bibnamefont {Lee}}, \bibinfo
  {author} {\bibfnamefont {H.}~\bibnamefont {Mantysaari}}, \bibinfo {author}
  {\bibfnamefont {B.~S.}\ \bibnamefont {Page}}, \bibinfo {author}
  {\bibfnamefont {B.}~\bibnamefont {Schenke}}, \bibinfo {author} {\bibfnamefont
  {T.}~\bibnamefont {Ullrich}}, \bibinfo {author} {\bibfnamefont
  {R.}~\bibnamefont {Venugopalan}}, \ and\ \bibinfo {author} {\bibfnamefont
  {P.}~\bibnamefont {Zurita}},\ }\bibfield  {title} {\enquote {\bibinfo {title}
  {{The Electron-Ion Collider: Assessing the Energy Dependence of Key
  Measurements}},}\ }\href@noop {} {\  (\bibinfo {year} {2017})},\ \Eprint
  {http://arxiv.org/abs/1708.01527} {arXiv:1708.01527 [nucl-ex]} \BibitemShut
  {NoStop}%
%%CITATION = ARXIV:1708.01527;%%
\bibitem [{\citenamefont {Boussarie}\ \emph {et~al.}(2014)\citenamefont
  {Boussarie}, \citenamefont {Grabovsky}, \citenamefont {Szymanowski},\ and\
  \citenamefont {Wallon}}]{Boussarie:2014lxa}%
  \BibitemOpen
  \bibfield  {author} {\bibinfo {author} {\bibfnamefont {R.}~\bibnamefont
  {Boussarie}}, \bibinfo {author} {\bibfnamefont {A.~V.}\ \bibnamefont
  {Grabovsky}}, \bibinfo {author} {\bibfnamefont {L.}~\bibnamefont
  {Szymanowski}}, \ and\ \bibinfo {author} {\bibfnamefont {S.}~\bibnamefont
  {Wallon}},\ }\bibfield  {title} {\enquote {\bibinfo {title} {{Impact factor
  for high-energy two and three jets diffractive production}},}\ }\href
  {\doibase 10.1007/JHEP09(2014)026} {\bibfield  {journal} {\bibinfo  {journal}
  {JHEP}\ }\textbf {\bibinfo {volume} {09}},\ \bibinfo {pages} {026} (\bibinfo
  {year} {2014})},\ \Eprint {http://arxiv.org/abs/1405.7676} {arXiv:1405.7676
  [hep-ph]} \BibitemShut {NoStop}%
%%CITATION = ARXIV:1405.7676;%%
\bibitem [{\citenamefont {Boussarie}\ \emph {et~al.}(2016)\citenamefont
  {Boussarie}, \citenamefont {Grabovsky}, \citenamefont {Szymanowski},\ and\
  \citenamefont {Wallon}}]{Boussarie:2016ogo}%
  \BibitemOpen
  \bibfield  {author} {\bibinfo {author} {\bibfnamefont {R.}~\bibnamefont
  {Boussarie}}, \bibinfo {author} {\bibfnamefont {A.~V.}\ \bibnamefont
  {Grabovsky}}, \bibinfo {author} {\bibfnamefont {L.}~\bibnamefont
  {Szymanowski}}, \ and\ \bibinfo {author} {\bibfnamefont {S.}~\bibnamefont
  {Wallon}},\ }\bibfield  {title} {\enquote {\bibinfo {title} {{On the one loop
  $ {\gamma}^{\left(\ast \right)}\to q\overline{q} $ impact factor and the
  exclusive diffractive cross sections for the production of two or three
  jets}},}\ }\href {\doibase 10.1007/JHEP11(2016)149} {\bibfield  {journal}
  {\bibinfo  {journal} {JHEP}\ }\textbf {\bibinfo {volume} {11}},\ \bibinfo
  {pages} {149} (\bibinfo {year} {2016})},\ \Eprint
  {http://arxiv.org/abs/1606.00419} {arXiv:1606.00419 [hep-ph]} \BibitemShut
  {NoStop}%
%%CITATION = ARXIV:1606.00419;%%
\bibitem [{\citenamefont {Boussarie}\ \emph {et~al.}(2017)\citenamefont
  {Boussarie}, \citenamefont {Grabovsky}, \citenamefont {Ivanov}, \citenamefont
  {Szymanowski},\ and\ \citenamefont {Wallon}}]{Boussarie:2016bkq}%
  \BibitemOpen
  \bibfield  {author} {\bibinfo {author} {\bibfnamefont {R.}~\bibnamefont
  {Boussarie}}, \bibinfo {author} {\bibfnamefont {A.~V.}\ \bibnamefont
  {Grabovsky}}, \bibinfo {author} {\bibfnamefont {D.~{\relax Yu}.}\
  \bibnamefont {Ivanov}}, \bibinfo {author} {\bibfnamefont {L.}~\bibnamefont
  {Szymanowski}}, \ and\ \bibinfo {author} {\bibfnamefont {S.}~\bibnamefont
  {Wallon}},\ }\bibfield  {title} {\enquote {\bibinfo {title} {{Next-to-Leading
  Order Computation of Exclusive Diffractive Light Vector Meson Production in a
  Saturation Framework}},}\ }\href {\doibase 10.1103/PhysRevLett.119.072002}
  {\bibfield  {journal} {\bibinfo  {journal} {Phys. Rev. Lett.}\ }\textbf
  {\bibinfo {volume} {119}},\ \bibinfo {pages} {072002} (\bibinfo {year}
  {2017})},\ \Eprint {http://arxiv.org/abs/1612.08026} {arXiv:1612.08026
  [hep-ph]} \BibitemShut {NoStop}%
%%CITATION = ARXIV:1612.08026;%%
\bibitem [{\citenamefont {Chirilli}\ \emph
  {et~al.}(2012{\natexlab{a}})\citenamefont {Chirilli}, \citenamefont {Xiao},\
  and\ \citenamefont {Yuan}}]{Chirilli:2011km}%
  \BibitemOpen
  \bibfield  {author} {\bibinfo {author} {\bibfnamefont {Giovanni~A.}\
  \bibnamefont {Chirilli}}, \bibinfo {author} {\bibfnamefont {Bo-Wen}\
  \bibnamefont {Xiao}}, \ and\ \bibinfo {author} {\bibfnamefont {Feng}\
  \bibnamefont {Yuan}},\ }\bibfield  {title} {\enquote {\bibinfo {title}
  {{One-loop Factorization for Inclusive Hadron Production in $pA$ Collisions
  in the Saturation Formalism}},}\ }\href {\doibase
  10.1103/PhysRevLett.108.122301} {\bibfield  {journal} {\bibinfo  {journal}
  {Phys. Rev. Lett.}\ }\textbf {\bibinfo {volume} {108}},\ \bibinfo {pages}
  {122301} (\bibinfo {year} {2012}{\natexlab{a}})},\ \Eprint
  {http://arxiv.org/abs/1112.1061} {arXiv:1112.1061 [hep-ph]} \BibitemShut
  {NoStop}%
%%CITATION = ARXIV:1112.1061;%%
\bibitem [{\citenamefont {Chirilli}\ \emph
  {et~al.}(2012{\natexlab{b}})\citenamefont {Chirilli}, \citenamefont {Xiao},\
  and\ \citenamefont {Yuan}}]{Chirilli:2012jd}%
  \BibitemOpen
  \bibfield  {author} {\bibinfo {author} {\bibfnamefont {Giovanni~A.}\
  \bibnamefont {Chirilli}}, \bibinfo {author} {\bibfnamefont {Bo-Wen}\
  \bibnamefont {Xiao}}, \ and\ \bibinfo {author} {\bibfnamefont {Feng}\
  \bibnamefont {Yuan}},\ }\bibfield  {title} {\enquote {\bibinfo {title}
  {{Inclusive Hadron Productions in pA Collisions}},}\ }\href {\doibase
  10.1103/PhysRevD.86.054005} {\bibfield  {journal} {\bibinfo  {journal} {Phys.
  Rev.}\ }\textbf {\bibinfo {volume} {D86}},\ \bibinfo {pages} {054005}
  (\bibinfo {year} {2012}{\natexlab{b}})},\ \Eprint
  {http://arxiv.org/abs/1203.6139} {arXiv:1203.6139 [hep-ph]} \BibitemShut
  {NoStop}%
%%CITATION = ARXIV:1203.6139;%%
\bibitem [{\citenamefont {Altinoluk}\ \emph {et~al.}(2016)\citenamefont
  {Altinoluk}, \citenamefont {Armesto}, \citenamefont {Beuf}, \citenamefont
  {Kovner},\ and\ \citenamefont {Lublinsky}}]{Altinoluk:2015vax}%
  \BibitemOpen
  \bibfield  {author} {\bibinfo {author} {\bibfnamefont {Tolga}\ \bibnamefont
  {Altinoluk}}, \bibinfo {author} {\bibfnamefont {N\'{e}stor}\ \bibnamefont
  {Armesto}}, \bibinfo {author} {\bibfnamefont {Guillaume}\ \bibnamefont
  {Beuf}}, \bibinfo {author} {\bibfnamefont {Alex}\ \bibnamefont {Kovner}}, \
  and\ \bibinfo {author} {\bibfnamefont {Michael}\ \bibnamefont {Lublinsky}},\
  }\bibfield  {title} {\enquote {\bibinfo {title} {{Heavy quarks in
  proton-nucleus collisions - the hybrid formalism}},}\ }\href {\doibase
  10.1103/PhysRevD.93.054049} {\bibfield  {journal} {\bibinfo  {journal} {Phys.
  Rev.}\ }\textbf {\bibinfo {volume} {D93}},\ \bibinfo {pages} {054049}
  (\bibinfo {year} {2016})},\ \Eprint {http://arxiv.org/abs/1511.09415}
  {arXiv:1511.09415 [hep-ph]} \BibitemShut {NoStop}%
%%CITATION = ARXIV:1511.09415;%%
\bibitem [{\citenamefont {Iancu}\ \emph {et~al.}(2016)\citenamefont {Iancu},
  \citenamefont {Mueller},\ and\ \citenamefont
  {Triantafyllopoulos}}]{Iancu:2016vyg}%
  \BibitemOpen
  \bibfield  {author} {\bibinfo {author} {\bibfnamefont {E.}~\bibnamefont
  {Iancu}}, \bibinfo {author} {\bibfnamefont {A.~H.}\ \bibnamefont {Mueller}},
  \ and\ \bibinfo {author} {\bibfnamefont {D.~N.}\ \bibnamefont
  {Triantafyllopoulos}},\ }\bibfield  {title} {\enquote {\bibinfo {title} {{CGC
  factorization for forward particle production in proton-nucleus collisions at
  next-to-leading order}},}\ }\href {\doibase 10.1007/JHEP12(2016)041}
  {\bibfield  {journal} {\bibinfo  {journal} {JHEP}\ }\textbf {\bibinfo
  {volume} {12}},\ \bibinfo {pages} {041} (\bibinfo {year} {2016})},\ \Eprint
  {http://arxiv.org/abs/1608.05293} {arXiv:1608.05293 [hep-ph]} \BibitemShut
  {NoStop}%
%%CITATION = ARXIV:1608.05293;%%
\bibitem [{\citenamefont {Caron-Huot}(2015)}]{Caron-Huot:2013fea}%
  \BibitemOpen
  \bibfield  {author} {\bibinfo {author} {\bibfnamefont {Simon}\ \bibnamefont
  {Caron-Huot}},\ }\bibfield  {title} {\enquote {\bibinfo {title} {{When does
  the gluon reggeize?}}}\ }\href {\doibase 10.1007/JHEP05(2015)093} {\bibfield
  {journal} {\bibinfo  {journal} {JHEP}\ }\textbf {\bibinfo {volume} {05}},\
  \bibinfo {pages} {093} (\bibinfo {year} {2015})},\ \Eprint
  {http://arxiv.org/abs/1309.6521} {arXiv:1309.6521 [hep-th]} \BibitemShut
  {NoStop}%
%%CITATION = ARXIV:1309.6521;%%
\bibitem [{\citenamefont {Bondarenko}\ \emph {et~al.}(2017)\citenamefont
  {Bondarenko}, \citenamefont {Lipatov}, \citenamefont {Pozdnyakov},\ and\
  \citenamefont {Prygarin}}]{Bondarenko:2017vfc}%
  \BibitemOpen
  \bibfield  {author} {\bibinfo {author} {\bibfnamefont {S.}~\bibnamefont
  {Bondarenko}}, \bibinfo {author} {\bibfnamefont {L.}~\bibnamefont {Lipatov}},
  \bibinfo {author} {\bibfnamefont {S.}~\bibnamefont {Pozdnyakov}}, \ and\
  \bibinfo {author} {\bibfnamefont {A.}~\bibnamefont {Prygarin}},\ }\bibfield
  {title} {\enquote {\bibinfo {title} {{One loop light-cone QCD, effective
  action for reggeized gluons and QCD RFT calculus}},}\ }\href {\doibase
  10.1140/epjc/s10052-017-5208-8} {\bibfield  {journal} {\bibinfo  {journal}
  {Eur. Phys. J.}\ }\textbf {\bibinfo {volume} {C77}},\ \bibinfo {pages} {630}
  (\bibinfo {year} {2017})},\ \Eprint {http://arxiv.org/abs/1708.05183}
  {arXiv:1708.05183 [hep-th]} \BibitemShut {NoStop}%
%%CITATION = ARXIV:1708.05183;%%
\bibitem [{\citenamefont {Hentschinski}(2018)}]{Hentschinski:2018rrf}%
  \BibitemOpen
  \bibfield  {author} {\bibinfo {author} {\bibfnamefont {Martin}\ \bibnamefont
  {Hentschinski}},\ }\bibfield  {title} {\enquote {\bibinfo {title} {{The Color
  Glass Condensate formalism, Balitsky-JIMWLK evolution and Lipatov's high
  energy effective action}},}\ }\href@noop {} {\  (\bibinfo {year} {2018})},\
  \Eprint {http://arxiv.org/abs/1802.06755} {arXiv:1802.06755 [hep-ph]}
  \BibitemShut {NoStop}%
%%CITATION = ARXIV:1802.06755;%%
\bibitem [{\citenamefont {Lipatov}(1997)}]{Lipatov:1996ts}%
  \BibitemOpen
  \bibfield  {author} {\bibinfo {author} {\bibfnamefont {L.~N.}\ \bibnamefont
  {Lipatov}},\ }\bibfield  {title} {\enquote {\bibinfo {title} {{Small x
  physics in perturbative QCD}},}\ }\href {\doibase
  10.1016/S0370-1573(96)00045-2} {\bibfield  {journal} {\bibinfo  {journal}
  {Phys. Rept.}\ }\textbf {\bibinfo {volume} {286}},\ \bibinfo {pages}
  {131--198} (\bibinfo {year} {1997})},\ \Eprint
  {http://arxiv.org/abs/hep-ph/9610276} {arXiv:hep-ph/9610276 [hep-ph]}
  \BibitemShut {NoStop}%
%%CITATION = HEP-PH/9610276;%%
\bibitem [{\citenamefont {Caron-Huot}\ \emph {et~al.}(2017)\citenamefont
  {Caron-Huot}, \citenamefont {Gardi}, \citenamefont {Reichel},\ and\
  \citenamefont {Vernazza}}]{Caron-Huot:2017zfo}%
  \BibitemOpen
  \bibfield  {author} {\bibinfo {author} {\bibfnamefont {Simon}\ \bibnamefont
  {Caron-Huot}}, \bibinfo {author} {\bibfnamefont {Einan}\ \bibnamefont
  {Gardi}}, \bibinfo {author} {\bibfnamefont {Joscha}\ \bibnamefont {Reichel}},
  \ and\ \bibinfo {author} {\bibfnamefont {Leonardo}\ \bibnamefont
  {Vernazza}},\ }\bibfield  {title} {\enquote {\bibinfo {title} {{Infrared
  singularities of QCD scattering amplitudes in the Regge limit to all
  orders}},}\ }\href@noop {} {\  (\bibinfo {year} {2017})},\ \Eprint
  {http://arxiv.org/abs/1711.04850} {arXiv:1711.04850 [hep-ph]} \BibitemShut
  {NoStop}%
%%CITATION = ARXIV:1711.04850;%%
\bibitem [{\citenamefont {Del~Duca}\ \emph {et~al.}(2018)\citenamefont
  {Del~Duca}, \citenamefont {Druc}, \citenamefont {Drummond}, \citenamefont
  {Duhr}, \citenamefont {Dulat}, \citenamefont {Marzucca}, \citenamefont
  {Papathanasiou},\ and\ \citenamefont {Verbeek}}]{DelDuca:2018hrv}%
  \BibitemOpen
  \bibfield  {author} {\bibinfo {author} {\bibfnamefont {Vittorio}\
  \bibnamefont {Del~Duca}}, \bibinfo {author} {\bibfnamefont {Stefan}\
  \bibnamefont {Druc}}, \bibinfo {author} {\bibfnamefont {James}\ \bibnamefont
  {Drummond}}, \bibinfo {author} {\bibfnamefont {Claude}\ \bibnamefont {Duhr}},
  \bibinfo {author} {\bibfnamefont {Falko}\ \bibnamefont {Dulat}}, \bibinfo
  {author} {\bibfnamefont {Robin}\ \bibnamefont {Marzucca}}, \bibinfo {author}
  {\bibfnamefont {Georgios}\ \bibnamefont {Papathanasiou}}, \ and\ \bibinfo
  {author} {\bibfnamefont {Bram}\ \bibnamefont {Verbeek}},\ }\bibfield  {title}
  {\enquote {\bibinfo {title} {{The seven-gluon amplitude in multi-Regge
  kinematics beyond leading logarithmic accuracy}},}\ }\href@noop {} {\
  (\bibinfo {year} {2018})},\ \Eprint {http://arxiv.org/abs/1801.10605}
  {arXiv:1801.10605 [hep-th]} \BibitemShut {NoStop}%
%%CITATION = ARXIV:1801.10605;%%
\end{thebibliography}
\end{document}